 \documentclass[final,3p,times,authoryear]{elsarticle}

\usepackage{hyperref}
\hypersetup{colorlinks=true,citecolor=blue}
\usepackage{amssymb}
\usepackage{graphicx,amsmath,multirow}
\usepackage{ifthen}
\usepackage{endnotes}
\usepackage{sidecap}
\usepackage{breakurl}
\usepackage{rotating}
\usepackage{color}
\def\beq{\begin{equation}}
\def\eeq{\end{equation}}
\def\bea{\begin{eqnarray}}
\def\eea{\end{eqnarray}}
\def\nn{\nonumber}

\def\degC{$^{\circ}$C$\,$}

\biboptions{round,comma,sort&compress}

\journal{Climatic Change}

\begin{document}

\begin{frontmatter}

\title{Integrated assessment modelling as a positive science: private passenger road transport policies to meet a climate target well below 2\degC \\ }

\author[RU,CEENRG]{Jean-Fran\c{c}ois Mercure \corref{cor1}}
\ead{J.Mercure@science.ru.nl}
\cortext[cor1]{Corresponding author: Jean-Fran\c{c}ois Mercure}
\author[UM,CEENRG]{Aileen Lam}
\author[CE]{S. Billington}
\author[CE]{H. Pollitt}

\address[RU]{Department of Environmental Science, Radboud University, PO Box 9010, 6500 GL Nijmegen, The Netherlands}
\address[UM]{Department of Economics, Faculty of Social Sciences, Humanities and Social Science Building, University of Macao, E21, Avenida da Universidade, Taipa, Macao, China}
\address[CEENRG]{Cambridge Centre for Environment, Energy and Natural Resource Governance (C-EENRG), University of Cambridge, 19 Silver Street, Cambridge, CB3 1EP, United Kingdom}
\address[CE]{Cambridge Econometrics Ltd, Covent Garden, Cambridge, CB1 2HT, UK}

\begin{abstract}

Transport generates a large and growing component of global greenhouse gas emissions contributing to climate change. Effective transport emissions reduction policies are needed in order to reach a climate target well below 2\degC. Representations of technology evolution in current Integrated Assessment Models (IAM) make use of systems optimisations that may not always provide sufficient insight on consumer response to realistic policy packages for extensive use in policy-making. Here, we introduce FTT:Transport, an evolutionary technology diffusion simulation model for road transport technology, as an IAM sub-component, which features sufficiently realistic features of consumers and of existing technological trajectories that enables to simulate the impact of detailed climate policies in private passenger road transport. Integrated to the simulation-based macroeconometric IAM E3ME-FTT, a plausible scenario of transport decarbonisation is given, defined by a detailed transport policy package, that reaches sufficient emissions reductions to achieve the 2$^\circ$C target of the Paris Agreement.

\end{abstract}


\end{frontmatter}

\setcounter{tocdepth}{2}
\tableofcontents
\newpage

\part{Main Article}
\section{Introduction}

Road transport emits 17\% of global greenhouse gas (GHG) emissions, a flow of carbon that has grown historically by 2-3\% every year over the past 20 years \citep{IEACO22015}. Transport also uses a major proportion of oil produced worldwide: 48\% of global oil extraction powers one form or another of motorised road transport \citep{IEAWEB2016}.\footnote{158~EJ of oil was produced in 2013 and transformed into many products of which 42~EJ was gasoline and 51~EJ was diesel, of which 34~EJ and 42~EJ were used in road transportation. } While developed economies (e.g. USA, Japan) typically have low transport activity growth, middle-income nations (e.g. Brazil, China, India) have fast growth rates \citep{Euromonitor}. Policy for transforming the environmental impact of transport is a key area to model in detailed Integrated Assessment Models (IAMs). 

Traditionally, IAMs with high detail in energy end-use technologies have been based on system cost-optimisation or maximisation of the utility of the representative agent\footnote{Cost-optimisation models include PRIMES\citep{PRIMES}, MESSAGE \citep{MESSAGE}, REMIND \citep{REMIND}, AIM-enduse \citep{AIM-enduse}, TIMES \citep{TIMES}, TIAM \citep{TIAM}, GET \citep{GET}; utility maximisation (general equilibrium) models include GEM-E3 \citep{GEM-E3}, IMACLIM \citep{IMACLIM}, GEMINI \citep{GEMINI}; bottom-up technology models based on discrete choice theory include IMAGE/TIMER \citep{IMAGE}, IMACLIM \citep{IMACLIM}, CIMS \citep[][]{Rivers2006}. Simpler IAMs such as DICE \citep{DICE} and FUND \citep{FUND} are also optimisations.}. The optimisation methodology used in IAMs is useful from a normative perspective as it helps map out feasible space and determine what are desirable configurations from a societal point of view \citep[e.g. see][]{ETSAP}. For instance, optimisation can be useful in a context of agenda setting. The carbon price is typically used as a \emph{control} parameter that internalises the climate externality, which moves the solution in technology space towards decarbonisation. 

However, optimisation interpreted in a strictly positive scientific sense implies assuming consumers with infinite information about the whole system and no preferences tied to the social context. In that work philosophy, such a representation may be deficient, as it seems unlikely, from a behavioural science point of view, that choices of consumers could be incentivised and coordinated by the chosen policy signal (the externality price) in exactly the way that results from an optimisation calculation \citep{Mercure2016}. Optimisations interpreted as positive descriptions may not be reliable for use for impact assessments of policy scenarios, particularly if the modelled behaviour of agents is not sufficiently well informed. 

There are two major issues with current optimisation-based IAMs \citep{Wilson2015, McCollum2016, Pettifor2017a, Mercure2016}:
\begin{enumerate}
\item Many IAMs are employed using typically one single policy lever for decarbonisation: the carbon price (through assumed emissions trading), which is applied to all emitting sectors including road transport. Real-world climate policy, however, features a much richer diversity of sector-specific incentives, particularly in transport, where carbon pricing is generally not used.
\item The collective response of agents to policy incentives (and their degree of access to/interest in reliable relevant information) is assumed to be coordinated in such a way that a system cost minimum or utility maximum is realised. In the real world, however, agents are far from being coordinated in a total system cost perspective, but instead, act according to specific behavioural features that do not usually feature in IAMs. For instance, no real decision-maker anywhere faces the global energy system cost and related presumed trade-offs.
\end{enumerate}

The question we ask then is, what kind of methodology could solve these problems, that could be used at the scale of IAMs? Would using a different model structure enable to model more detailed and multiple policy instruments, including their interactions? Can we make model projections more consistent with recent technology diffusion data? To address these questions, we introduce a new type of evolutionary model that simulates the diffusion of transport technology, FTT:Transport, as a sub-module of the IAM named E3ME-FTT-GENIE \citep[see][ for details of the IAM itself]{Mercure2018, Mercure2018b}. FTT models the diffusion of innovations calibrated on recent diffusion data and observed cost-distributions as a representation of consumer heterogeneity. It offers a highly detailed set of possible policy packages. Its strong path-dependence and high policy resolution allows to assess policy interactions explicitly, with a modelling horizon of 2050. 

In section \ref{sect:model}, we summarise the theoretical background and empirical basis of the model. In section \ref{sect:results}, we show plausible endogenous projections of low-carbon vehicle diffusion as a result of specific transport policies for fast decarbonisation consistent with a target well below 2\degC. We conclude with a methodological recommendation for policy-relevance. We provide a detailed model description and its parameterisation in the Supplementary Information (SI).

\section{Background, model and method \label{sect:model}}

\subsection{Behavioural information}

Work is now developing to improve behavioural representations in IAMs \citep{Wilson2015, McCollum2016, Pettifor2017a, Pettifor2017b}. However, in order to effectively inform policy-making, it is also crucial to clearly delineate normative (i.e. ``tell me what are the components and I will tell you the best way to organise the system") from positive (i.e. ``tell me the context and I will predict what people will choose") modelling philosophies. 

\begin{figure*}[t]
	\begin{center}
		\includegraphics[width=1\columnwidth]{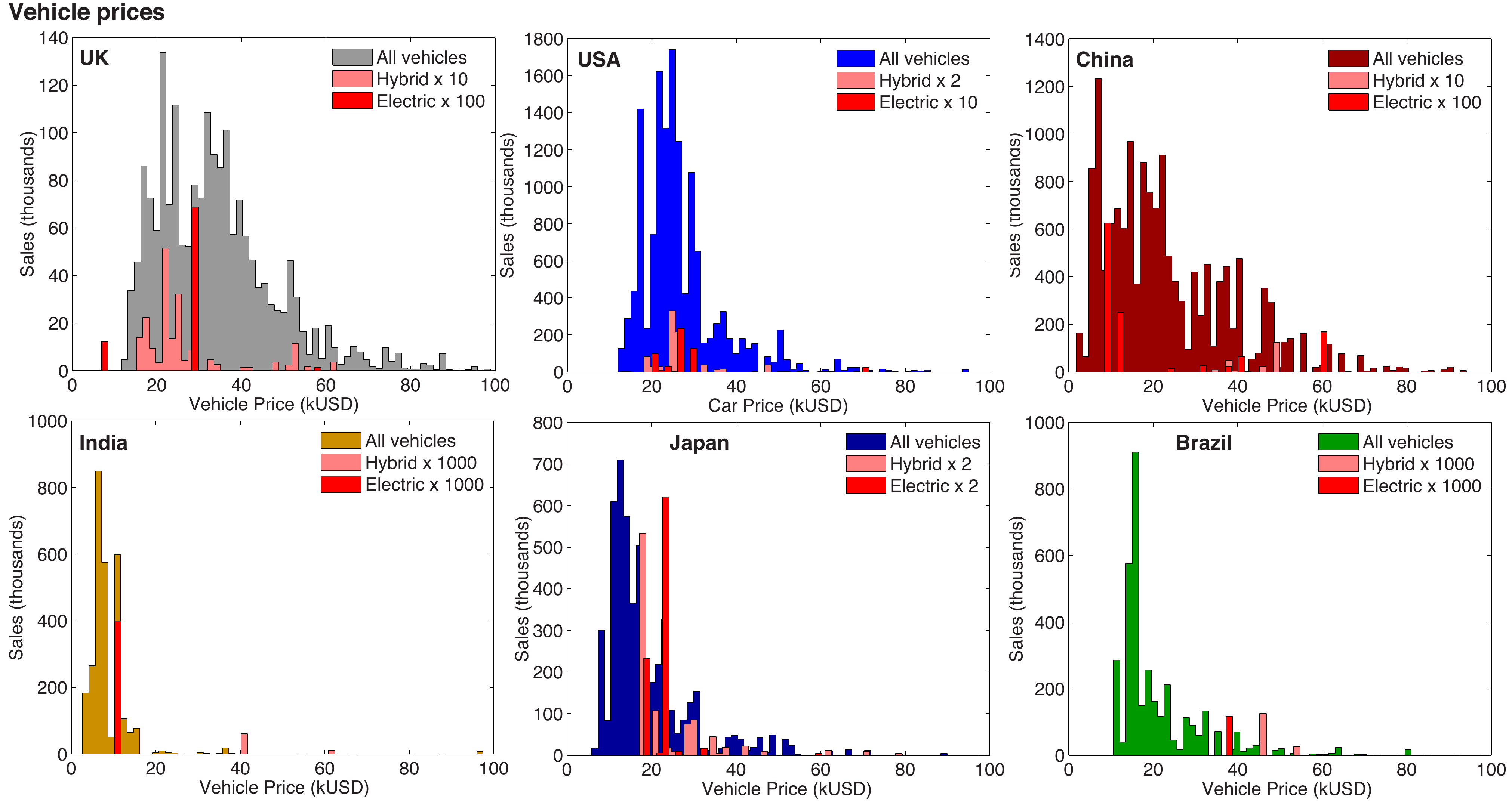}
	\end{center}
	\caption{Price data for vehicles in six major economies, reproduced from \cite{MercureLam2015}.}
	\label{fig:Figure1}
\end{figure*}

Of interest here, passenger road transport is not normally covered by a carbon price, but many other policy types are used \citep[regulatory, push and pull policies, see e.g. ][]{ICCT2011}. The `cost' of vehicles as mitigation options in the traditional modelling sense is not very well defined since the (lognormal) frequency distribution of vehicle prices spans a range often much larger than its average \citep[see figure~\ref{fig:Figure1} and the data in ][and more data in SI section~5.1]{MercureLam2015}. The heterogeneity of vehicle consumers is large. 

To better understand this requires using tools and knowledge from behavioural economics as well as marketing research, which has been largely overlooked by IAM modellers. \cite{Wilson2015, McCollum2016, Pettifor2017a, Pettifor2017b} make a compelling argument for the inclusion of significantly more behaviourally relevant information and functionality in existing IAMs, including a particular emphasis on heterogeneity, social influence, and the number of policy instruments represented. This led to the development of a new behaviourally rich model \cite{Pettifor2017a,Pettifor2017b}. This plea applies to markets for private vehicles, where the heterogeneity of consumers is high \citep{MercureLam2015} and social influence dynamics, typically not modelled, may well have as much explanatory power as prices \citep[e.g. ][]{McShane2012, Wilson2015, Pettifor2017b}. However, existing technological trajectories are also important to consider, due to their momentum \cite[see e.g.][]{Geels2005}. 

Whether agents are believed to minimise costs or not may not be the issue to resolve: the result of agents individually optimising their costs and benefits \emph{does not} necessarily lead to a cost optimum at the system scale, i.e. to an optimising representative agent/social planner \citep{Kirman1992}.\cite{Mercure2016, Mercure2018c} show that, in a quantitative (as well as qualitative) social theory, as soon as agents interact with one another and \emph{value} the behaviour of other agents when making consumption decisions, fads, fashions, band-wagon effects arise that violate the premise of systems optimisation. These effects break the connection between cost/utility-optimisation at the individual agent level, and optimisation at system level, partly because the representative agent cannot exist.\footnote{In other words, if agents behave following their peers to any degree, the aggregate macro-behaviour of a group of cost/utility-optimising agents does not result in an optimum at the system level (and conversely), since that optimal point ceases to exist.} Multi-agent influence instead \emph{leads to diffusion dynamics} (as in \citealt{Rogers2010}; see also the work on information contagion by \citealt{Arthur1993, Lane1997}). 

\cite{Wilson2015} and \cite{Pettifor2017b} review an extensive body of knowledge on social influence in vehicle choice. It generates a system in which the state of diffusion is not a simple function of input prices, but instead depends on the order of past events \cite[path-dependence, see][]{Arthur1989,Arthur1987}. In systems optimisation, such increasing returns generates multiple solutions and model instabilities \citep{Gritsevskyi2000}: which ones are the `correct' optimal outcomes? Broadly speaking, social influence `attracts' agents towards the adoption of popular innovations and away from unpopular ones, despite absolute costs and benefits \citep[][]{Arthur1989,Arthur1993,Brock2001,Brock2001b}\footnote{We acknowledge that heterogeneity in principle includes varying characteristics across adopter groups, which also means that receptivity to social influence is itself heterogenous \cite[see e.g.][]{Donnelly1974}. This is, however too detailed for the present study.}. The value ascribed by consumers to the choices of others can be as influential to their choices as  the sum of the characteristics of the goods themselves. 

This effect is not only generated by social influence: if one allows that the availability of technology to agents may be restricted by existing market shares (availability follows the size of the industry), which grow with technology diffusion (the more technologies diffuse, the more agents can access them), then market shares partly determine the pace of diffusion, a recursive problem of the same form. In this interpretation, each agent has a different set of knowledge of technology options, stemming from access to a different set of options. This makes the use of a representative agent impossible. Technology producers expand production capacity following demand growth, and demand grows following technology availability. In fact, in a model, it is not straightforward to empirically attribute the effects of social influence, technology diffusion and industry growth dynamics. Whichever the root source, these dynamical effects are mathematically crucial to represent correctly, as they determine whether a model assumes that agents behave in a perfectly coordinated way or not (i.e. whether a representative agent exists or not).

It is therefore potentially insufficient to better parameterise basic optimisation models with additional consumer behavioural information, if the methodology remains tied to optimisation at the system level, which precludes bandwagon effects by construction. It is noteworthy that the same recursive effects arise in animal population ecology \citep{Kot2001}, and in fact one finds that evolutionary modelling methods can achieve realistic consumer representations with behavioural heterogeneity and social influence \citep{Young2001}, without the use of representative agents. We describe one here, deployed at IAM scale.

\subsection{Technology diffusion as band-wagon effects}

The FTT model uses a modified version of discrete choice theory in the form of an evolutionary theory. It uses observed distributions of costs to represent agent heterogeneity (a form of revealed preferences). It is based a on dynamical systems approach as opposed to systems optimisation, minimising perceived costs at a bounded rational agent level as opposed to minimising whole system costs in standard models. After providing some theoretical background, we describe the methodology in this section.

Discrete choice theory \citep[DCT,][]{Anderson1992, Domencich1975, Ben-Akiva1985} is the main workhorse to regress choice by non-interacting heterogeneous agents. Agents in a DCT model are assumed to have knowledge of, and have access to all options available in the market (perfect information). The resulting multinomial logit (MNL) can also be mathematically derived from a problem of utility maximisation under a budget constraint when utility follows a constant elasticity of substitution (CES) model (\citealt{Anderson1992}), where the elasticity is related to the heterogeneity of agents. MNLs, CES and optimisation models\footnote{Computable general equilibrium (CGE) and cost-optimisation (partial equilibrium).} thus share a common theoretical foundation, in which agents do not interact with one another, and base choices on infinitely detailed information. These models do not endogenously generate diffusion profiles consistent with what is observed \citep[S-shaped curves, e.g. see][]{Nakicenovic1986, Grubler1999}, unless externally constrained to (e.g. by just the right carbon price\footnote{Many partial equilibrium (cost-optimisation) models can be described as `moving equilibrium' models  \citep[][]{Young2009}, in which diffusion is driven by appropriately chosen external parameters producing the right profiles.}).

In diffusion problems, it is specifically the case that agents \emph{do not} have (or wish to have) access to or knowledge of all existing options in the market, since some options are largely unknown/untried innovations with small market shares and small production capacities, and thus limited access. Production capacity for new technologies are not expanded instantaneously to respond to changes in consumer demand; they co-evolve over time. Widely used products have a higher capacity for diffusion, as they are more visible and they have a larger producing industry \citep{Bass1969, Fisher1971, Mansfield1961, Sharif1976, Marchetti1978}. These properties are core elements of innovation diffusion theory \citep{Rogers2010}.\footnote{These are also common concepts in marketing research carried out by firms placing products, going back to \cite{Smith1956,Bass1969}.} Band-wagon effects are a key component of transitions theory \citep[e.g. ][]{Geels2002, Geels2005, Rotmans2001, Turnheim2015}, and Agent-Based Models \citep[ABM, see][]{Kohler2009, Holtz2011, Holtz2015}.

Including interactions between agents (agents learning from each other, i.e. social influence) in a discrete choice model leads to diffusion dynamics of products in markets (\citealt{Mercure2018c}, see also \citealt{Arthur1993}; and SI section 3.3). Here, we use the so-called `replicator dynamics', a mathematical system used in evolutionary theory to describe the selection process in evolutionary problems \citep[evolutionary game theory, ][]{Hofbauer1998}, derived in detail in SI section~3.3, summarised here (see \citealt{Mercure2018c,Mercure2015,Safarzynska2010}, and \citealt{Young2009, Young2001} for broader discussions). Its dynamical behaviour is consistent with empirical diffusion observations \citep[e.g.][]{Mansfield1961, Fisher1971, Nakicenovic1986}. 

\subsection{A bounded-rational discrete choice model with heterogenous agents}

Consumers in vehicle markets are highly heterogenous, and this heterogeneity varies by country, shown in fig.~\ref{fig:Figure1} (see SI section~5.1 and \citealt{MercureLam2015}). This heterogeneity can be observed, amongst many other ways,\footnote{See for instance \cite{Aini2013, Baltas2013}} through differentiated prices, which typically increase exponentially with linearly increasing engine sizes (vehicle power, \emph{ibid}). Taking account of this heterogeneity is crucial in models to quantify the impact of pricing policies on rates of adoption (e.g. subsidies). Indeed, if the distribution of prices spans an order of magnitude, then purchase and/or fuel tax schemes will generate widely different levels of incentives in different market segments, and the diffusion of new technologies often starts in more affluent segments of the population. This can be modelled by using distributed variables. 

In this work, heterogeneity is `observed' from the market (Fig.~\ref{fig:Figure1}) because markets, consumers and regulation co-evolve: entrepreneurs strive to better match the differentiated tastes of consumers, while consumer tastes are influenced by how the market evolves. Observed distributions of prices reflect consumer taste heterogeneity,  related to a myriad of socio-economic contextual variables (income, geography, culture, etc), which change over time.

It is not necessary here to track every individual agent or agent type in order to represent heterogeneity: DCT statistics can be used. ABMs do so, but using DCT is computationally faster. However, in our bounded rational model, agents do not know every vehicle model type in the market (i.e. we reject perfect information)\footnote{E.g. in \cite{MercureLam2015}, we reviewed the characteristics of over 8000 different individual vehicle models in registration data for the UK, from \cite{DVLA}, and we are fully convinced from that experience that consumers in the UK do not carry out such an exhaustive search when choosing a vehicle.} but, rather, consumers choose within various subsets of the market. This means that every agent has a different set of knowledge, which violates the premise of the standard MNL. Instead, modelling this is done using chains of binary logits, with pair-wise comparisons of options, each \emph{weighted according to the number of agents} carrying out these comparisons. These weights are the market shares of each vehicle type, reflecting the probabilities of consumer learning events, for example through visual influence \citep[as in][]{McShane2012}.

In such chains of binary logits, agent preferences between pairs are treated as distributions of the perceived costs and benefits of technologies (the generalised cost), and compared at every time step of the model. When faced with a choice between vehicle categories $i$ and $j$, a fraction of agents making the choice will prefer technology $i$, denoted $F_{ij}$, while the rest will prefer $j$, denoted $F_{ji}$, where $F_{ij} + F_{ji} = 1$. Denoting that option $i$ is perceived by that subset of agents to have a generalised cost $C_i$ that follows a frequency distribution $f_i(C-C_i)$, and cumulative distribution $F_i(C-C_i)$, with mean $C_i$ and standard deviation $\sigma_i$ (and similarly for option $j$), the fraction of agents making the choice preferring $i$ over $j$ is:
\beq
F_{ij}(\Delta C_{ij}) = \int_{-\infty}^\infty F_j(C) f_i(C-\Delta C_{ij}) dC, \quad \Delta C_{ij} = C_i-C_j,
\eeq
which, if $f_i$ is a double exponential Gumbel distribution (as in standard DCT), yields the classic binary logit \citep[see][]{Domencich1975}. The standard deviation is treated using the standard error propagation method:
\beq
F_{ij} = {1 \over 1 + \exp\left(\Delta C_{ij} / \sigma_{ij}\right) }, \quad \sigma_{ij} = \sqrt{\sigma_i^2 + \sigma_j^2}.
\eeq
This is a logistic function of the ratio of the mean cost difference to the width of the price distribution (SI section~3.2). Any noticeable changes in aggregate preferences requires any perceived cost difference to be larger than the combined standard deviations. This is how, in choice models, rates of diffusion relate to heterogeneity, and is one way to model heterogenous agents that cost-minimise individually, within their context, under social influence, without using any systems optimisation algorithm. Price distributions such as in fig.~\ref{fig:Figure1} are used for parameterising $f_j(C)$. FTT is thus parameterised by cross-sectional datasets (SI section~5.1). 

\subsection{The replicator dynamics equation of evolutionary theory}

We take $S_i$ as the market share of option $i$ (the number of units of type $i$ in the fleet, with respect to the total). We evaluate exchanges of market shares between technology categories as time goes by, the magnitude of which is determined by preferences $F_{ij}$, while the rate originates from the fleet turnover. At each time step $dt$, the amount of shares flowing away from category $i$ into category $j$ is proportional to the number of vehicles of type $i$ requiring replacement, itself proportional to the market share $S_i$. The number of agents replacing vehicles of type $i$ exploring the possibility of purchasing a vehicle of type $j$ is a subset of all agents who have access or have reliable knowledge of option $j$, which is proportional to the market share of option $j$ \citep[see][]{Mercure2015}. Being probabilistic, shares flow simultaneously in opposite directions but with typically unequal magnitude (if preferences are exactly 50\%/50\%, then the net flow is zero). The expression that results for the net flow is the replicator dynamics equation (also called Lotka-Volterra, SI section~3.3):
\beq
\Delta S_{j \rightarrow i} = S_i S_j {F_{ij} \over \tau_i} \Delta t,\quad \Rightarrow \quad \quad {dS_i \over dt} = \sum_j S_i S_j \left( {F_{ij} \over \tau_i} - {F_{ji} \over \tau_j} \right).
\label{eq:replicator}
\eeq
This is a dynamical equation that is path-dependent and hysteretic \citep{Mercure2018c}.\footnote{Meaning that shares are not single valued functions of perceived costs, they also depend on configurational history: many sets of $S_i$ can occur with each set of $C_i$, depending on what $S_i$ and $C_i$ have been in the past.} Costs and policy incentivise agents to make choices that orient the trajectory of diffusion, and the trajectory has momentum.\footnote{By momentum we mean that the system has some degree of inertia that prevents it from changing direction very rapidly.} Costs are influenced by learning curves, typically stronger for new technologies, reinforcing diffusion and path-dependence. The mathematics describe a system in perpetual flow without equilibrium, and indeed, problems of technology diffusion do not have steady states.\footnote{This can be expressed in a myriad of ways, e.g. from the network structure of technology evolution \citep{Grubler1998}, from an evolutionary perspective \citep{Young2001, Young2009, Hofbauer1998}, from the scaling dynamics of innovation \citep{Arthur2006}, or from the presence of multi-agent interactions \citep{Mercure2018c}.} Innovations come and go, as the popularity of novelty products rises and later declines. This equation is derived in detail in two distinct ways in the SI section 3.3.

\subsection{Cost distributions database and micro-model of vehicle consumer choice}

Price distributions for private vehicles are typically log-normally distributed \citep[see fig.~\ref{fig:Figure1},][and SI section 5.1]{MercureLam2015}. Cost-comparisons in the FTT binary logit are thus made between cost distributions in logarithmic space, using an appropriate transformation (SI section 3.2). Consumer decisions are not made solely based on vehicle prices; future operation and maintenance costs are taken into account, with a discount rate, as well as non-pecuniary benefits. It can never be fully clear what intuitive or quantitative evaluations are carried out by vehicle consumers when taking decisions (and evaluation methods may differ across the population). For modelling tractability, we require a suitably general, statistical and flexible micro-model that can encompass all sorts of heterogenous behaviour. We use comparisons of the net-present values in log scaling, which we denote as the Levelised Cost Of Transportation $LCOT$. It expresses a discounted cost of generating a unit of transport service:\footnote{In dollar per person-kilometre (\$/pkm).}
\beq
\log \left[ \sum_{t=0}^\tau {I_i + VT_i + CT(\alpha_i) + Fu_i(t)FT(\alpha_i,t) + MR_i + RT_i(t) \over (1+r)^t} \bigg/  \sum_{t=0}^\tau 1/ (1+r)^t \right] +\gamma_i, 
\label{eq:LCOT}
\eeq
where time $t$ refers to moments in a hypothetical future at which agents expect costs to take place during vehicle type $i$'s lifetime $\tau$ (i.e. not real time), $r$ is the consumer discount rate, $I_i$ is the vehicle price, $VT_i$ is a vehicle specific one-off registration tax rate, $CT(\alpha_i)$ is a registration tax based on the fuel economy $\alpha_i$, $Fu_i$ is the expected fuel costs, $FT$ is the fuel tax, $MR_i$ is repair costs and $RT_i$ is a yearly road tax. The $LCOT_i$ is the mean of the combined distributions of these cost components,\footnote{Every term is distributed; however the distribution of car prices dominates variations.}. It is paired with its standard deviation $\Delta LCOT_i$, calculated using the root of the sum of the squares of all variations. Phase-out regulations are approximated by setting $F_{ij} = 0$, i.e. overriding consumer choices, preventing further sales of a particular vehicle category (see SI section~3.5 for details on our policy representations).

The costs explicitly represented in the above equation are not sufficient to realistically model technology diffusion, since many other pecuniary and non-pecuniary costs are valued by agents, as we find empirically, for which we have no explicit data, to explain observed technological trajectories. An adjustment to this equation is necessary in order for FTT:Transport to match diffusion trajectories observed in recent years (see our global historical diffusion database, SI section~5.2). Since FTT is a path-dependent simulation, its formulation would be inconsistent if it suggested a change of diffusion trajectory at the start of the simulation. Indeed, to be self-consistent, historical data \emph{must} determine the diffusion trajectory in the first few years or decade of the simulation.

An additional parameter is determined empirically, $\gamma_i$, which represents all unknown constant pecuniary and non-pecuniary cost components, and policies in place, that are not explicitly represented or included in eq.~\ref{eq:LCOT}, needed to match the modelled diffusion trajectory to the observed trajectory, in order to ensure consistency with diffusion theory.  $\gamma_i$ has the unique value set that makes the diffusion rate ($dS_i/dt$) continuous across the transition from historical data to simulated data for $S_i$ at the start year of the simulation. $\gamma_i$ is determined with a methodology described in SI section~5.5. As with econometric parameters, $\gamma_i$ is assumed not to change over the simulation period. This is not necessarily fully satisfactory; however, there exists no reliable scientific basis upon which to predict distant future changes in $\gamma_i$, which we consider best of current knowledge. 

\subsection{The FTT:Transport database}

Data gathering for the FTT:Transport vehicle price database is described in detail in \cite{MercureLam2015} and SI section~5. Light duty vehicle types were classified as petrol and diesel, compressed natural gas (CNG), hybrid, electric vehicles (EV) and motorcycles. Each category was sub-divided into three consumer classes: economic (\emph{Econ}, below 1400cc), mid-range (\emph{Mid}, between 1400cc and 2000cc) and large engine vehicles (\emph{Lux}, above 2000cc),\footnote{We show in \cite{MercureLam2015} that engine sizes strongly relate to prices, hence this classification.} each of which has its own vehicle price distribution as an explicit representation of agent heterogeneity (see SI section~2 for detailed UK data). We stress that it is not the engine size classification that we ascribe to heterogeneity, but rather, the fact that prices are distributed, whereas the engine size classification mainly serves presentational purposes. Motorcycles were divided as either above or below 125cc. Hypothetical future higher efficiency vehicle categories are added using scenario defined fuel efficiencies based on current targets.\footnote{Due to lack of detailed or reliable shares data worldwide, plug-in hybrids are not represented explicitly but are instead lumped together with EVs. Due to lack of shares and cost data, fuel cell vehicles are not currently included but may be included in the future as a dominant design forms and reliable costs can be obtained.}

2012 data for new registrations per vehicle model type were obtained from either national statistics or from \cite{Marklines} and matched, model by model, to recent prices obtained online \citep{MercureLam2015}. Vehicle price data were matched to sales numbers for 18 representative regions, used as proxies for 53 out of E3ME's 59 regions based on economic and regional similarities, following data availability. Data for other countries were used by proxy based on market similarities (SI section~5). Historical total yearly distances driven nationally and total numbers of vehicles registered in national fleets were obtained from \cite[][]{Euromonitor, Eurostat2015}. Historical shares per vehicle category for 53 E3ME regions were obtained by merging several datasets \citep{Euromonitor, Marklines, Eurostat2015}, and cover 2004 to 2012, while total fleet sizes and yearly sales cover 1990 to 2012 (detailed procedure given in SI section~5.2, the historical data itself provided separately in the Suppl. Excel data file). 

\subsection{Projecting vehicle sales, fuel use and emissions with E3ME}

FTT:Transport is built as a sub-module of E3ME \citep[see][]{E3MEManual}, itself able to calculate global emissions and coupled to the climate model GENIE1 \citep{Holden2013}, making it a fully detailed IAM \citep[see][for a full model description]{Mercure2018,Mercure2018b}. E3ME is a non-equilibrium macroeconometric simulation model based on a demand-led Post-Keynesian structure \citep{Pollitt2017}, theoretically coherent with the evolutionary simulation basis of FTT. The degree to which vehicles are used is assumed not to depend strongly on their types of engines, and is calculated by regressing total vehicle use (in veh-km/y) with respect to fuel prices and income, and projecting these to 2050, using fuel prices and income endogenously determined by E3ME. The number of vehicles purchased does not strongly depend on vehicle type composition of the fleet, and thus vehicle sales are regressed and projected against income and average vehicle prices, the first endogenously determined by E3ME.\footnote{These regressions did not include variables such as the extent of road network, congestion, urban vs rural population ratios, omitted due to the difficulty of obtaining such data consistently for 59 E3ME regions worldwide. Note that these variables change only slowly over time and therefore would not significantly improve the reliability of our parameters.} Elasticities from the literature were used to constrain regression parameters and avoid spurious results. Fleet sizes are calculated using projected sales and a survival function derived from \cite{DVLA} data (SI section~4.3).\footnote{FTT calculates vintage effects due to the age of vehicles and the fact that fuel efficiencies were lower in the past, based on evidence from our UK dataset \citep{DVLASurvey}.} 

Resulting demand profiles vary substantially across regions. As a general rule, fast growing economies with fast growing fleets (e.g. China, India, Brazil) have a higher response to price changes than slow-growing developed economies where fleets don't grow (e.g. UK, USA), which applies to both the demand for vehicles and the demand for travel (SI section~5.4).

FTT is fully integrated to E3ME with several dynamical feedbacks to the global economic simulation. In E3ME, income, prices, fuel use, investment, employment, and more quantities are calculated endogenously globally, in 59 regions, 70/44 sectors (EU/non-EU countries), 23 fuel users and 12 fuels. E3ME calculates global fuel use and combustion emissions, where fuel use for electricity generation is simulated using the sister model FTT:Power. Thus, the combination of FTT:Power, FTT:Transport, FTT:Heat and E3ME provides a relatively high definition dynamical coverage of global fossil fuel use and emissions. Disposable income is calculated based on wages, GDP, price levels and employment. Fuel prices are derived from endogenous dynamical fossil fuel depletion and cost calculations \citep[see our model in][]{MercureSalas2013}. Fuel use from road freight transport is accounted for, but there technological change is not modelled in as much detail; biofuel mandates form the main freight decarbonisation mechanism (see SI section 3.4).

\subsection{Summary of improvements over incumbent models}

We summarise here the novel improvements that FTT:Transport provides over standard methods:
\begin{enumerate}
	\item {\it FTT endogenously projects current diffusion trends with a path-dependent diffusion profile (S-shaped);} 
	\item {\it Diffusion is driven by choices of endogenously modelled heterogenous consumers under bounded rationality and social influence, not a representative consumer;} 
	\item {\it The diffusion trajectory is tied to recent historical data but does not strongly depend on technological assumptions;}
	\item {\it A bounded-rational choice framework enables to model many forms of policy instruments and composite packages (currently 8 different policy levers are implemented), and strong policy interaction is observed;}
	\item {\it Diffusion trends cannot be made discontinuous by a sudden change or break in the policy regime, due to endogenous diffusion inertia.} 
\end{enumerate}

\section{Policy strategy and model results \label{sect:results}}

\subsection{Policies for decarbonising private personal transport}

Policies for transport decarbonisation currently take four forms: (1) improving the efficiency of conventional ICE vehicles, (2) promoting technological change towards lower emissions vehicles with alternate engine types, including kick-starting new markets, (3) substituting the fuel for lower carbon content alternatives (biofuel blends), and (4) policies to curb the amount of driving. In order to reach the 2\degC target with over 66\% probability, global CO$_2$ emissions must be reduced to well below 5.5~GtC in 2050 \citep{Meinshausen2009,Zickfeld2009,Rogelj2012}. Since road transport emissions make roughly 17\% of emissions, transport emissions must likely be reduced to well below 1~GtC in 2050, starting from 1.5~GtC in 2016. This necessitates at least a partially electric composition of vehicle fleets, since calculated biofuel potentials are not guaranteed sufficiently large to replace the whole current use of $\simeq$170~EJ of liquid fossil fuels \citep{Hoogwijk2009, MercureSalas2012}. Efficiency policies for conventional ICE vehicles are not likely sufficient to meet the 2\degC target. Using a combination of technology push, pull and regulatory policies appears \emph{a priori} to be a reasonable strategy. 

Efficiency standards are traditionally imposed using regulatory policy. In the model, this corresponds to controlling the nature of substitutions in new vehicle sales, leaving existing vehicles in the fleet to operate until the end of their statistical lifetime. This can be used in the model to force phase-in of a number of existing environmental innovations to existing conventional technologies, for instance targeting the fuel economy and phasing out older technologies (SI section~3.5). 

Purchase taxes or rebates are often used as a demand-pull policy to level the corporate playing field, and create space in the market for new, more expensive low-carbon technologies. Registration taxes can also re-allocate purchases along the price-engine size axis \citep{MercureLam2015}. If taxes applied to the vehicle price are made proportional to vehicle rated emissions, a `carbon tax' results on future expected lifetime emissions of the vehicle. Meanwhile, a tax on fuels matches more closely an actual carbon tax, but may be less effective per dollar paid at influencing the type of vehicles purchased, depending on consumer time preference. 

Promoting diffusion in markets where particular types of low-carbon vehicles do not exist, using price policies, does not typically work if manufacturers and infrastructure is not present to allow it. In this case, large institutions (e.g. government) can kick-start markets, where for example, public or private institutions purchase or impose the purchase of a fleet of a particular type (e.g. natural gas buses, electric municipality vehicles or taxis), jump-starting later diffusion, which would not happen otherwise. Such strategies are common in many countries (SI section~3.5). 

In FTT, policy formulations currently take 8 possible forms: regulations, standards, registration/fuel/road taxes, subsidies, biofuel mandates, and public procurement (kick-start). As an example, we used several of these types of policies to create one possible coherent framework that achieves worldwide decarbonisation, with the following strategy (detailed numbers given in the Suppl. Excel data file):

\begin{enumerate}
\item Setting the fuel efficiency standard of new liquid fuel vehicles to amongst the best currently available, in each vehicle engine size class, with near term compliance deadlines;
\item Phasing out by regulation the sale of low efficiency liquid fuel vehicles starting in 2018;
\item Introducing electric vehicles in all markets in which they don't exist (in our historical data), in all consumer classes, with procurement policies by 2020;
\item Aggressively taxing the registration of new liquid fuel vehicles proportionally to rated emissions, in order to re-orient consumer choices (here we used 100\$/(gCO2/km) in constant 2012USD), starting in 2020;
\item Increasing taxes on fossil liquid fuels to acquire better control of the total amount of driving (here we used a value increasing from 0.10\$ to 0.50\$ per litre of fuel between 2018 and 2050 in constant 2012USD);
\item Increasing biofuel blend mandates gradually until they reach up to 70\% all regions in 2050.
\end{enumerate}

One advantage of using a non-optimisation diffusion model is that policy interactions can be assessed explicitly, and synergies between instruments can be observed. Here, each of these layers of policy plays a specific role, and none of them can achieve decarbonisation task on their own; they influence the effectiveness of each other. Thus, they only work when applied simultaneously in a coordinated manner. For example, taxing registrations of vehicles based on emissions will drive consumers to the best available, and a key opportunity would be missed if only marginally higher efficiency vehicles were available for purchase. In this case, kick-start policies for EVs take a crucial role to enable the full effectiveness of taxes at reducing emissions, especially in developing countries. Furthermore, the biofuel mandate can only be increased to large values if the liquid fuel consumption of the fleet declines, otherwise the demand for biofuels could imply future issues of excessive land-use changes for biofuel production \citep[e.g. see][]{Searchinger2008, Fargione2008}. 

\subsection{Exploring the impact of policy strategy by layers}

\begin{figure*}[t]
	\begin{center}
		\includegraphics[width=0.9\columnwidth]{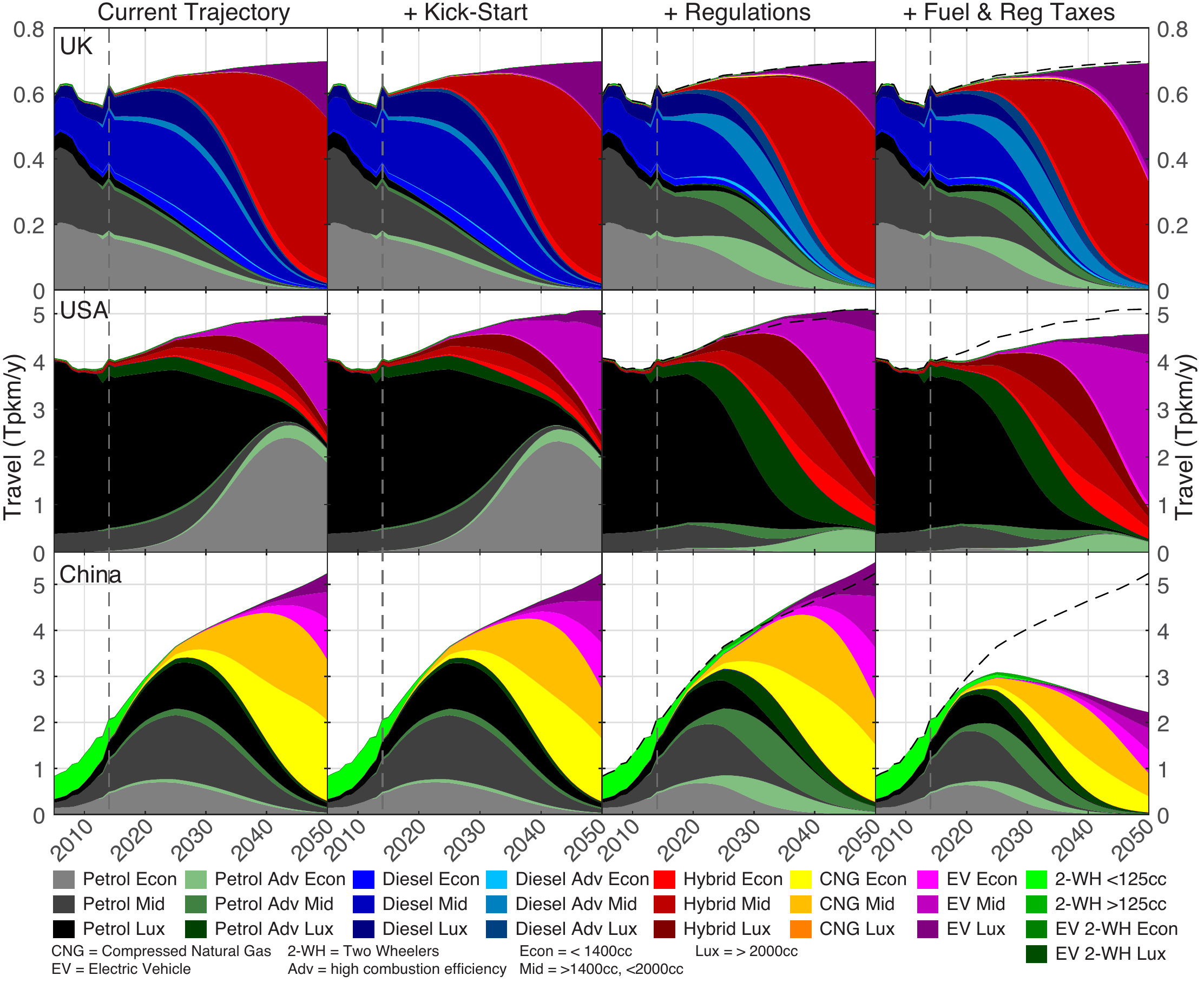}
	\end{center}
	\caption{FTT transport generation (in Tera person-kilometres per year, Tpkm/y) by five technology types in three engine size classes in the UK, China and the USA. The simulation starts in 2012. Prior to this, historical data is shown. Differences in totals arise with tax policy, where consumers drive less. The black dashed lines reproduce the baseline total for comparison. Data to the left of the dashed lines are historical. Differences between panels with respect to the `Current Trajectory' are given in SI section 6.1.}
	\label{fig:Figure2}
\end{figure*} 

We explore in this section how the 6 steps above can deliver sufficient cuts, focusing on the UK, the USA and China. The UK fleet has a significant number of diesel vehicles, a growing fleet of hybrids and a nascent diffusion of electric vehicles. China, dominated by petrol (gasoline) engines, sees its large fleet of motorcycles decline and an emerging diffusion of CNG vehicles. The USA is dominated by large conventional petrol (gasoline) engines, with growing hybrid and electric fleets. These trends, observed in our historical data, continue in the baseline scenario of FTT, in which a slowdown of consumption of liquid fuels already takes place due to existing diffusion dynamics of alternative engine vehicles having already acquired momentum. These baseline diffusion profiles lead to a globally peaking liquid fuel consumption in the 2030s, leading to stranded fossil fuel assets worldwide \citep[see][but not due to biofuels; described in the next section]{Mercure2018}.

Substantial efficiency changes are currently taking place in vehicle fleets around the world, due to efficiency changes and the gradual adoption of hybrid, CNG and EV drivetrains. In FTT:Transport, this is projected to reduce current emissions by 56\%, 65\% and 72\% in 2050 in the UK, China and the USA respectively, in the baseline. It is to be noted that these changes are mostly the result of technological trajectories observed in recent historical data, which the model projects into the future, as no new policies are explicitly included in the FTT:Transport baseline. Many policies currently being adopted or adopted recently will alter these trajectories. For instance, the rise of CNG in China is likely to become replaced by EVs with the support of new policies \citep{Ou2017}, but this is not included in our baseline.\footnote{A detailed review of all existing policies in 59 regions, representing a substantial challenge, is in progress but has not yet been completed nor integrated to our baseline, and it is clear that very recent policies not explicitly included here could affect our projected technological trajectories. This doesn't affect the validity of our methodology.} To accelerate that, policy for decarbonisation described above first involves regulations to phase out from the market less efficient engine types and force in emission standards across engine size classes (steps 1-2). Without other policies, this contributes additional reductions of 0\%, 7\% 13\% over the baseline trends in 2050 for the UK, USA and China respectively, modest additional impacts effectively due to the modest efficiency targets achievable with ICE engines.

Tax policies are applied (policy steps 4-5) to both (1) rated emissions and (2) fuel consumption. Fuel taxes do comparatively little to incentivise changes of technology, mainly due to our average consumer discount rate of 15\%.\footnote{Consumer discount rates in vehicle purchases are controversial \citep{Busse2013,OECD2010} and could lie anywhere between 5\% and 40\%. However, the fact that some studies identify high discount rates signals that some consumers take relatively little consideration of future fuel savings when purchasing a vehicle. At 15\%, the incentive of a fuel tax, per unit of carbon taxed, is comparatively much smaller than that for a tax at registration time \citep[e.g. see the supplementary information in][]{MercureLam2015}.} However, they contribute to curbing driving.

Taxes on registration of vehicles proportional to emissions per kilometre have a higher impact on guiding consumer choices towards low-carbon vehicles, in particular as they become more available through their diffusion: in FTT, the more they are adopted, the more the tax becomes effective at incentivising their adoption. EVs take considerable time to diffuse, and what is observed is that an intermediate layer of diffusion of intermediate emissions vehicles arises. In the UK and the USA, they are hybrids, while in China and India, they are CNG. The tax also incentivises changes of engine category; however, this is limited, as consumers can typically save more tax money by changing engine type rather than engine size, while their preference for vehicle class remains (due to the $\gamma_i$ parameters). With registration tax policies, the strategy must involve providing choice to consumers, as otherwise it only achieves raising tax income without sufficient change in emissions, particularly where EVs are not widely available. Note that similar results could be achieved using tax/feebate combinations. 

It is useful, and possibly necessary in many regions, for the authorities to kick-start the EV market, by sectoral regulation or public procurement, where the industry and infrastructure is absent.\footnote{In the model, in many regions, small and mid-size EVs have zero market shares (zero sales in 2012, e.g. in India, China, Brazil), and thus, policy step 3 involved exogenously introducing non-zero shares. Our assumption is that in 2020, 0.01\% shares are purchased by governments to kick-start the EV markets. We did not include infrastructure costs.} We note that a kick-start policy nucleates simultaneously (i) a market, (ii) a network of supporting industry, and (iii) a social diffusion process, which subsequently co-evolve with the diffusion process itself. In many regions, sales of economic EVs are non-existent in the data, but this will not remain so indefinitely. In the model, mass diffusion of EVs takes-off after 2040, at which point the fuel consumption of the whole fleet declines substantially.  

Remaining fossil fuel use is reduced further by the use of biofuel mandates at 70\%(100\%) by volume. Altogether, these combined policies lead to 88\%(98\%), 96\%(99\%) and 91\%(91\%) emissions reductions based on 2016 levels for the UK, the USA and China.

\subsection{Global road transport decarbonisation, fuel use and emissions}

\begin{figure*}[t]
	\begin{center}
		\includegraphics[width=1\columnwidth]{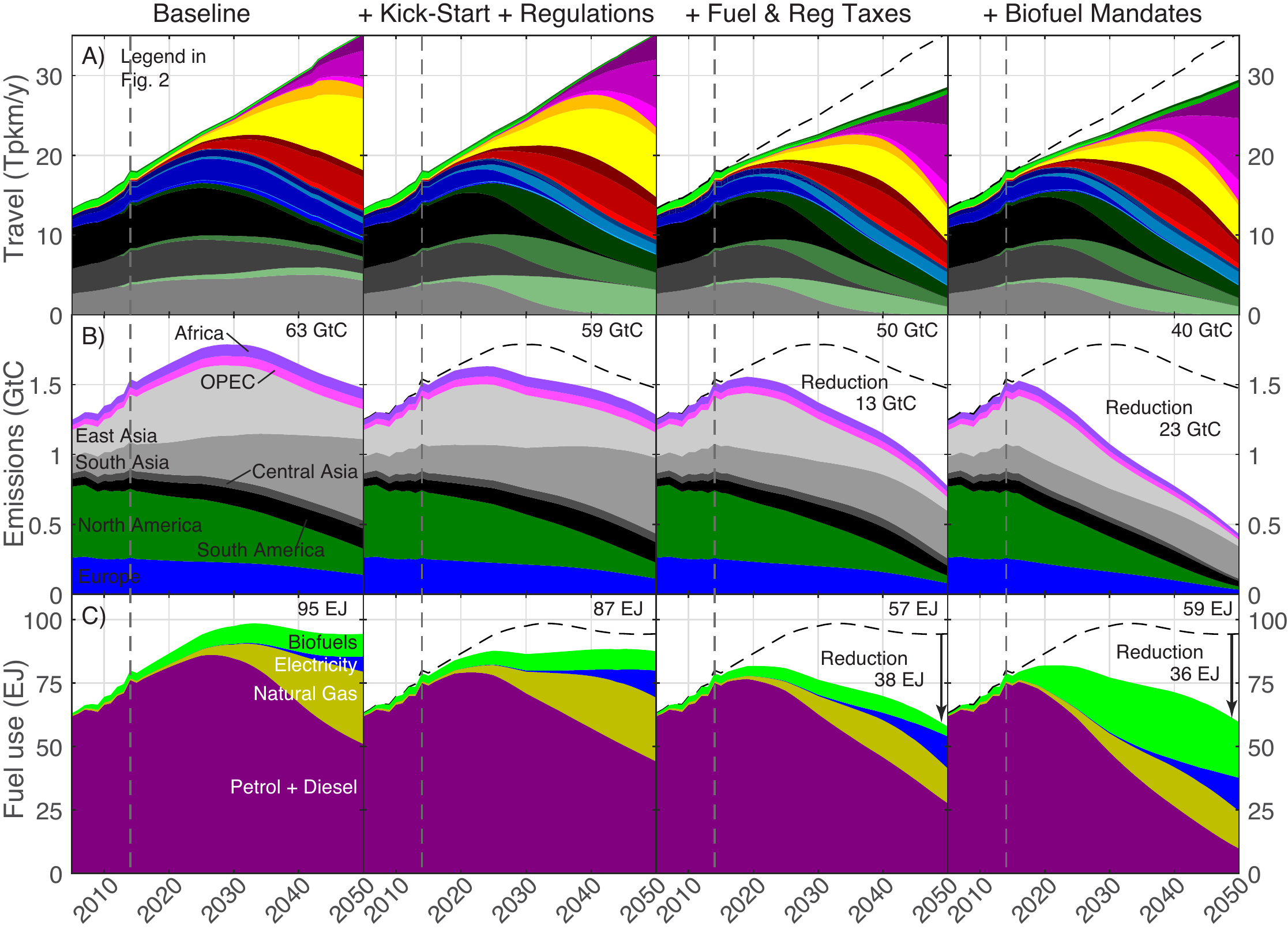}
	\end{center}
	\caption{Global distance driven per vehicle category (in Tpkm/y, top row), global private passenger vehicle emissions in eight regions (in GtC/y) and global road transport fuel use per E3ME fuel type including freight. The black dashed lines reproduce the baseline total for comparison. Central Asia includes Russia and the former Soviet bloc excluding those in the EU; South Asia includes India and Indonesia; East Asia includes China and Japan. Data to the left of the dashed lines are historical.}
	\label{fig:Figure3}
\end{figure*} 

The composition of the global fleet is given in figure~\ref{fig:Figure3}, top row. Given fleet turnover rates and existing trends, it is unlikely that emissions can be reduced with the diffusion of EVs alone sufficiently by 2050 to reach a climate target well below 2\degC, or a 1.5\degC target. 

Instead, emissions are reduced with successive waves of diffusion of innovations, of ever lower carbon intensity. Policy step 6 involves the use of relatively high biofuel percentage blends (70\%, 20\% is in the baseline) in liquid fuels, a policy that has been controversial in Europe and elsewhere \citep[e.g.][]{Searchinger2008, Fargione2008}. Indeed, a high biofuel mandate does not appear realistic in the baseline. However, in a 2$^\circ$C scenario, by 2030, total liquid fuel use declines significantly due to the diffusion of more efficient combustion technologies, including hybrids, as well as CNG and EVs displacing conventional engines. Even when including freight transport, biofuel blend percentages can in fact be increased to 70\% in 2050, while maintaining global liquid biofuel use for transport below 27~EJ.\footnote{For reference, around 50~EJ of bioenergy is currently used globally for traditional heating and cooking \citep{IEAWEB2016}.} This enables to decarbonise road transport to below 0.43~GtC/y by 2050 (72\% of 2016 global transport emissions, Fig.~\ref{fig:Figure3} middle row, 83\% with 100\% biofuel mandates).\footnote{Note furthermore that total use of biofuels declines post-2050 due to increasing diffusion of electric vehicles and the gradual phase out of ICE engines. We do not include in the transport sector land-use change emissions. Land-use modelling is required to estimate what an additional 27~EJ of biofuel production means for land-use change and agriculture emissions.} This is consistent with at least the 2$^\circ$C target, possibly even the 1.5$^\circ$C target, depending on emissions from other sectors.

Fig.~\ref{fig:Figure3}, bottom row, shows that the use of middle distillates peaks in the baseline, reflecting existing technology diffusion trends. With regulations, taxes and biofuel blends, the use of liquid fossil fuels for road transport declines to below 10~EJ/y (70\% biofuel blends) or 1~EJ (100\% biofuel blends) in 2050 (86\% below the 2016 value with 70\% biofuels, 99\% with 100\% biofuel blends), leading to drastic reduction of demand for crude oil \citep{Mercure2018}. Biofuel use due to the biofuel mandate remains below 27~EJ with baseline blends, and below 38~EJ with 100\% blends, peaking in around 2040 due to fleet efficiency improvements and displacement of the combustion engine by other technologies. The use of natural gas remains comparatively low ($<18$~EJ), due to a relatively low global share of CNG. The use of electricity, in a scenario where electric cars make up 33\% of the fleet in 2050, remains comparatively small at 14~EJ, (with respect the E3ME 2050 total electricity demand of 140~EJ), due to the very high conversion efficiency of EVs. The result is that transport electrification significantly reduces global energy use, and does not imply an excessive or unmanageable new load for the power sector.\footnote{Total electricity demand was of order 80~EJ in 2016 and may increase to between 130~EJ (2\degC) and 170~EJ (baseline) in 2050, according to E3ME; more electricity is saved through decarbonisation than what is demanded by transport. We do not consider changes in intermittency of power demand due to EVs, which could be substantial.} Reaching the 2\degC target remains, however, contingent on power generation and land-use decarbonising.\footnote{Due to relative combustion efficiencies between power plants and internal combustion engines, fleet electrification reduces emissions even if the power sector is coal intensive. Thus we do not quote full life-cycle emissions from transport. Baseline power generation emissions from FTT:Power in our model increase by 1\% when decarbonising transport alone, while total emissions go down by 8\%. However we do not consider likely that stringent decarbonisation policies would be adopted for transport but not for power generation.}. 

\subsection{Reflections on the model}

Standard cost-optimisation models are normative and search parameter space for a system state that achieves a set of system and political objectives. Meanwhile, a positive model offers a representation that attempts to guess what future states of an existing system may look like, given its present state and evolution trajectory, and decisions taken to alter that trajectory. In the world of IAMs, almost all models are of the normative optimisation type, and often have relatively low policy resolution. However, the development of new climate policy requires, in most national policy processes, impact assessment of detailed policy frameworks. This unavoidably demands the use of positive models that model complex policy packages and can give policy-makers indications of current trajectories and potential outcomes of the various policy options considered. 

Here, we have shown that this can be achieved, but with a different type of modelling framework, in comparison to standard methods. We used a model without representative agent, based on dynamical systems without equilibrium, to explore the evolution of the global road passenger vehicle fleet, based on trajectories observed in our historical database. We found that indeed, results are different from those using standard methods. For example, EVs diffuse faster than in optimisation models, even when these include substantial amounts of behavioural information \citep{Pettifor2017a} in both the current trajectory, and in a decarbonisation scenario (more comparisons to other models given in SI section~6.2). Furthermore, oil demand for transport peaks in the current trajectory, substantial efficiency changes are already taking place due to the popularity of new technologies such as hybrids. Thus, we can expect that using this type of method can provide a critical lens with which to look at all types of models used for advising policy-making, in particular IAMs. 

Perhaps the key advantage of this model is that outcomes are more dependent on observed technological trajectories, and less reliant on technological assumptions, such as costs, in comparison to optimisation models. We demonstrate this in SI section~6.1 with an extensive sensitivity analysis. We observe that changing technological parameters generate outcome variations generally of lower magnitude than the parameter variations introduced. In particular, varying the $\gamma_i$ values for non-pecuniary costs has relatively low impacts on results (outcome changes $<<$ parameter variations). Adding or removing new or hypothetical technological options also has a relatively low impact on outcomes. This is a reflection of strong model path-dependence as opposed to parameter dependence. This is further explained in SI section~3 on model theory. 

The converse of this property regards the model's validity range in time. With a dynamical systems model, one can quantify the time span over which one should expect projections to be valid, with a cone of uncertainty that increases with time from the present day. Given that this model takes part of its parameterisation from recent technological trajectories, the further we model in time, the less reliable projections become. The validity range is determined by the degree of systemic inertia, which in this case is of about 30 years. We discuss the validity range in time of the model in SI section 6.3, where we explain why a maximum modelling horizon of 2050 is appropriate.

The model also enables a relatively easy method to implement and analyse a large range of policy instruments fairly closely to their actual legal definitions, from regulatory instruments to some types of push and pull strategies. Here, we currently have eight types of policies and used several of them to construct one possible global decarbonisation scenario. It is clear that many other such scenarios can be designed, and assessed alongside one another, each with pros and cons. We note, however, that it is most likely not possible to find an `optimal' policy package when one does not have a representative agent, but has a huge policy space.\footnote{Optimising all possible scenarios that can be generated by FTT, with its huge policy parameter space, appears challenging with our computing power, simply due to its degree of non-linearity, which we argue is quite representative of the real world.} Meanwhile, we also find that strong policy interaction arises in the model, through the fact that it is non-linear and based on a diffusion/bandwagon effect theoretical basis. While this complicates policy analysis, we believe that it is closer to reality. A comparison to other model results is given in SI section~6.2.

We note, however, that we do not achieve the degree of detail of most other IAMs in other important domains, such as infrastructure \citep{Waisman2013}, travel time budgets \cite{Daly2014}, rural/urban splits and range anxiety \citep{McCollum2016, Pettifor2017a}, other non-pecuniary costs and behavioural features \citep{Pettifor2017a, Pettifor2017b}, while we represent modal shift and freight only partially (other studies reviewed in SI section~2). Furthermore, to model more accurately technological trajectories, we would need to review and include explicitly, for all 59 regions, all transport policies that have been implemented between the start date of the simulation and the present day, a substantial challenge. These are areas that are under development or that can be improved in future work, in a more mature version of the model. We note however that including some of these could conflict with our own methods (possible implicit double-counting). For instance, we consider the provision of infrastructure (e.g. for EVs) part of the diffusion process, where for example, kick-start programs imply infrastructure developments. Similarly, rural/urban splits are implicitly accounted for in our distributed cost data; however they may generate constraints that we do not represent. The difference in model results that some of these would imply are not fully clear to us, for instance where modal shifts reduce the number of road vehicles, or freight electrification, which could reduce emissions substantially. 

\section{Conclusion}

Emissions reductions consistent with the 2\degC target of the Paris Agreement have been extensively demonstrated to be technically feasible \citep{IPCCAR5WGIII}. However, policy frameworks to reach these goals are not yet clearly established across the world, even where emissions targets are the most stringent. Existing IAMs, rich in technology options, have been used to explore the technically feasible parameter space for decarbonisation. However, the representation of specific policy instruments or realistic portfolios, and their representation of behaviour in agent decision-making in their current use has not been extensive, leaving a gap for advising policy-making.

Here, we presented a model that overcomes many of these issues, with a global transport simulation model that projects the diffusion of innovations based on historical data and choices of heterogenous agents making individual choices, which is part of a global IAM. Instead of optimising a whole system, this model projects its evolution based partly on observed trajectories of technology diffusion, partly on a representation of consumer choices that includes agent heterogeneity, social influence and non-pecuniary aspects. This model type enables a finer representation of specific transport policy instruments that are pecuniary, regulatory or of the technology push type. 

We used this model to assess the impacts of a chosen portfolio of transport policies that leads to emissions reductions consistent with a policy target of 2$^\circ$C, and possibly even 1.5$^\circ$C. We find that in such a non-optimisation representation of agent decision-making, policies interact and enable each other. This opens a door to finer model-based analysis of composite transport policy packages, while remaining focused on climate change and global emissions. 

We conclude by suggesting that decreasing returns are now emerging with cumulative efforts at mapping the feasible decarbonisation parameter space by modelling optimisations of the transport sector, while demand is increasing for finer detailed impact assessment of possible policy packages. This potentially requires to alter modelling methodologies that are used for analysing climate policy. It also demands to clearly delineate normative analysis, in which one identifies policy objectives, to positive analysis in which the goal is impact assessment of proposed policies, both of which play a different role in the policy cycle. However, this exercise also highlights the limited validity range that decreases in time of any non-prescriptive modelling strategy. In an effective science-policy bridge, IAMs must attempt to assess the impacts of possible composite policy packages that are currently considered by policy-makers. We argue that this is only possible through the use of positive behavioural science and models, and showed that this is possible with a relatively simple non-optimisation modelling framework. 

\section{Acknowledgements}

The authors thank the C-EERNG centre and J. Vinuales for support, as well as N. Edwards, P. Holden, C. Wilson, H. Pettifor and D. McCollum for informative discussions. This work was supported by a fellowship of the UK Engineering and Physical Sciences Research Council (EPSRC), no EP/ K007254/1 (JFM), a grant from the UK's Natural Environment Research Council (NERC) no NE/P015093/1 (HP and JFM) and a grant from the UK's Economic and Social Research Council (ESRC) no ES/N013174/1 (JFM, HP). JFM thanks L. Turner for enduring support during treatment for a life-threatening health situation at the time of revision of this work. 

\section{Author contributions}
JFM designed the theoretical model, co-designed the scenarios and wrote the text. AL parameterised the model,  co-designed the scenarios and contributed to the text. SB and HP designed and executed the econometric specifications for transport demand, maintained the E3ME model and contributed to the text. JFM and AL have equal contributions.

\setcounter{section}{0}
\newpage
\part{Supplementary Information}


\section{Introduction}

FTT:Transport, in its 2016 version, is a global model of technological change in road transport. It is based, conceptually, in parts on previous work for the power sector, FTT:Power \citep{Mercure2012}, using the same evolutionary economics approach and the replicator dynamics equation. Vehicle choice for passenger transport, however, is much more complex to model, in our view, than technology choice in industry. This is simply because the reasons coming into play in consumer choices are more heterogenous and varied than in firms. For example, some vehicles sell at \$15,000, some vehicles sell at \$150,000, while both apparently supply the same mobility services. However, just as in the clothing industry, extensive sociological studies exist, and are continuously carried out, to better understand how this choice is made. Here we present the details of the FTT:Transport model, which attempts to model these choices in a way detailed enough to produce useful insight for transport policy design with the purpose of reducing road transport emissions.

FTT:Transport is dynamically integrated to the E3ME macroeconometric model, in the FORTRAN language, maintained and operated by Cambridge Econometrics Ltd (see \burlalt{http://www.e3me.com}{www.e3me.com}), but also implemented in a separate stand-alone version as an opensource MATLAB code and graphical user interface, that can be obtained by contacting the authors. When a model exists in two different implementations, it is a challenge to maintain both such that they continually produce identical results. While the authors make extensive efforts at maintaining the two models identical, this cannot be guaranteed. The version as integrated into E3ME features dynamical feedbacks with the global economic model, and is used more frequently, thus considered the official version. The matlab version, however, benefits from an easy-to-use graphical user interface and is opensource. The overall simulation-based integrated assessment model E3ME-FTT-GENIE1 is described in \cite{Mercure2018, Mercure2018b}.

\newpage
\section{Review of the modelling context \label{sect:LittRev}} 
This section presents a literature review of all the main aspects surrounding our research objective and identifies knowledge gaps that lead to our research. In our climate change mitigation perspective, transport modelling is seen through the lens of energy systems, since we are focusing on modelling transport emissions as opposed to many other aspects of transport systems. In this context transport will often be modelled within larger energy system models. Studying transport emissions however involves changing technologies, which thus invokes technology transitions theory in order to understand the process of diffusion. But technological change takes place only if decisions to change occur, and this requires understanding consumer choices, for which we bring in discrete choice theory. Thus this section is subdivided into three parts: a review on the existing energy system models, a review on the technology transitions literature and a review on choice modelling for energy system models.

\subsection{Energy system models}
Energy system models are valuable mathematical tools that have been used to help understand how to plan, operate and coordinate complex networks of energy supply and demand technologies. With the issue of climate change taking increasing prominence within policy discussions, energy models have taken a central for advising climate policy-making with the goal of reducing GHG emissions. Transport producing a significant fraction of global emissions, the generation of mobility services has often been included in energy models. Since studying climate change requires knowledge of the global amount of GHGs emitted, models used for climate policy are often global models. This is the type of models reviewed here.

A variety of modelling techniques have been utilised for studying energy and emission projections. They vary in terms of data requirements, technology specifications and computing demands. Following several existing studies (see e.g. \citealt{Nakata2011,grubb2002, Herbst2012,loschel2002}), we divide energy system models into five main non-exclusive categories, according their underlying methodologies. This is summarised in Table~\ref{table:models}.

 \subsubsection {Energy sector optimisation models}
Energy-oriented models are designed to consider the energy sector and emissions from energy production and consumption in detail. One of the most adopted approaches in modelling the energy sector is optimisation. The most well known examples that utilise the optimisation approach include MARKAL and MESSAGE. Both have been employed extensively to model transport and energy systems \citep{mccollum2013, gul2009,yeh2011}. MARKAL is a linear programming optimisation model developed by the Energy Technology Systems Analysis (ETSAP) of the IEA \citep{IEA/ETSAP}, which has been adapted into numerous variants across the globe. Related to MARKAL, the TIMES Model (\textbf{T}he \textbf{I}ntegrated \textbf{M}ARKAL-\textbf{E}FOM \textbf{S}ystem) is a bottom-up technology rich optimisation model generator \citep{TIMES}. MESSAGE  is  is a dynamic linear programming model that calculates cost minimal supply structures over a given time horizon \cite[see][]{Messner}. It provides core inputs for major international assessment and scenario studies, such as the Intergovernmental Panel of Climate Change (IPCC), the World Energy Council (WEC) and the Global Energy Assessment \citep[GEA][]{IIASAWeb}. Other energy sector models that operate with an optimisation framework include LEAP \citep{heaps2012}, REDGEM70 \citep{takeshita2008}, MERGE \citep{kypreos2005}, REMIND \citep{leimbach2010}, BEAP \citep{grahn2007}, GET \citep{grahn2007}, OSEMOSYS \citep{Osemosys2015} and PRIMES \citep{PRIMES}. 

The strength of optimisation frameworks is their ability to find cost minimal energy pathways, as normative planning tools. They feature however potential convergence problems due to non-convexities when including learning curves, which can lead to multiple solutions that then need to be sorted \citep{kohler2006}. The traditional optimisation model assume a social planner with perfect foresight, where ideal scenarios are sought that achieve particular goals, which are `internalised' to the optimisation (e.g. energy security, climate change, health and other externalities). However, for a purely descriptive purpose, information for the full time frame is not in reality available immediately to agents responsible for energy sector developments, making perfect foresight inconsistent with positive modelling goals \citep{greene2010,anderson2011}. To this end, the new myopic (or limited foresight) modelling approach incorporated as an option in some optimisation frameworks, such as MESSAGE and GET-LFL, allows for the analysis using different time horizons for decision making (see~\citealt{MyopicMess,hedenus2005,nyqvist2005}). While the myopic approach improves upon traditional perfect foresight approach by providing a framework for exploring some degree of `path dependency' in the energy system (see below), it has been argued that \emph{bounded rationality} \citep{simon1972} embedded in decision making involves factors not considered in these approaches, such as adaptive behaviour or network (bandwagon) effects, which we discuss below.

\begin{table}[t] 
	\small
			\begin{tabular}{p{4cm}p{3cm}p{3.8cm}p{1cm}p{1.5cm}}
			\hline\hline
			Model name 				&  Type                           &Methodology/tools  &ETC &Global transport \\                
			\hline 
			MARKAL, TIMES, GET, MESSAGE, REDGEM70, BEAP, REMIND, LEAP   &Energy sector models     &Linear optimisation  &Yes        &Yes \\  
			\\
			PRIMES 					&Energy sector model               &Non-linear optimisation   &Yes      &Yes \\
			\\
			TREMOVE, WEM			&Simulation model               &Simulation and optimisation   &Yes   &Yes\\
			\\                     
			IMAGE/TIMER, ASTRA, CIMS, GLADSTE         &Simulation model                 &System dynamics and \mbox{non-optimisation} &Yes  &Yes\\
			\\
			POLES					&Simulation model                 &Simulation and partial \mbox{equilibrium}   &Yes         &Yes \\
			\\
			TAPAS					&Simulation model                 &Agent-based model      &Yes           &Yes\\
			\\
			MESSAGE-MACRO, MARKAL-MACRO, RICE, DICE, GEM-E3            &Macroeconomic model           &Optimisation framework     &Yes     &No\\
			\\
			E3MG					&Macroeconomic model           &Non-optimising dynamic \mbox{simulation} approach  &Yes   &No\\
			\\
			MERGE, FUND, CETA, WIAGEM			&IAM		&Optimisation hybrid models &Yes     &yes\\
			\\			
			\\
 			CIMS, GREEN, NEMS, GEM-E3, WITCH		&CGE		&Equilibrium structure   &No       &yes\\
			\hline
			\hline
		  \end{tabular}
	\label{table:models}
    \caption{Examples some major energy models. \emph{ETC} stands for endogenous technical change. \emph{Global transport} indicates whether models have a component to model the global transportation system. }
\end{table}

\subsubsection{Macroeconomic models} 
Macroeconomic models focus on the entire economy of a society, considering the energy sector as one subcomponent \citep{Nakata2011}. Macroeconomic models are usually top-down models that have higher sectoral aggregation and better characterisation of impacts of economic growth \citep{hourcade1993}. Although they reflect greater details regarding macroeconomic feedbacks, they provide a less detailed description of technology and technological change. 

Most macroeconomic models use the Computable General Equilibrium (CGE) approach. This includes GEM-E3 (\textbf{G}eneral \textbf{E}quilibrium \textbf{M}odel for \textbf{E}nergy-\textbf{E}conomy-\textbf{E}nvironment interactions, \citealt{capros1997}) and NEMS (\textbf{N}ational \textbf{E}nergy \textbf{M}odelling \textbf{S}ystem, \citealt{NEM}). CGE models have been criticised for a general lack of detailed technological information regarding the energy system \citep{bohringer1998}. In particular, CGEs face considerable difficulties in incorporating Endogenous Technological Change (ETC). This is because linear programming is generally more suited to solving problems with a single maximum, and the introduction of a increasing returns to scale associated to ETC can generate several equilibria \citep{kohler2006}, a consequence of the path-dependent nature of ETC and the accumulation of knowledge. Considerable improvement has been made regarding the incorporation of ETC into equilibrium models \cite{IMCP2006}, which however involves complex searching methods for finding the `true' system equilibrium.

Traditional macroeconomic models typically assume that there exists an autonomous energy efficiency improvement (AEEI), and, thus, technological change is exogenous to the model \citep{grubb2002}. The exogenous technical growth assumption with AEEI has been criticised for neglecting the interactions between policy options and technological change (see \citealt{weyant1999, kohler2006, gillingham2008}). This can significantly bias the policy assessment because policies induce technical change (ITC) and cost reduction within the system \citep{kohler2006,van2002,goulder1999}.

In response to criticisms on modelling technological change, hybrid models were that incorporate both a detailed energy component (bottom up) and a neoclassical optimal growth economic structure (top-down). This includes CIMS (\textbf{C}anadian  \textbf{I}ntegrated  \textbf{M}odelling  \textbf{S}ystem, \citealt{CIMS}), GREEN (\textbf{G}eneral \textbf{E}quilibrium \textbf{E}nvironmental Model, \citealt{burniaux1992}) and WITCH (\textbf{W}orld \textbf{I}nduced \textbf{T}echnical \textbf{C}hange \textbf{H}ybrid \citealt{bosetti2006}). Hybrid models bridge the gap between conventional top-down and bottom-up modelling approaches \citep{hourcade2006}. Other hybrid models include MESSAGE-MACRO and MARKAL-MACRO, MERGE \citep{kypreos2003, kypreos2005} and REMIND-D \citep{leimbach2010}. Of particular interest here, the E3ME-FTT model is a hybrid non-equilibrium macroeconometric model based upon a Post Keynesian demand-led economic view \citep{barker2006, barker2008, Barker2012}, with bottom-up non-optimisation models of technology (FTT, see \citealt{Mercure2012, Mercure2014}; see \citealt{Mercure2018b} for a full  description of E3ME-FTT).

\subsubsection{Simulation models}

Quantitative simulation models can provide important insights about the effects of energy policies because they allow for the assessment of consequences of policy and economic measures and a straightforward incorporation of increasing returns and path-dependence (see below), which contrasts with optimisation models. Well known examples of simulation models include the World Energy Model (WEM) \citep{WEM}, POLES \citep{criqui1999} and TREMOVE for the transport sector in the EU \citep{van2005}. This section discusses two major approaches in simulation modelling: systems dynamics and agent-based. 

\vspace{8pt}
\emph{Systems dynamics approach (SD)}:

SD is `the study of information-feedback characteristics of industrial activity to show how organisational structure, amplification and time delays interact to influence the success of enterprise' \citep{Forrester1961}.  It combines technology and market-behaviour frameworks into one in order to represent the causal structure of the system \citep{martinsen2008}. It is used to analyse the wider impacts of policies being tested \citep{xavier2013} which, in terms of methodology, contrasts with the search for optimal scenarios done with optimisation models \footnote{Unless one accepts a debatable premise that energy systems are perpetually optimal, a subject that requires a separate discussion. See \cite{Mercure2014}.}. While SD offers clear benefits to modelling energy systems characterised by a large number of interactions between several variables \citep{Armenia}, it is not widely applied to energy system models. In this class, IMAGE/TIMER (The \textbf{T}argets \textbf{IM}age \textbf{E}nervy \textbf{R}egional, \citealt{de2002targets}) analyses the long-term dynamics of the energy system. With a focus on transport, both ASTRA (\textbf{AS}sessment of \textbf{TRA}nsport Strategies, \citealt{fiorello2010}) and GLADYSTE (Global Scale System Dynamic Simulation Model for Transport, \citealt{fermiglobal}) are system dynamics models at a European scale for the strategic assessment of policy scenarios. Technology vintage models also fall within this class since they track technology units from their construction to their dismissal, which includes CIMS (see above). 

We note however that any simulation models based on Multinomial Logits (MNL) or constant elasticity of substitution (CES) functions use the assumption of perfect information and no interactions between agents, but instead, have a representative agent. The MNL can be converted into a CES using a basic optimisation calculation. This means that they have a theoretical basis closely related to optimisation in general (for instance, CGE models, even without foresight in recursive dynamic mode, remain utility-optimisation models). 

Within its own class which also involves systematic feedbacks, amplification and time delays, the FTT family of models, which includes FTT:Transport, can be seen as SD models with a technology vintage component. However, unlike many SD models that use the Vensim system with multiple nodes, the Lotka-Volterra approach of the FTT family of models naturally includes more dynamics and interactions not represented in traditional SD models, while it assumes bounded rationality and multi-agent interactions (which implies that it rejects the notion of a representative agent).

\vspace{8pt}
\emph{Agent-based modelling approach (ABM)}

Social science often involves and emphasises heterogenous human agents with diverse preferences, making diverse choices. ABM is a computerised simulation of a number of interacting decision-makers. Each assesses its situation and makes decisions on the basis of a set of rules \citep{bonabeau2002}. Since ABMs simulate learning at the individual level and in the model of innovation networks \citep{gilbert2008}, it is a way to model the system dynamics and complex adaptive properties.\footnote{A complex adaptive system is defined as being composed of population of adaptive agents whose interactions result in complex non-linear dynamics \citep{brownlee2007}.}

ABMs have clear advantages in modelling consumer decisions and agent diversity. They have been applied to modelling technology adoption in transport systems. For example, TAPAS (\textbf{T}ransportation \textbf{A}nd \textbf{P}roduction \textbf{A}gent-based \textbf{S}imulator, \citealt{holmgren2012}) is a micro-level model for the assessment of different types of transport-related policies. \cite{eppstein2011} has developed an ABM of vehicle consumers that incorporates spatial and social effects. \cite{kohler2009} used ABM to produce a representation of the Multi-Level Perspective on technology transitions, which features strongly non-linear effects with a representation of technology diffusion consistent with observed S-shaped diffusion patterns, described below. 

Unlike other global energy system models, the ABMs comprise primary research models applied to particular study areas, or to restricted geographical scope \citep{wegener2004}. This is because complete input data for ABM above national levels is difficult. Furthermore simulating the behaviour of large numbers of agents can also be computationally intensive \citep{bonabeau2002}. Thus, detailed ABMs are less suitable for modelling transport emissions at the global scale required for climate change-related research. 
 
\subsection{Modelling technology transitions} 

Within many of the approaches and models listed above for modelling GHG emissions remains a general lack of an \emph{endogenous} representation of technological change. Within optimisation models, technological change takes place in such a way that given changes of policy context, the energy system remains at a cost minimum. The rate of diffusion is limited exogenously however, constraining the results of the optimisation. These diffusion rates are are therefore among the most critical assumptions for assessments of long term emissions and energy issues. In optimisation models, diffusion rates were found to be possibly pessimistic in comparison with observed patterns of diffusion \cite{Wilson2013}. Beyond overcoming pessimism, added realism would require that, as empirical work tells us, diffusion should be represented as context and time dependent, and this effectively opens Pandora's box. 

Missing in many models is a representation of how technologies come to gradually appear and diffuse to widespread use, which is not instantaneous and involves building up new markets and new production capacity for new technologies. Technology diffusion is known in textbooks to follow S-shaped curves, which is supported by a large body of literature. Ideally, added realism in technology models would include increasing amounts of elements found there. We thus review here the empirical and theoretical literature on technology diffusion and transitions theory.

\subsection{Technological transition in the transport sector} 
The socio-technical system describes systems that involve complex interactions between human, machine and environment \citep{baxter2011socio}. A transition of socio-technical regime (STR) is a set of processes that lead to a fundamental shift in the socio-technical regime \citep{kemp1994technology,geels2007typology}. STR involves technological changes, user practices, regulations, and industrial networks, with a range of actors and over a period of time \citep{Geels2005}. 

The modern automobile industry is deeply embedded into legal, social, cultural and economic practices. The lock-in mechanisms imply that socio-technical regimes are extremely rigid in the automobile market. There are substantial sunk investments in plants for IC engines, skills and fuel infrastructure. Personal mobility practices have also become dominated by petrol-based cars, in turn shaping urban infrastructure. The majority of cars on the street are internal combustion engines and there are strong path dependencies on automobile consumption \citep{geels2011automobility}. Consumers do not only optimise cost when they choose a car, instead, consumers take into account the social trend, the availability of infrastructure and the models available in the market that match their preferences when they purchase a car \citep{tanaka2014consumers}.

\subsubsection{S-shaped curves of technology diffusion}

Technology diffusion is often described by S-shaped sigmoid curves. S-curves illustrate the fact that technology diffusion is a gradual process, with a slow initial growth rate, followed by accelerated growth as as new markets are reached, and slowing down again \citep{barreto2008}. One of the earliest classic studies was conducted by Griliches, who found that the penetration of corn seeds followed logistic curves \citep{griliches1957}. \cite{Mansfield1961} described how the diffusion innovations follows a behaviour similar to the spread of an epidemic. 

\cite{fisher1972simple} subsequently also proposed a `technological substitution model', which describes the penetration process of new technologies replacing old ones with S-shape curves. \cite{marchetti1980dynamics} expanded the Fisher-Pry model into a model involving more than two technologies, while \cite{Nakicenovic1986} explored specifically the early 20th century transition in transport. Empirically, analysis found that diffusion and substitution of transport technologies historically followed S-shape curves (see e.g. \citealt{wilson2012, grubler1990rise} for further reviews and analysis).   

The logistic diffusion process is consistent with long wave theory \citep{freeman2002}, a socio-technical approach to transitions \citep{kemp2001} and evolutionary economics \citep{nelson1982} that highlight co-evolution and multi-dimensional interactions between technology and society. In particular, \cite{Metcalfe2004} argues how Mansfield's work can be interpreted as an expression of evolutionary economic behaviour. \cite{geels2002technological} and \cite{elzen2004} propose adopting a dynamic multi-level perspective on system innovations. The process in the multi-level perspective consists of four phases.  The first two represent the emergence of technology niches and the development of technical trajectory through `probing and learning'. The third describes the wider diffusion of the technology and in the four, how the new technology replaces the old technology.

In technology diffusion models, it is common to represent technology diffusion with an S-curve. Dominant models of explanation of this pattern have been epidemic and neoclassical models of diffusion \citep{nill2008}. Diffusion in epidemic models is determined by the `epidemic spread of information among potential adopters'. Thus, adoption is a function of the product of the uninfected numbers and the share of population that is already infected \citep{sarkar1998}. While this approach picks up information contagions of users, it is often criticised by economists as having neglected the economic aspects of diffusion \citep{geroski2000}. Neoclassical models base their explanation of diffusion on heterogenous adopters. While they provide useful insights into the differences between potential adopters, criticisms are mainly directed at the assumption of the equilibrium approach \citep{sarkar1998}. In essence, both epidemic and neoclassical models might have missed important features of technological evolution, such as the dynamics of technological competition \citep{nill2009} and complex systems driven by co-evolutionary interactions \citep{garnsey2006}.  

The evolution paradigm of technological change has its root in Schumpeter's theory \citep{schumpeter1942}, which analyses innovation as a historical process and technological substitution as a process of `creative destruction' of prior technologies \citep{schumpeter1942,Levinthal1998}. To use an analogy with biology, when an invasive species proves to posses better `fitness' to environmental conditions, this species may come dominate at the expense of others by expanding in ecological space. Similarly, the growth of any technology depends on its fitness to markets in comparison to competitors. If the parallel with population ecology holds, then diffusion should depend on population sizes as well as fitness and follow standard population dynamics.

A number of studies have shown that Lotka-Volterra population competition equations (LVC), a set of coupled differential equations, can be applied to model technological diffusion \citep[e.g.][]{saviotti1995, grubler1990rise, marchetti1980dynamics, bhargava1989generalized, morris2003analysis} and organisational change \citep{lee2009}. Equivalent to the replicator dynamics equations commonly used in evolutionary economics and evolutionary game theory \citep{Hofbauer1998, Safarzynska2010}, the LVC effectively represents the growth and decline of technologies competing in a market according to the size of their industries and their fitness. In his argument bridging diffusion work to evolutionary economics, \cite{Metcalfe2004} suggests that S-curves, corresponding to the two-technology case, are only a subset of possible profiles of diffusion, themselves governed by a more general law, the replicator dynamics or equivalently the LVC. In other words, in a problem where three or more units interact, diffusion does not follow simple S-curves but instead is described by the LVC. 

Given that, in all previous work, to our knowledge, no clear method has been suggested for parameterising an LVC system. The problem is obvious: in a two technology case, because it is a non-linear system, one requires diffusion data up to beyond the inflexion point of the logistic curve in order to be able to determine the diffusion timescale (the parameter of the logistic function). This means that it can only be determined for technologies that have already diffused past half of the total market. However, to be useful for forecasting technology diffusion, it is in the cases where diffusion is

\subsubsection{Path dependence} 

The argument that technological change is path dependent was advanced by \cite{Arthur1989, arthur1994, arthur1987} and through the work \cite{david1985,david2007} conducted on the economic history of technology. Path dependence arises when some economic processes have increasing returns to scale, or, say, technology adoption has increasing returns to adoption (i.e. positive feedbacks).  As Arthur demonstrates, in the presence of increasing returns to adoption, technology lock-ins can arise as a result of small `historical accidents', which may seem to make little economic sense from a rational perspective. 

Path dependence in energy systems arises, when positive feedbacks are present, from differences in initial conditions, engineering traditions, policy choices and historical accidents leading to differences in infrastructures and consumption patterns \citep{grubler2008}. Four types of processes will lead to path dependent behaviour: increasing financial returns to adoption (learning curves), and coordination benefits, the latter arising when technical constraints increase the usefulness of a technology the more it is adopted. Technology investments are also highly irreversible, with decision rules chained to the context \citep{saviotti1991}. Within the automotive industry, while visions of what future mobility could be exist, agents in the system can remain very stubborn facing change \citep{howey2010, wells2010, wells2012}, for reasons that the modeller, or the theorist, cannot know or exhaustively enumerate. However, as is known historically, after the mobility transition of the early 20th century, once economies and societies were `locked in' to petrol vehicles, then huge increasing returns were seen from petrol car production and infrastructure. Social life also became irreversibly locked into that mode of mobility \citep{urry2004}. Thus from this historical perspective, it is difficult to justify theoretical structures that lack path dependence, despite that it implies a higher level of complexity: several or no equilibrium points, uncertainty propagation and the importance of small events.

Recognising the importance of path dependence, technological change modelling has gone through major improvements, especially in the field of climate policy, with Endogenous Technical Change (ETC) incorporated \citep{IMCP2006}. There are two main concepts employed: knowledge capital at the sectoral scale and learning curves at the technology level \citep{kohler2006}. Essentially, knowledge can be generated through investment in R\&D, while with learning-by-doing, the costs of specific technologies decrease with experience and cumulative investment. These however are only two of the possible ways in which path dependence arises, those associated to prices changing with adoption. Possible coordination benefits unrelated to prices, for example, are missing in these models. 

Using the replicator dynamics, the FTT family of technology models has a natural representation of path dependence that stems from feedback structures and self-reinforces the dynamic system of equations. It describes the interactions between the populations of organisation and captures theories (including irreversibility, hysteresis and non-linearity) associated to path dependence \citep{ebeling2001}. With cumulative causation endemic in technology adoption decisions represented in by the replicator equation \citep{jacobsson2004, carrillo2006}, and economic factors considered by a dynamic simulation model of the economy (E3ME), FTT enables to simulate energy systems in the presence of strong path dependence. 

\subsection{Choice modelling}

\subsubsection{Choice heterogeneity}

While industrial dynamics accounts for technology diffusion capabilities, consumer demand drives the rate and direction of innovation \citep{nemet2009}. The adoption rate is related to the perceived benefits received by the user and the costs of adoption \citep{hall2003}. Even though `non economic factors', such as tastes and needs, are unique to heterogenous consumers, traditional technology models (e.g. MESSAGE, LEAP) often approximate the system to representative agents and assume a homogenous society with perfectly coordinated choices that keep technology systems at overall cost minima \citep{jebaraj2006}. 

More recently, some energy system models (e.g TIMER, CIMS, GLADYSTE, TREMOVE, IMACLIM) have integrated consumer heterogeneity with discrete choice models. Discrete choice theory is a prolific field of economics that enable to parameterise models that predict the choices of heterogenous groups of people using information either from surveys, from socioeconomic data or from revealed preferences \cite[see for instance][]{ben1985}. With such a parameterisation, a discrete choice model can predict how the choices of a group of people is likely to change given a change of context (e.g. prices, value, income, distance, etc), summarised by the MNL (and equivalently CES). This has been used extensively to model transport mode choice in cities using surveys carried out over, for example, a population of commuters, situations in which all agents have the same knowledge and do not influence each other. However, while parameterising a discrete choice model using surveys at the country scale is quite challenging, it is simply impossible at the global scale. This work proposes a method to parameterise a discrete choice model based on market revealed preferences, which gives a reasonable approximation to parameters without using surveys. But furthermore, the assumption of agents with the same knowledge clearly breaks down at a national scale and when modelling the whole transport fleet. Clearly, agents do not analyse the whole vehicle market (which features thousands of model variants), but instead clearly influence each other. In this case, an assumption of bounded rationality with social interactions seems unavoidable, and thus we argue that the MNL is not suitable.

`Diversity' is an attribute of any system whose elements may be apportioned into categories \citep{leonard1989}. According to \cite{stirling1994,stirling2007}, diversity concepts display some combinations of three basic properties, namely, `variety', `balance' and `disparity'. 

Variety is the number of variants to which system elements are apportioned. For this work, vehicles technologies are represented according to the distribution of revealed-preference data to people's actual choices. Balance is a function of the pattern of apportionment of elements across categories. Referred to as evenness and concentration, balance is analogous to statistical variance \citep{pielou1977}. Thus, instead of utilising a mean cost comparison \citep{schaefer2005,grahn2009}, this work uses a probabilistic treatment with variance reflecting of heterogeneity of products (see \citealt{Mercure2012}). Disparity refers to the manner and degree in which the elements may be distinguished. It is implicitly implied in the representations used to characterise variety and balance \citep{stirling2007}. The degree of disparity between categories of products (or technologies) determines cross elasticities and the degree of substitutability between products and, thus, is consistent with discrete choice theory. 

Product diversity can be interpreted as different consumers using different varieties, or as diversification on the part of each consumer \citep{dixit1977}.  The basic premise is that product variety reflects the requirements of market segments \citep{smith1956,adner2001, horrace2009}. Product differentiation is concerned with bending the will of demand to the will of supply \citep{smith1956}. Thus, the diversity in products, represented by cost distribution in FTT: Transport, reflects choices diversity. 

Evolutionary thinking in technology has long argued that competitive selection can only operate if there is sufficient economic diversity of behaviour \citep{turner1992, metcalfe1994}. The reasons are two-fold. Firstly, the difference between consumer choices drives competition between varieties, leading to a selection process that determines the rate of technological change \citep{basalla1988,saviotti1995}. Secondly, the basic premise of the adoption and diffusion of technology is that there are different categories of adopters \citep{Rogers1985,Rogers2010,slater2006}. Diversity of choices is therefore crucial in determining the rate of technology diffusion. 
 
\subsection{Social influence and consumer choices} 

Social influences and consumer choices are consistently found to be important in energy modelling. Within the transport sector, behaviour adaptation, network interactions and diversity of consumer preferences are central to the understanding of technology adoption \citep{mueller2009,McShane2012}. The rate of technological update is influenced by the diversity of the agent's perception towards car purchases. The difference between consumer choices drives competition between product varieties, leading to a selection process that determines the rate of technological change \citep{basalla1988}. Based on \cite{Rogers2010}, diversity is responsible for the gradual adoption of innovations and technology diffusion. 

Within the transport sector, consumer choices for certain vehicle technologies take place within contexts of distributed income that span several orders of magnitude \citep{mercure2015effectiveness}. The fact that consumers are diverse implies that car technologies will not diffuse into the market instantly in the presence of a change in price or policy reform. Instead, car technologies diffuse gradually according to consumers' choices and the heterogeneity within the population. 

General equilibrium and partial equilibrium (optimisation) models do not sufficiently account for agent diversity \citep{mercure2016modelling}. People and firms in these models are represented by a representative agent with rational expectations. There are various inherent limits for informing policy making if the model does not take into account agent heterogeneity. In the equilibrium model, agents all respond in the same way to changes in government policy. Without social influence, the market shares for the technology will change only if the incentives change \citep{Mercure2018c}. 

Existing studies have shown the important influence of behavioural assumptions on policy-relevant outcomes in the diffusion of low emissions vehicles \citep{mau2008neighbor,li2017actors,Pettifor2017a,Pettifor2017b}. This is because individual decisions are strongly affected by social norms and customs when choosing a car \citep{McShane2012}. Instead of instantaneous change in shares when an incentive is imposed, the technology diffusion will be shaped by a new diffusion trajectory. However, in the optimization model, as long as solutions remain with the multidimensional box of constraints, without considering any self-reinforcing effects, the shares for technologies rise instantaneously when an incentive is imposed and this is not consistent with the diffusion trajectory of technology (i.e. market shares only change when incentives change for the optimisation model, while in a diffusion process, the market shares change without the need for a change in incentives).

Capturing behavioural realism in consumer preferences of passenger cars in global IAM increases their usefulness to policy makers. Modellers have attempted to incorporate some behavioural realism in existing global IAMs. For instance, \cite{wilson2015improving, Pettifor2017a} represents heterogeneous consumer groups for vehicle choices with varying preferences for vehicle range and variety in the MESSAGE model. While this approach considers consumer heterogeneity by segmentation, the MESSAGE model remains an optimisation model without self-reinforcing effects present in the diffusion of technologies \citep{McCollum2016}. 

The socio-MARKAL model integrates technological, economic and behavioural contributions to the environment in a few cities (e.g Nyon) \citep{nguene2011socio}. Similarly, \cite{Daly2014} incorporates travel behaviour into the TIMES model by accounting for individual travel budget constraint for Canada and Ireland. \cite{bunch2015incorporating} incorporates behavioural content from MA3T (Market Acceptance of Advanced Automotive Technologies) into the TIMES model. The MA3T simulates vehicle market behaviour over time, where the core behavioural model is a nested multinomial logit discrete choice model that yields market shares of competing technologies for a large number of consumer segments. However, the MA3T is limited to projecting the behaviour of the vehicle market in the US under alternative policy scenarios.

While the above research does improve the behavioural realism of the global IAMs, it is possible to improve further the representation of consumer behaviour with the FTT-Transport model. Firstly, The E3ME-FTT-Transport model is an attempt to model consumer preferences change over time as a result of changes in trends, fashion and income. Secondly, in the E3ME-FTT-Transport model, consumer behaviour is integrated into the IAM in a global scale. This is different to the existing equilibrium models where the research mostly focus on a few countries and regions. Thirdly, with a improved representation of consumer behaviour, the model allows a higher numbers of policy levers in the model for the analysis of detailed policy incentives.

\subsubsection{Modelling choice with intangibles} 

People's purchasing behaviour is not only affected by financial costs, but also by comfort, luxury, practicality and aesthetics \citep{axsen2012,McShane2012}. However, since these behavioural parameters are difficult to quantify, few attempts have been made to include intangible costs in technology models. With few exceptions, SOCIO-MARKAL modelled behaviour through sociological surveys in order to capture the perception of the population regarding energy consumption \citep{nguene2011}. Similarly, the CIMS user can specify an intangible cost factor to characterise estimated real-world consumer preferences \citep{jaccard2003}. Focusing on residential energy projection, both the MESSAGE-Access model and the IMAGE-REMG model introduce `inconvenient costs' that capture some of the non-monetary aspects of households preferences \citep{pachauri2013pathways,farsi2007fuel,maconachie2009descending}. 

Similarly, in FTT: Transport, intangible costs are included in the consumer decision function to represent all the perceived costs (or benefits) of a technology that are not derived from its financial attributes. Since survey data is not available on a global scale, the intangible costs are empirically determined from historical data for each technology. 

\subsection{Policy assessment for emissions from the transport sector} \label{sec:EngPol}
 
Energy models are useful for policy makers to assess the impact of policy incentives on the emissions from the transport sector in the long term. At a national level, a number of models have been applied to analyse the effectiveness of policy incentives on emissions reduction. For instance, \cite{kloess2011simulating} investigates the effect of various tax incentives and technological progress on the Austrian passenger car fleet. Using the UK Transport model, \cite{brand2013accelerating} assesses the long term scenario of low carbon fiscal policies and their effects on transport demand. \cite{mccollum2009achieving} examines the potential of deep cuts in US transportation using Long-term Evaluation of Vehicle Emission Reduction Strategies (LEVERS) model. \cite{alam2017assessment} modelled passenger car fleets in Ireland from 2015 to 2050 to assess the impact of current and potential greenhouse gas mitigation policies using the integration of the COPERT model and the W2W model to assess the impact of current and potential greenhouse gas mitigation policies.

Most models for detailed policy analysis have been focused on a particular region or country. The shortcoming of studying one country is that such models have limited use in terms of the transport policy incentives in the global climate change context.  Global IAMs are coupled with land use, agricultural and climate change models, and have been used to assess the medium to long term impact of transport policy on global climate change. With the policy analysis tool linked to a global IAM, it is possible to derive insights on the systematic consequence of technology policy options in the transport sector.

In order for global IAMs to be useful for policy assessment, it is crucial to feature at least a few policy instruments that realistically represent real world climate policy. While in reality, in the transport sector, climate policies feature a wide range of different types of incentives, most IAMs feature a few, sometimes only a single, policy lever for decarbonisation, a carbon price that is applied to all sectors targeted by the climate policy. As argued by \cite{grubb2014planetary}, it is likely that a carbon price alone will not be sufficient to achieve the climate target. Moreover, all Emissions Trading Schemes to date exclude road transport, so there is little reason to study the impact of a carbon price on road transport. To be realistic, models needs to study the impact of at least vehicle taxes, road taxes, fuel taxes, regulations, standards and biofuel mandates, since these are very common across the world, often all used simultaneously. Some recent studies recognise this limitation of current IAMs and integrate additional policies, including fuel taxes, registration or road use \citep{deetman2013deep,yin2015china, zachariadis2005assessing};  Technology subsidies and mandates \citep{bertram2015complementing, deetman2013deep,yin2015china, zachariadis2005assessing}; efficiency policies \citep{deetman2015regional,siskos2015co,yin2015china,zachariadis2005assessing}, biofuel blends and mandates \citep{calvin2014eu}. The FTT model includes eight policy instruments (vehicle taxes, registration taxes, fuel taxes, vehicle subsidies, regulations, fuel economy standards, biofuel mandates and possible kick-start programs) and can be used to model policies for each individual countries.

Policy assessment in the transport sector requires coping with a large number of decision makers, involved complex interaction between consumers and vehicle markets. However, the existing global IAMs are predominantly optimization models, assume a social planner and no interactions between consumers. In some interpretations, this implies that agents respond to policy incentives in a collective manner based on system-wide cost-minimisation or utility-maximisation criteria. However, in reality, no-one in society faces the system cost, each individual faces his/her own costs and benefits in his/her own social context. Therefore, while a social planner perceives tradeoffs between, for example, vehicle fleet carbon costs and power generation carbon costs, and can trade them for one another under the prevailing carbon price, in reality investment happens independently by different people in different sectors, and different sectors use different ranges and types of policies. 

\newpage
\section{Detailed theoretical model description for FTT:Transport}

The theory is presented here. A list of variable definitions is given in appendix~\ref{sect:AppendixListVar}.


\subsection{Basic equations \label{sect:basicequations}}
\subsubsection{Transport demand}

The model starting point is with an exogenous total transport service demand $D(t)$, in million vehicle-kilometres per year (denoted Mvkm/y), and a fleet size $N(t)$, in thousands of vehicles (denoted kv), derived from vehicle sales, $\xi(t)$. In FTT:Transport, we consider the choice of purchasing a vehicle, which happens every few years, quite independent from the decision to use a vehicle, which is a daily decision. Thus, depending on conditions, people may purchase vehicles that are later not used as much as expected (e.g. say due to higher than expected fuel prices), or more than expected. Transport services are also generated for people who do not own vehicles. Hence we assume that the vehicle owner and the vehicle user are different agents, where, to keep the model concept and equations clear, the vehicle owner sells a transport service to the vehicle user, even when both are the same person.

For that reason, the model features two independent econometric specifications as part of the macroeconometric model E3ME, one for the demand for transport $D(t)$ and one for the demand for new vehicles (i.e. sales) $\xi_{tot}(t)$ (in kv/y), extrapolated from historical data using a number of socio-economic parameters (we come back to this in section \ref{sect:econometrics}). $D$ is obtained using chosen econometric relationships, function of macroeconomic variables: economic growth, employment, fuel prices, etc, while sales $\xi(t)$ are determined from disposable income and vehicle prices and operation costs. The role of FTT:Power is primarily to determine which transport technologies supply these demands, populating vehicle fleets, and which type and quantity of energy the fleets use.

Vehicles survive for a number of years statistically, something that we discuss in more detail in section \ref{sect:survival}, giving a total number of vehicles $N(t)$ that is endogenous and dependent on the sales $\xi(t)$ (we assume the survival rate constant):
\beq
N(t) = \int_0^\infty \xi(t-a) \ell(a) da,
\eeq
where $\ell(a)$ is the survival function, the probability of a vehicle to still be operational on the road at age $a$, i.e. $a$ years after purchase (see sections~\ref{sect:survival} and \ref{sect:survival-Data} for information on the survival function). Thus the fleet is composed of vehicles with a mixture of ages that depends on how many vehicles were purchased in each past year and their probability of having survived to the present \citep[see][for an exposition on technology survival]{Mercure2015}.

We assume here that sales and the demand for transport services is independent of the type of engine that vehicles have. This may not always be true, if for example, infrastructure for electric vehicles is not as extensive as that for liquid fuel Internal Combustion Engine (ICE) vehicles, or motorcycles which not everyone may like. Nevertheless, this makes our model simple, tractable and useable at a global scale with the data that we have, and we consider it good enough for the purpose at hand.

The relationship between the two independent variables $D(t)$ and $N(t)$ generates an endogenous average capacity factor $\overline{CF}(t)$, which expresses to which degree, or intensity, vehicles are used (in Mvkm/y/kv = k-km/y),
\beq
D(t) = \overline{CF}(t) N(t).
\eeq
We have designed FTT:Transport to calculate transport generation in person kilometres rather than vehicle kilometres, enabling to include vehicles with different numbers of seats, notably motorcycles. Road transport technologies (e.g. petrol, electric, motorcycles) by class (economic, mid-size, luxury), indicated with a subscript $i$, generate each a transport service component $G_i$ (in million person-kilometres per year, Mpkm/y), using a total seating capacity $U_i(t)$ (in thousands of seats, ks),
\beq
G_i(t) = CF_i(t) U_i(t) = FF_i N_i d_i(t) U_i(t),
\eeq
where $CF_i(t)$ is now technology dependent, and further subdivided into an average filling factor (occupancy rate) $FF_i$ (number of passengers per seat, p/s), determining how full vehicles are on average when they travel, which we consider constant over time, as well as an average distance travelled per year $d_i(t)$ (km/y), which is itself function of the fraction of time vehicles are used and at what average speed. $CF_i(t)$ varies mainly between vehicles and motorcycles, which on average travel shorter distances per year.

In the order of the calculation, we first determine $U_i(t)$ in order to obtain $G_i(t)$. For this, we define the transport technology market share:
\beq
S_i(t) = {U_i(t) \over N(t)}, \quad N = \sum_i U_i.
\eeq

We follow the lines of the main paper and of section~\ref{sect:replicator}, referring the reader to earlier work for a highly detailed treatment \citep{Mercure2015}. Shares, first determined from historical data, evolve over time following replicator dynamics equation, which is explained in section \ref{sect:replicator}. The replicator dynamics is the source of dynamical behaviour of FTT models, and as we see further down, represents technology diffusion following S-shaped curves. 
\beq
\Delta S_i = \sum_j S_i S_j \left( {F_{ij} \over \tau_i} - {F_{ji} \over \tau_j} \right) \Delta t.
\label{eq:replicator1}
\eeq
The average capacity factor $\overline{CF}$ is obtained from shares,
\beq
D(t) = N(t) \overline{CF}(t), \quad\overline{CF}(t) = \sum_i S_i(t) FF_i N_i d_i(t).
\eeq
This enables to write the capacity in terms of the demand and shares quite conveniently:
\beq
U_i(t) = {S_i(t) \over \overline{CF}(t)} D(t).
\eeq
In this equation, the number of vehicles in a category can change for three independent reasons: (1) the total demand can change, (2) the technology composition can change and (3) the efficiency at which vehicles are used can change. This is summarised by the following differential form:
\beq
\Delta U_i = {S_i \over \overline{CF}}\Delta D +{D\over \overline{CF}} \Delta S_i - {S_i D \over \overline{CF}^2} \Delta \overline{CF}.
\eeq
These are three chosen independent variables that influence the evolution of the numbers of vehicles by category; all other variables are by construction functions of these three.

\subsubsection{Fuel use and emissions}

One of the main outputs of this model are the emissions of greenhouse gases and other pollutants of type $j$ (in t/y, e.g. tCO$_2$/y):
\beq
E_i^j = J_i \alpha_i^j,
\eeq
where $J_i$ is the amount of fuel used by vehicles of type $i$, while $\alpha_i^k$ is the emissions factor for pollutant of type $k$ emitted by technology of type $i$, a property specific to the fuel used by that technology. In the case of petrol and carbon dioxide, $\alpha_i^k$ has units of tCO$_2$/GJ. However in the case of electric vehicles, no emissions of pollutants occur at the vehicle level. 

Fuel use is function of vehicle use, and in most cases does not vary by large amounts whether the vehicle is used at capacity or not. It does not scale proportionally to the filling factor, but is a rather complicated function of that variable, depending the speed profiles, purpose of driving and driving habits, where more acceleration means higher fuel use for larger filling factors. Such information is not readily available, and therefore, as an approximation, fuel use is assumed independent of the filling factor of vehicles. Vehicle of type $i$ uses fuel of type $k$ (in PJ/y), 
\beq
J_i^k(t) = {G_i(t) \over FF_i} \beta_i^k,
\eeq
where $\beta_i^k$ is the fuel consumption per kilometre driven (in MJ/vkm) for technology $i$, a property of the technology.

\subsubsection{Investment, learning and cost reductions}

One of the reasons for calculating the capacity $U_i$ is that costs evolve with the cumulative production of vehicles due to learning by doing. Each year a certain amount $\xi_i$ of transport capacity of type $i$ comes out of factories and is registered, sold at cost $IC_i$, generating investment $I_i = IC_i \xi_i$. Registrations correspond to positive increases in numbers of vehicles plus the replacement of scrapped vehicles, while negative changes correspond to destructions that are not replaced. 
\beq
\xi_i(t) = \left\{ \begin{array}{ll}  {dU_i \over dt} + {U_i \over \tau_i},  & {dU_i \over dt} > 0 \nn\\
{U_i \over \tau_i} &  {dU_i \over dt} \leq 0 \end{array}, \right.
\eeq
where the term $U_i / \overline{\tau}$ refers to vehicle replacements, with $\overline{\tau}$ a technology life expectancy \citep{Mercure2015}. 

Learning curves are expressed as cost reductions that occur with the cumulative production of units of technology, or in our case, cumulative sales of vehicles. The usual form in which learning curves are expressed is
\beq
IC_i(t) = IC_{0,i} \left({W_i(t) \over W_{0,i}}\right)^{-b_i},
\eeq
where $W_i$ is the cumulative investment and the pair $IC_{0,i}$, $W_{0,i}$ are corresponding costs and cumulative sales at a particular point in time, taken here as the start of the simulation and $b_i$ is the learning exponent, related to the learning rate.\footnote{The learning exponent is $b_i = \ln(1-LR_i)/\ln(2)$, where $LR_i$ is the learning rate.} Learning however happens on a component level rather than at the technology level (e.g. engines, batteries, materials), which may be used in more than one type of technology, and therefore sales in one technology category may induce learning in other categories. A spillover matrix $B_{ij}$ is thus defined, mixing the learning:
\beq
W_i(t) = \sum_j B_{ij} \int_0^t \xi_j(t') dt', \quad W_{0,i}(t) = \sum_j B_{ij} \int_0^{t_0} \xi_j(t') dt'.
\eeq

\subsection{The decision-making model core }
\subsubsection{Perceived costs and decision-making}

We detail further here our model of decision-making in the context of diverse agents. For a model of technology diffusion, we require an aggregate representation of decision-making when agents are diverse, and costs have variations. This core model component is evaluated at every time step, and the decision-making determines the composition of new technology units purchased, which in time gradually changes the overall transport fleet. 

Diversity stems from different perceptions from agents when they take a decision, which may originate from a large set of particular preferences and constraints that is impossible to enumerate in a model. We summarise this by distributions. We assume that choice is made on the basis of a single quantity, a generalised cost, evaluated by agents for each option they see as available to them, and this value features a quantification of all possible aspects that weigh in the decision-making balance. 

We postulate here that \emph{distributions of perceived costs correspond to distributions of observed costs}, with a possible constant offset between them. People, we assume, when considering purchasing a vehicle, most likely choose something they have seen being purchased, perhaps by someone they know such that they were able to gather information (i.e. they most likely do not choose something they know nothing of, and they gather reliable information predominantly through observations of their peers). Their observations of the fleet is a subset of what is on roads, and every agent observes something slightly different from every other. This may be due to their belonging to a particular social group and social class, and they are most likely to choose amongst what their peers have previously chosen, which itself is a subset of what the whole market has to offer (e.g. poor rural households perhaps purchase different types of vehicles to rich suburban families, which itself is different than single middle-class persons, i.e. their peers are a subset of the population and their observations are a subset of all observations). Thus we assume \emph{restricted technology/information access}, in other words, agents do not choose what they do not know, and they do not know all of the market (or even perhaps do not care for all of the market). \emph{Importantly, this means that FTT does not have a representative agent}. Choices of particular social groups endure through peer observation and visual influence, which has been demonstrated is the case for vehicle purchases in \cite{McShane2012}. 

The frequency of observations of a particular model (by consumers shopping), sample of an ensemble of such events, corresponds to the frequency of recent sales of that model (purchases by their peers). We then postulate that the probability of choosing a particular model is proportional to this frequency of observation, and thus these preference distributions, associated to circumstances and constraints of consumers difficult to enumerate and unknown to the modeller, are relatively stable. These combined frequencies form a generalised cost distributions of sales (see for example the sales distribution in Figure \ref{fig:UKCarDistributions}). In this perspective, \emph{the generalised cost distribution of recent sales is a representation of the diversity (hegerogeneity) of choices}. We go further and say that we can use the measured heterogeneity of sales and interpret it in terms of the heterogeneity of agents. Furthermore, in such a perspective, vehicle sales by vehicle model reinforce the sales of those very models, consistent with sociological evidence (see e.g.\cite{McShane2012}). 

\subsubsection{Diversity is crucial}
\begin{figure}[t]
	\begin{minipage}[t]{0.5\columnwidth}
		\begin{center}
			\includegraphics[width=1\columnwidth]{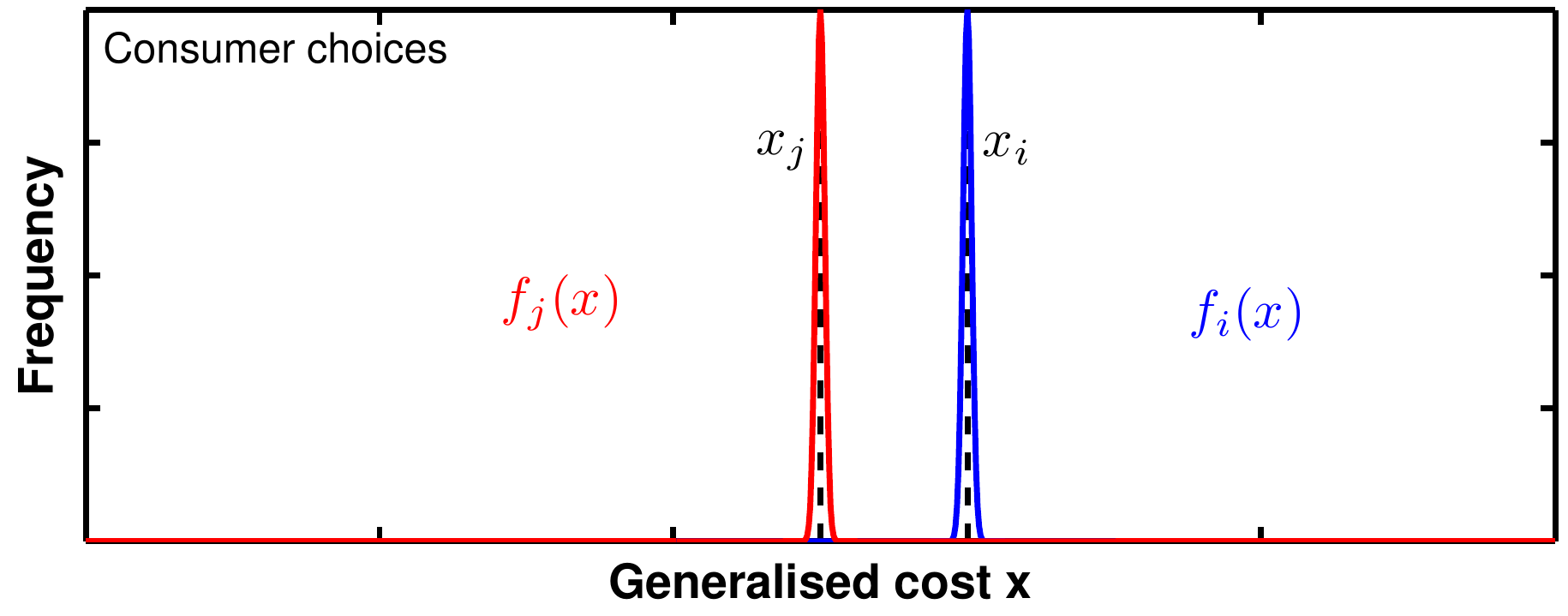}
			\includegraphics[width=1\columnwidth]{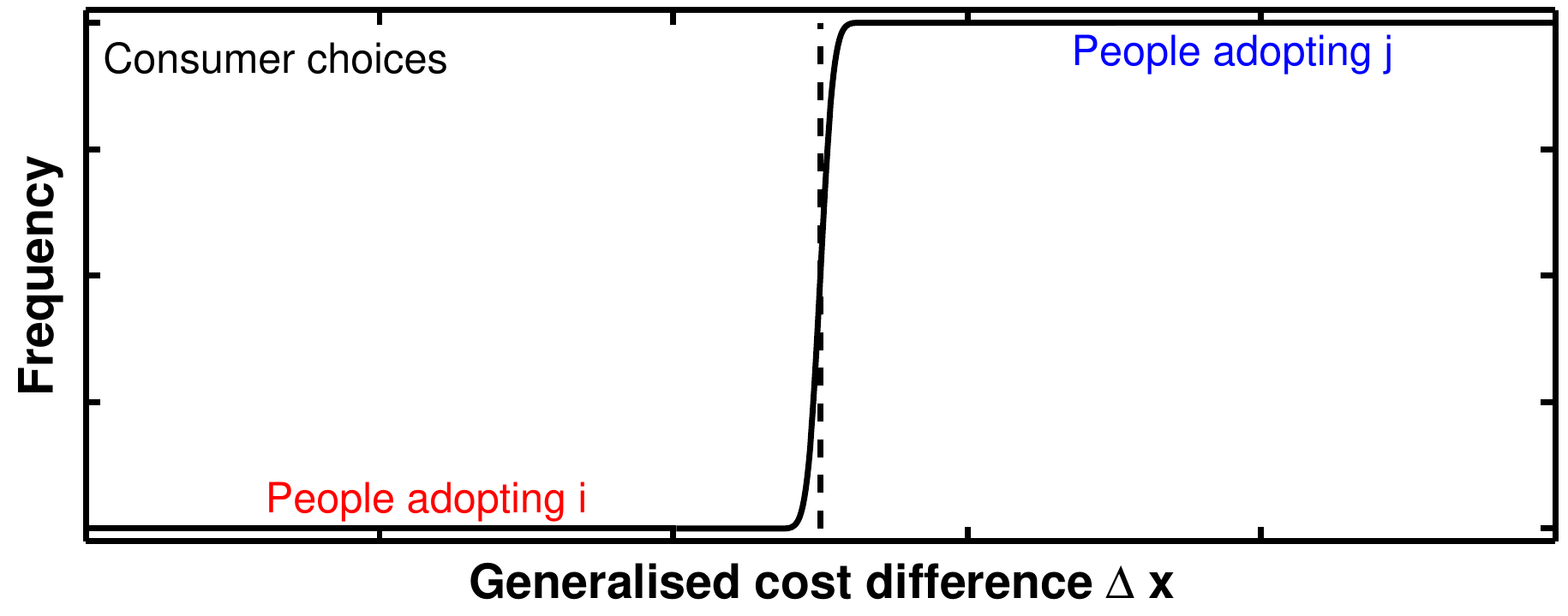}
		\end{center}
	\end{minipage}
	\hfill
	\begin{minipage}[t]{0.5\columnwidth}
		\begin{center}
			\includegraphics[width=1\columnwidth]{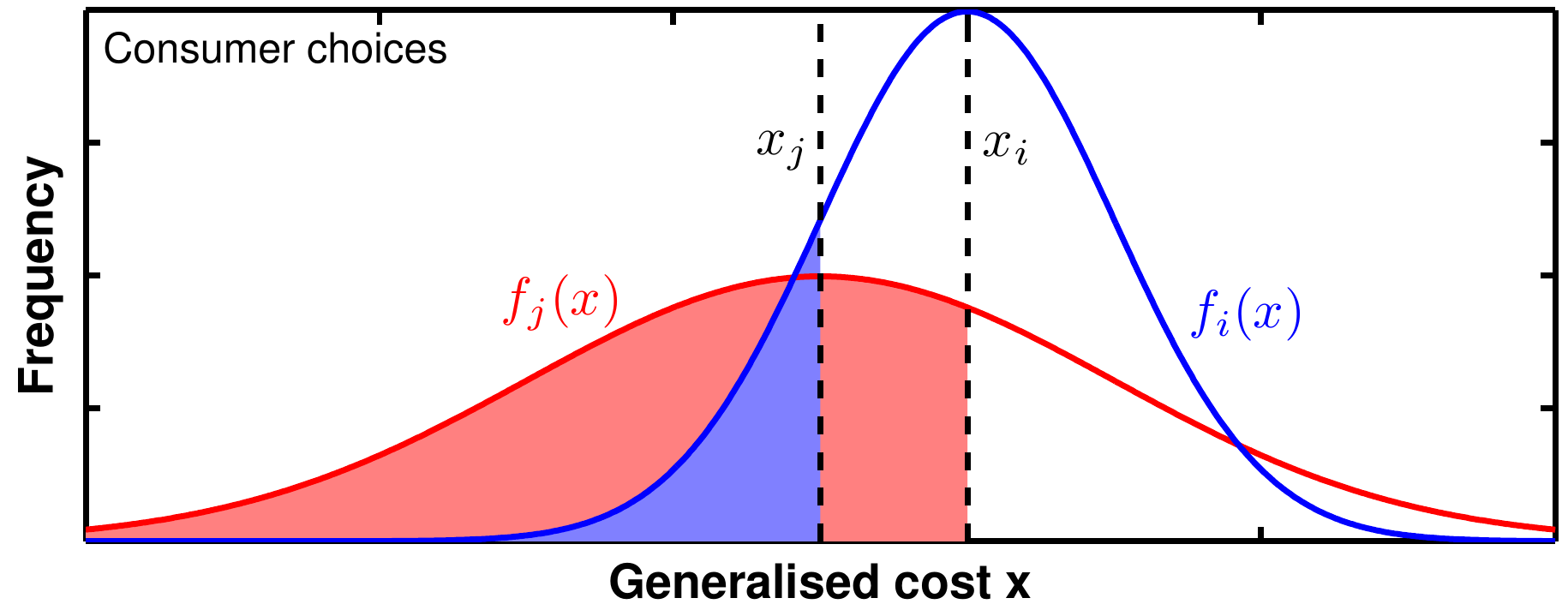}
			\includegraphics[width=1\columnwidth]{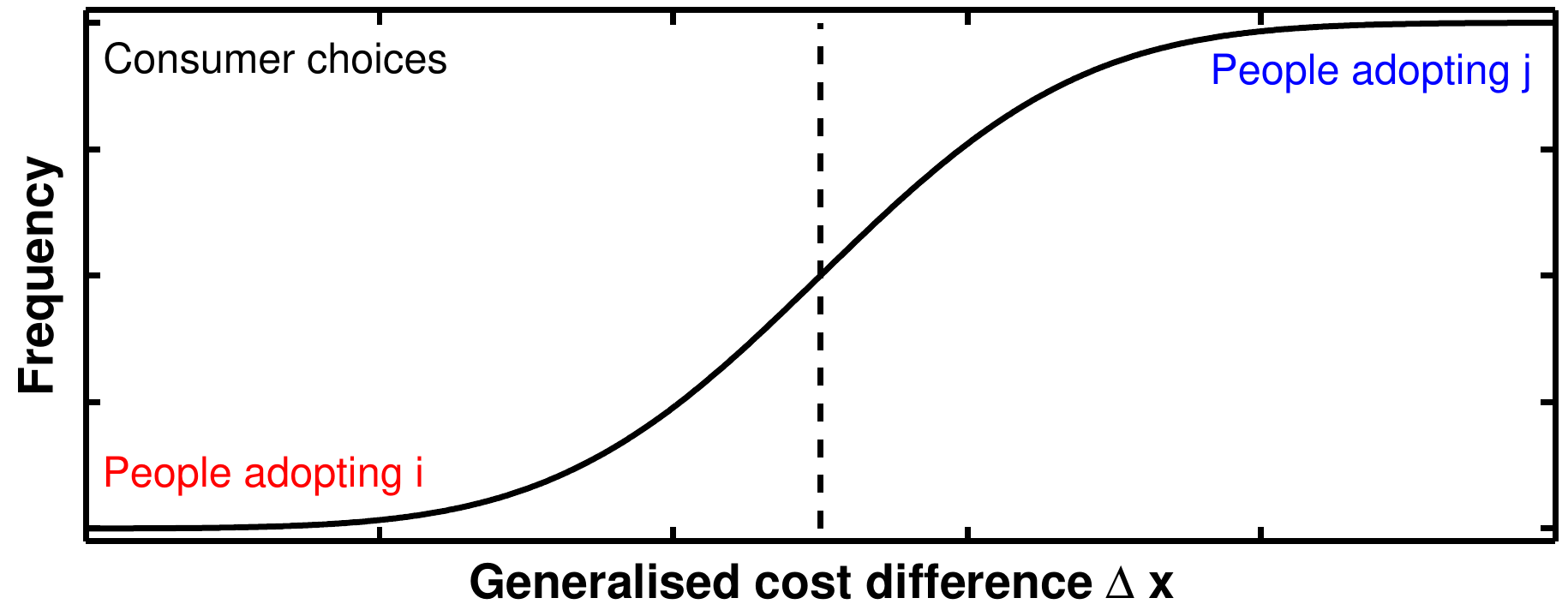}
		\end{center}
	\end{minipage}	
	\caption{Illustration of the process of decision-making under diversity of agents between two technologies. The blue curve represents the distribution of perceived generalised costs for one technology, and the red curve that of the other. In the left panel, if diversity is very low, choices can flip very abruptly as average costs cross. This corresponds to the representative agent case. In the right panel, introducing significant diversity makes choices distributed and choices change very gradually as costs cross.}
	\label{fig:Choices}
\end{figure}

The importance of diversity must be emphasised. Its is often taught in basic diffusion theory that the various parts of the diffusion curve belongs to different types of consumers: early adopters, middle adopters, followers, laggards, etc. This crude picture is useful here, as it connects the notion of diversity to a rate of technology adoption. We show below that it is a very important assessment of the problem, as follows.

Vehicle purchases are distributed in time, by consumers that take different decisions at different points in time for different reasons. If we were to imagine, temporarily, that agents had identical preferences and constraints, were there to be only two or more options for vehicles in the market with non-identical properties, they would nevertheless always all choose the same technology. We could furthermore temporarily imagine that agents know all of the market perfectly, and thus know about all these vehicle models that no one buys, and finally consider (by construction) that the vehicle model that everyone keeps buying is the less expensive in terms of generalised cost. Then, if for one reason or another, the generalised cost difference between of one of the unused technologies and the common one was to cross zero, then from then onwards all agents would simultaneously change their preference and the adoption of the new vehicle type would be instantaneous were it not for probable industrial supply problems. If we depict this situation using cost distributions for technologies and a cumulative probability distribution for choice in terms of a cost difference, we obtain the representation shown on the left panel of fig.~\ref{fig:Choices}. We know that this is an unrealistic representation. When we make abstraction of industrial growth dynamics, the diffusion of technology is a process that results from consumer heterogeneity.

We take a \emph{diverse} group of technologies purchased by a \emph{diverse} group of consumers. Comparing these technologies based on consumer choices (i.e. on the basis of their generalised cost) leads to a comparison of frequency distributions, shown in the right panel of fig.~\ref{fig:Choices}. That these distributions have unequal means signifies that one is on average less expensive than the other. However this does not mean that this is the case in all individual situations where a consumer makes a choice, in his own perspective, but it should be the case a major fraction of decisions. Thus if the generalised cost difference, say the difference between the means, gradually decreases to zero, at each value of this difference a larger fraction, but not all, of consumers will choose the technology which is on average the less expensive, until when means are equal exactly half of consumers choose each. As the cost difference crosses zero, this fraction decreases gradually below 50\%. The resulting profile of adoption is then a very gradual one, the steepness of which depends on the widths of the distributions (the degree of heterogeneity), as we show next, and this results from all consumers having slightly different perspectives.

\subsubsection{Pairwise choices comparisons of distributed choices \label{sect:distributions}}

FTT:Transport operates by using chains of binary logits, which is made clear in section~\ref{sect:replicator}. We describe here the binary logit itself, i.e. preferences in a group of heterogenous agents for every possible pair of options. Combining them, with the unequal frequencies at which they take place, will later yield the replicator equation.

We assume that we have two cost distributions for two vehicle types for what we assume are the relative numbers of situations where agents, stating their individual preference between technologies $i$ and $j$, face different situations and state different choices. By counting how many agents prefer which technology in each pair, one can state what the probabilities of preferences between these two technologies are for future situations where choices are to be made (e.g. 70\% of agents choose $i$ and 30\% $j$). 

We denote these generalised cost distributions $f(C,C_i,\sigma_j)dC = f_i(C-C_i)dC$ and $f(C,C_j,\sigma_j)dC = f_j(C-C_j)dC$, where $C_i,C_j$ are the mean generalised costs and $\sigma_i,\sigma_j$ are their standard deviations, for technologies $i$ and $j$. These distributions can be of any kind, but they require to have a single well defined maximum and variance (e.g. they cannot have two maxima\footnote{In which case we would need to subdivide such a technology category into two.}). For FTT:Transport, we found that these are lognormal, and therefore we make cost comparisons in log space.

We evaluate the probability of choosing $i$ over $j$ using the following. First, we calculate the probability of choosing $i$ in all cases where $j$ has an arbitrary cost $C$. Our central assumption is that the fraction of agents for whom the generalised cost of $j$ is $C$ and for whom the cost for $i$ is lower than $C$ will choose technology $i$ over $j$ if given a choice, and this fraction is equal to the cumulative probability distribution $F_i(C-C_i)$. But this situation occurs a fraction $f_j(C-C_j)$ of the time, giving a total probability
\beq
P(C_i < C | C_j = C) = F_i(C-C_i) f_j(C-C_j)dC,
\eeq
while the converse is
\beq
P(C_j < C | C_i = C) = F_j(C-C_j) f_i(C-C_i)dC.
\eeq
In order to know how often the cost of technology $i$ is lower than that of technology $j$, and the converse, we must sum over all possible values of $C$. For simplicity, we use as variables $C' = C-C_j$ and $C'' = C - C_i$, with the mean cost difference $\Delta C = C_i - C_j$:
\beq
F_{ij}(\Delta C) = P(C_i < C_j) = \int_{-\infty}^{+\infty}  F_i(C'-\Delta C) f_j(C')dC',\nn
\eeq
\beq
F_{ji}(\Delta C) = 1- F_{ij} = P(C_j < C_i) = \int_{-\infty}^{+\infty} F_j(C''+\Delta C) f_i(C'')dC''.
\eeq
This appears difficult without further knowledge of the distribution type, however we can use a simple reasonning: we take a derivative with respect to $\Delta C$, which makes the integral a convolution of the two distributions
\beq
{d F_{ij} \over d\Delta C}  = -\int_{-\infty}^{+\infty} f_i(C'-\Delta C) f_j(C')dC' = - f_{ij}(\Delta C) = \int_{\infty}^{-\infty} f_i(C'') f_j(C''+\Delta C)dC'' = - f_{ji}(-\Delta C) = {d F_{ji} \over d\Delta C}.
\eeq
As derived in appendix~\ref{sect:AppendixA}, this convolution yields a new distribution $f_{ij}(\Delta C) d\Delta C$ of which the standard deviation is $\sigma_{ij} = \sqrt{\sigma_i^2 + \sigma_j^2}$. This is the probability distribution of technology switching in terms of $\Delta C$. The difficult integral having been computed, this distribution can be integrated again as a function of $\Delta C$ to yield a cumulative probability distribution that technology $i$ is less expensive than $j$ (and conversely):
\beq
F_{ij}(\Delta C)  = \int_{-\infty}^{+\infty} f_{ij}(\Delta C)d\Delta C = 1 - \int_{-\infty}^{+\infty} f_{ji}(\Delta C)d\Delta C = 1-F_{ji}(\Delta C).
\eeq
Thus given a choice between technologies $i$ and $j$, the fraction $F_{ij}$ of agents tends to choose technology $i$ and the fraction $F_{ji}$ chooses $j$, these fractions being functions of the generalised cost difference, and this function has a standard deviation that follows the sum of the squares $\sigma_{ij} = \sqrt{\sigma_i^2 + \sigma_j^2}$. Note that this calculation is independent of probability distribution type; however $F_{ij}(\Delta C)$ should have roughly the shape of a `smooth' step function, its `smoothness' determined roughly by the combined root mean square widths of \emph{both} cost distributions.

Note that if there is a transformation under which $f_i(C-C_i)dC$ is a normal distribution, then $f_{ij}(\Delta C)d\Delta C$ is also normally distributed (the convolution of normal distributions together yields normal distributions, see appendix~\ref{sect:AppendixA}). In the case of FTT:Transport, costs are lognormally distributed, and therefore so is $f_{ij}(\Delta C)d\Delta C$. However the logistic function is a relatively close approximation of the normal distribution, but much faster to calculate than the error function, and thus in FTT:Transport, we compare logistic distributions of the log of the generalised cost.

\subsubsection{Relationship to discrete choice modelling and the binary logit model}

This model is equivalent in many respects to a binary logit model, as initially derived by McFadden \citep[see][]{Domencich1975,Ben-Akiva1985}. In the binary logit model, one assumes that heterogenous consumers maximise their individual utility (or minimise their cost), which is defined in terms of several factors including prices but also other non-monetary factors. The cost associated with choices $i$ and $j$ for a population is written in terms of an average value and a distributed value $\epsilon_i$ of mean zero and standard deviation $\sigma)$, which represents consumer diversity (i.e. differences of perception, situations and constraints, which are unknown to the modeller):
\beq
C_i = \overline{C}_i + \epsilon_i \quad C_j = \overline{C}_j + \epsilon_j.
\eeq
The difference in cost $C_i-C_j$ is thus also distributed. The assumption is taken that $\epsilon$ varies following a Gumbel distribution located at the mode value $C_i$
\beq
f_i(C) = {1 \over \sigma} \exp\left(-e^{C - C_i \over \sigma}\right)e^{C - C_i \over \sigma} \quad F_i(C) = \exp\left(-e^{C - C_i \over \sigma}\right).
\eeq
This distribution is common whenever a distribution is made of the maxima of underlying distributions, as in extreme value theory. Among others, the Gumbel distribution has the property that a distribution of the extreme values of Gumbel distributions follows a Gumbel distribution. It also has the property that the sum of two Gumbel distributed values follows a logistic distribution, an aspect used here. Thus the probability of consumers with distributed generalised costs (or utilities) choosing technology $i$ over technology $j$ is
\beq
P_i(\overline{C}_i + \epsilon_i > \overline{C}_j + \epsilon_j) = P_i(\epsilon_i - \epsilon_j > \overline{C}_j - \overline{C}_i ),
\eeq
which involves calculating the difference between both distributions $\epsilon_i - \epsilon_j$. This corresponds to a convolution of $f_i(C)$ with $f_j(C)$ (as done above), equal to a logistic distribution:
\beq
f_{ij}(\Delta C_{ij}) = \int_{-\infty}^{\infty} f_i(C'-\Delta C_{ij}) f_j(C') dC' = {\exp\left( \Delta C_{ij} \over \sigma \right) \over \left( 1 + \exp\left( \Delta C_{ij} \over \sigma \right) \right)^2},
\eeq
with a cumulative distribution which is a logistic curve:
\beq
F_{ij}(\Delta C_{ij}) = {1 \over 1 + \exp\left( \Delta C_{ij} \over \sigma \right)} = {\exp(C_i/\sigma) \over \exp(C_i/\sigma) + \exp(C_j/\sigma)}.
\label{eq:binarylogit}
\eeq
This result is nearly identical to the one above, with the difference that the result above is more general with technology specific diversity parameters which follow standard error propagation, $\sigma_{ij} = \sqrt{\sigma_i^2 + \sigma_j^2}$, and arbitrary distribution forms, while in the logit model it is explicitly assumed that the widths and shapes of the distributions are identical and that the distributions are Gumbel, leading to a logistic choice function of the difference in generalised cost (or utility). 

\subsubsection{Correspondence between consumer diversity and elasticities of substitution}

The probability of change of choice given a change in context, for instance prices, is generally expressed with elasticities and cross elasticities of substitution, defined in this context as
\beq
{{\partial F_{ij} \over F_{ij}}\over{\partial C_i \over C_i} } = {\partial \log F_{ij} \over \partial \log C_i } = {\partial F_{ij} \over \partial C_i}{C_i \over F_{ij}} \quad \text{and} \quad {\partial F_{ij} \over \partial C_j}{C_j \over F_{ij}}
\eeq
This can be computed using eq.~\ref{eq:binarylogit}:
\beq
\lambda_{ii} = {\partial F_{ij} \over \partial C_i}{C_i \over F_{ij}} = - {C_i \over F_{ij}} {1\over \sigma} {\exp\left( \Delta C_{ij} \over \sigma \right) \over \left(1 + \exp\left( \Delta C_{ij} \over \sigma \right)\right)^2} = -{C_i \over \sigma} F_{ji},
\label{eq:elast1}
\eeq
\beq
\lambda_{ij} = {\partial F_{ij} \over \partial C_j}{C_j \over F_{ij}} =  {C_j \over F_{ij}} {1\over \sigma} {\exp\left( \Delta C_{ij} \over \sigma \right) \over \left(1 + \exp\left( \Delta C_{ij} \over \sigma \right)\right)^2} = {C_j \over \sigma} F_{ji}.
\label{eq:elast2}
\eeq
These elasticities can also be calculated using time series data, if available, of $F_{ij}$ against parameters that include $C_i$ and $C_j$:
\beq
\Delta \log F_{ij} = \lambda_{ii} \Delta \log C_i + \lambda_{ij} \Delta \log C_j.
\eeq
This generates a connection between elasticities of substitution, as calculated from measured time series, and the diversity of consumer behaviour, as measured using price distributions. If we define a diversity parameter as the standard deviation of consumer revealed preferences for technology $i$, measured using sales data as done in this work, normalised by the average cost of that technology, $2 \sigma / C_i$, then when costs are similar, the logistic form of $F_{ji}$ in eqns.~\ref{eq:elast1}-\ref{eq:elast2} can be linearised
\beq
\lambda_{ii} \simeq -{C_i \over \sigma} \left({1\over 2} + {\Delta C_{ij} \over 4 \sigma}\right) \simeq {C_i \over 2 \sigma}, \quad \lambda_{ij} \simeq {C_j \over 2 \sigma}.
\eeq
We thus find that \emph{the elasticity is inversely proportional to consumer diversity}. This can be understood as follows: when diversity is low, consumers tend to all act similarly simultaneously, and this results in price changes having an impact on the whole population, leading to important changes of preference. Meanwhile, when the diversity is high, price changes may have an impact only on a subset of the population, leading to small changes of preference. 


\subsubsection{The levelised cost of transportation}

For the decision-making component of this model, we separate the \emph{investor} in transport technology from the \emph{consumer} of transport services. We think of them as separate entities for clarity, even though in some cases they might happen to be the same person. Whether the roles are fulfilled by the same actors or not, they  are quite distinct, where the \emph{investor} purchases a vehicle to sell a transport service to the \emph{consumer}. This is done in order to clarify the distinction between technology investment and associated market competition, and the consumption of the service technologies produce. It also allows for that even when a person purchases a car, he/she can still travel by train/plane (i.e. not use the car he purchased): the mode choice is distinct from the technology choice, even when performed by the same person. 

The cost of the vehicle, as perceived by the investor purchasing a vehicle or unit of transport technology, must be taken to include all components relevant to the decision making. Many of the components are easy to quantify from available data. Others are not straightforward, and we show here how this is done. When a vehicle is purchased, an initial investment is made, or a loan is obtained, for the capital cost, and henceforth fuel and maintenance costs are incurred for the lifetime of the technology. In addition to this taxes may be added either as a fixed initial cost or as a yearly fee, or both.  

Following the main text of the paper, we define, as a component of the decision-making process, the Levelised Cost of Transport services (LCOT), before policies are applied (in section~\ref{sect:Policies} we add policies):
\beq
LCOT_i = \sum_t {{I_i \over CF_i(t)} + {FU_i(t) \over \beta_i} + {MR_i \over FF_i} \over (1+r)^t} \bigg/  \sum_t 1/ (1+r)^t,
\label{eq:LCOT}
\eeq
where each term is identified in the main text of the paper (see the list of variables in appendix~\ref{sect:AppendixListVar}).

Several terms in eq.~\ref{eq:LCOT} are distributed, while others are single valued. Investment cost distributions can be assigned to a distribution of preferences, but variations can also arise in all other parameters. Since vehicles considered in each category are highly distributed in every one of their characteristics (emissions, price, engine size), most of these parameters are distributed, for example energy use. The discount rate could also be distributed, but we have not included this at this stage. It is to be kept in mind however that in a root mean square calculation, any dominating parameter rapidly makes smaller contributions negligible. Here, the vehicle price distribution dominates (it has the largest standard deviation), but we nevertheless keep energy use and maintenance parameters distributed.

As we show in appendix~\ref{sect:AppendixA}, the distribution of the sum of two distributions corresponds to their mutual convolution, and therefore the sum of several distributions corresponds to a chain of convolutions of all of these. As a result, means are added while the standard deviations are combined using the root of the sum of the squares of the individual standard deviations, as follows (using eq.~\ref{eq:propagation}), leading to
\beq
\Delta LCOT_i = \dfrac{\sum_t \dfrac{\sqrt{\dfrac{\Delta I_i^2 }{ CF_i^2} + \dfrac{I_i^2 }{CF_i^4}\Delta CF_i^2 + \dfrac{\Delta F_i^2}{ \beta_i^2} + \dfrac{F_i^2}{ \beta_i^4}\Delta \beta_i^2 + \dfrac{\Delta MR_i^2 }{ FF_i^2} + \dfrac{MR_i^2 }{ FF_i^4}\Delta FF_i^2} }{ (1+r)^t} }{ \sum_t\dfrac{1 }{ (1+r)^t}}.
\label{eq:dLCOT}
\eeq
This standard deviation of the generalised cost, $\Delta LCOT_i$, is our model representation of agent heterogeneity, parameterised by data. Policies are assumed not distributed; their possible distributional impacts will stem from other distributed parameters already specified here.

\subsubsection{Using log-normal distributions}

Costs, as we show in the data section, are generally distributed asymmetrically and are almost always well described by log-normal distributions, as are many economic processes. As we have just shown, the calculation of preferences $F_{ij}$ does not depend on a particular form of distribution. However, given this property of the data, the generalised cost comparison is better performed in logarithmic space than in real space. For this, one only needs to convert the means and the variances of the distributions measured in real (dollar) space into values measured in logarithmic space, with the following transformations:\footnote{See the wikipedia page on lognormal distributions.}
\beq 
\mu = \ln\left({m^2 \over \sqrt{v+m^2}}\right),
\eeq
\beq
\sigma = \sqrt{\ln \left(1 + {v \over m^2}\right)},
\eeq
where $\mu$ and $\sigma$ are the mean and standard deviation in logarithmic space, and $m$ and $v$ are the mean and standard deviation in normal dollar space. Thus a simple conversion can be made. 

In FTT:Transport, the cost comparison is made in logarithmic space, in other words $C_i$ above is the log of a cost in \$/pkm. Thus the LCOT and its standard deviation are calculated using eq.~\ref{eq:LCOT}~and~\ref{eq:dLCOT}, and subsequently converted using these transformations before carrying out comparisons. This is consistent with the fact that the income distribution is usually lognormally distributed as well.

\subsubsection{The generalised cost as a comparison measure}

As it is inferred from price distributions of sales, transport cost considerations are not the only elements of consumer decisions when purchasing a vehicle. Many additional aspects are valued by the consumer, of which we have little information beyond the price distribution of what they purchase. However, we do have existing trends of diffusion from historical data (section \ref{sect:HistoricalData}). We keep in mind that technologies have highly different pecuniary costs, particularly across engine size classes; and despite this, higher costs appear compensated by higher benefits, such that higher cost luxury vehicles maintain market shares.

Were we to simulate technology diffusion based on bare LCOT distribution comparisons, the lowest LCOT technologies would diffuse more successfully, which, as it turns out, is not consistent with our historical data. Clearly, components would be missing in the LCOT, for instance comfort, acceleration, style, that we may call the `intangibles'. We define these `intangibles' for this model as the difference between the generalised cost, which leads to observed diffusion, and the LCOT as calculated from pecuniary vehicle properties for which we have data. The value of the intangibles, denoted $\gamma_i$, is an empirical parameter that we obtain from making the FTT diffusion trajectory match the trajectory observed in our historical data, at the year of the start of the simulation. The theory goes as follows, while the practical method is described in section \ref{sect:gammapractice}.

The diffusion of technologies in a set takes place at the expense of one another in market share space. According to sections \ref{sect:distributions} the choice of investors is made based on pairwise comparisons of generalised cost distributions. Based on the transformations above, we use the following pair:
\beq
C_i = \ln \left({LCOT_i^2 \over \sqrt{LCOT_i^2 + \Delta LCOT_i^2}}\right) + \gamma_i,
\eeq
\beq
\Delta C_i = \sqrt{\ln \left(1 + {\Delta LCOT_i^2 \over LCOT_i^2}\right)}.
\eeq
When $\gamma_i = 0$, we obtain a rate of diffusion that does not normally match historical diffusion. One, and only one, set of $\gamma_i$ leads to the diffusion of technology in the simulation to have the same rate as the historical rate at the starting point of the simulation. We describe this in section \ref{sect:gammapractice}. The interpretation of the $\gamma_i$ parameters is that they ensure that FTT projects in the future a diffusion trajectory (the rate of change of shares) that is the same as what is observed in historical data, and represents all costs not explicitly specified as perceived by agents. The $\gamma_i$ are not distributed, and thus, are not to be associated with our representation of agent heterogeneity.

\subsection{Population changes as a result of decision-making }

\subsubsection{Survival rates and technology vintages \label{sect:survival}}

The connection between sales and actual vehicle numbers in a fleet depends on the length of time that vehicles survive for. This is described by standard survival (or reliability) analysis \citep[e.g.][]{Liu2012}. Vehicles come to the end of their useful life through various events or processes: accidents, failures or scrapping decisions. The rate of changes in the system depends on this length of time, which determines the size of the markets for second-hand and new vehicles. As we show in section~\ref{sect:UKsurvey}, observations of the UK fleet are available, making possible the parameterisation of vehicle survival rates. This requires a bit of theory on vehicle survival, as follows. This analysis enables to parameterise the rates of fleet turnover, a component of the rates of diffusion. It also determines at which rate a fleet \emph{can} physically be transformed. A detailed theoretical analysis is given in \cite{Mercure2015}.

Vehicles remain on roads for a length of time until they are scrapped, for one of the three reasons given above. For the vehicle fleet and its size, the nature of ownership, and the number of different owners of a vehicle along its life is not important in the perspective of this model. When vehicles are purchased, they remain in the system until they are scrapped, irrespective of their number of owners. What is important is the statistics of its survival: what its probability of making it to a certain age, which is in general very well defined. This is commonly termed the survival function. 

The fleet in year $t$ has an age distribution, which we denote $n_i(a,t')$, where $i$ denotes the technology type, $a$ is the age variable and $t'$ is the year of first registration. This distribution evolves due to ageing, where vehicles gradually move to older age brackets, or with scrapages, where vehicles are taken out. The distribution tends to decrease to zero as $a$ increases, particularly beyond 25-30~years. This change can be expressed in terms of an age dependent probability of scrapage $p_i(a)$, which increases with age. Therefore the change in the distribution for one year of ageing is proportional to the existing distribution:
\beq
\Delta n_i(a,t') = -p_i(a) n_i(a,t')\Delta t' \Delta a.
\label{eq:notSolved}
\eeq
This instantaneous probability of individual vehicles to be scrapped as they age translates to a probability of vehicles to survive up to a certain age, which is the solution to the previous differential equation:
\beq
n_i(a,t')\Delta t' = n_i(0,t')\ell_i(a) \Delta t', \ell_i(a) = \exp\left(-\int_0^a p_i(a')da'\right),
\label{eq:solved}
\eeq
where $\ell_i(a)$ is the survival function, or the fraction vehicles of type $i$ that survive up to age $a$. Assuming that the quality of the make of vehicles does not change with time (we see in section \ref{sect:UKsurvey} that this has been the case in the UK), its life expectancy $\mu_i$ can be derived:
\beq
\tau_i = -\int_0^\infty a {d\ell_i(a) \over da} da = \int_0^\infty \ell_i(a) da,
\eeq
where the second integral is obtained by integrating the first by parts.

The survival function can be obtained in either of two ways, depending in which `direction' one looks at the age distribution of a vehicle population, which is function time ($t$) and age ($a$). In age space at fixed time $t$, one notes from eq.~\ref{eq:solved} that 
\beq
1)\quad \left.{d n_i(a,t') \over da}\right|_t \Delta t' = n_i(0,t') {d\ell_i(a) \over da} \Delta t',
\label{eq:Survival1}
\eeq
and one looks how the distribution evolves between age brackets for a specific year of measurement. Here $n_i(0,t')$ corresponds to new registrations (the age zero population) at different years of make. Meanwhile, with age $a$ fixed, one notes from eq.~\ref{eq:notSolved} that
\beq
2) \quad \left.{d n_i(a,t') \over dt }\right|_a \Delta t' = n_i(a,t') p_i(a) \Delta t',
\label{eq:Survival2}
\eeq
where one looks at how the distribution evolves in time $t$ by comparing the change of population within specific fixed age brackets (e.g. specific years of make). Here $p_i(a)$ is the probability of making it to the next year given a certain age $a$.

In the first case, one can obtain the survival function $\ell_i(a)$ (or the probability of death $d\ell_i / da$) by dividing the distribution (or the age derivative of the distribution) for a particular year, e.g. 2011, by historical registrations, matching numbers and registrations by year of make. For instance, dividing out how many of the Citroen C3 2003 remain in 2011 by how many were initially registered in 2003. This procedure produces directly the survival function or its derivative but involves additional data, registrations of new vehicles.

In the second case, one looks at the change in numbers in each age bracket across years (e.g. 4 years old vehicles in 2011 compared to 3 years old vehicles in 2010, etc.). This then involves no additional data and yields $p_i(a)$, which has quite a different interpretation in comparison to $d\ell_i (a) / da$. Both analyses are done with UK data in section \ref{sect:UKsurvey}.

\subsubsection{Application: emissions factors of the fleet vs new vehicles \label{sect:TimeEmTh}}

As we show in the data section, it is easy to find out what are the emissions factors, or fuel efficiencies, of new vehicles currently in the market. However it is much more difficult to find out what the efficiency of the current fleet is, since it is composed of both old and new vehicles, and we expect that efficiencies may have improved over time. As we show below with data from the UK, for which we have the fortunate situation that a survey of the existing fleet has been carried out, this is effectively the case. The question arises then as to whether it is possible to work back fleet emissions from the distribution of emissions from newly purchased vehicles. 

This is difficult and can only be done very approximately. Emissions within the fleet are distributed, due to a very wide range of vehicle models with different power ratings and engine sizes, but also due to vehicles of different ages, with older technologies having been engineered with other primary objectives than fuel efficiencies. To calculate average fleet emissions, it is important to include the relative amount of new and older vehicles in order to weight the sum correctly. For instance, where numbers grow quickly as it does in China, the fleet is younger and average fleet emissions are closer to those of new vehicles than where numbers are stable, such as the UK. Using $E(a)$ as the average emissions from one age tranche of the fleet, average fleet emissions $\overline{E}_k$ of a particular world region $k$ are:
\beq
\overline{E}_k = \sum_i {N_{ik} \over N_{tot,k}} {\int_0^\infty E_{ik}(a) \xi_{ik}(a) \ell_{ik}(a) da \over \int_0^\infty \xi_{ik}(a) \ell_{ik}(a) da}
\eeq
We consider that technologies are produced by multinational car makers who apply the same technologies globally (international spillovers). Therefore, we do not assume that technology availability differs particularly significantly between regions, even though relative sales vary. However the rate of growth of car numbers does vary significantly between regions (e.g. China compared to the UK), making the ratio of new to old vehicles very different. We assume that international car makers standardise new fuel saving technologies in their models sold worldwide. For example, when a technology such as composite materials is developed, it eventually becomes applied to all models and is not subsequently removed unless it is superseded. This means that the time variation of $E(a)$ is not far from same globally, but that the efficiency of the fleet is determined by the relative numbers of vehicles with different emissions factors in different regions (e.g. vehicles with larger engines in the USA compared to the UK), $E_i(a) = E_i(0) f(a)$, where $f(a)$ is the relative variation of emission factors in time historically up to now, averaged across models. With this simplification the problem becomes separable:
\beq
\overline{E}_k = \sum_i {N_{ik} \over N_{tot,k}} {E_{ik}(0)}{\int_0^\infty f(a) \xi_{ik}(a) \ell_{ik}(a) da \over \int_0^\infty \xi_{ik}(a) \ell_{ik}(a) da} = \sum_i {N_{ik} \over N_{tot,k}} {E_{ik}(0)} \Theta_{ik}.
\eeq
Thus if $f(a)$ can be known for one region, by knowing the emission factors of new vehicles $E_{ik}(0)$ and the sales history by region, one can work back fleet emissions approximately for other regions. If furthermore emissions are dominated by petrol engines, we can furthermore approximate that the ratio emissions of old vehicles to new vehicles is technology independent, leading to
\beq
\overline{E}_k = E_k(0) \Theta_k.
\eeq
Using the survey of the UK car fleet, an approximate function $f(a)$ was determined, as shown below in section \ref{sect:UKEmTime}.

\subsubsection{Population dynamics \label{sect:replicator}}

Out of survival analysis or technology demography, with additional arguments concerning allocation of new sales, it is possible to derive population dynamics identical to that of competing species in an ecosystem, in other words a Lotka-Volterra set of differential equations (LVEs), sometimes called `Replicator dynamics', referred to in section~\ref{sect:basicequations}, eq.~\ref{eq:replicator1}. As opposed to many empirical works, the LVEs are not taken by assumption, they are derived from simple arguments of industrial dynamics, bandwagon effects and reliability theory, and its parameters have a meaning. This is done in detail in \cite{Mercure2015}, summarised here. This theory can be visualised in terms of flows of market shares between technology categories due to substitutions.

New vehicle purchases cover both replacements and increases in total population. Given that the global vehicle production capacity is large and that vehicles can be traded, we assume that sales are limited by the demand, not by the supply. Even in regions such as China, where growth is significant, sales leading to increases in population do not exceed sales for replacements, based on our data (section~\ref{sect:HistoricalData}). Therefore the rate of increase of production capacity is small in comparison to global production capacity \citep[see also][for a discussion of demand-led versus supply led assumptions in this population dynamics context]{Mercure2015}. 

During a time span $\Delta t$, out of a total $\xi_{tot}(t)$ of new registrations in a particular region, a certain fraction of sales is allocated to different technology categories according to consumer preferences $F_{ij}$ as derived above, and replacement rates, denoted by $1/\tau_i$. These parameters can be understood as determining the rate of influx and out-flux of shares of sales in and out of technology categories $i$ and $j$ in a set of $n$ possibilities. Using the variable $N_i$ for the vehicle population in category $i$, increases in $N_i$ due to purchases being allocated into $i$ related to the replacement of vehicles scrapped in category $j$ (i.e. substitutions of $i$s for $j$s at the time of scrappage) corresponds to:
\beq
\Delta N_{j\rightarrow i} = \left[ \begin{array}{ll} \text{\small Fraction of} \\ \text{\small prod. capacity} \\ \text{\small belonging to \emph{i}} \end{array}\right]_i
\left[ \begin{array}{ll} \text{\small Consumer} \\ \text{\small preferences} \end{array}\right]_{ij}
\left[ \begin{array}{ll} \text{\small Fraction of} \\ \text{\small destructions} \\ \text{\small belonging to \emph{j}} \end{array}\right]_j
\left[ \begin{array}{ll} \text{\small Number of} \\ \text{\small destructions} \end{array}\right]_{tot},
\label{eq:Words}
\eeq
where destructions of vehicles in $j$ are allocated to categories according preferences, which direct flows of units between categories. Meanwhile, the number of vehicles purchased that are not replacements are 
\beq
\Delta N_i^{\uparrow} = {1\over n}\sum_j^n \left[ \begin{array}{ll} \text{\small Fraction of} \\ \text{\small prod. capacity} \\ \text{\small belonging to \emph{i}} \end{array}\right]_i
\left[ \begin{array}{ll} \text{\small Consumer} \\ \text{\small preferences} \end{array}\right]_{ij}
\left[ \begin{array}{ll} \text{\small Population} \\ \text{\small increase} \end{array}\right]_{tot}.
\label{eq:Words2}
\eeq
The numbers of vehicles and vehicle destructions follow directly from the sum of the sales  time series, multiplied with the survival function, over all ages (numbers), or its derivative (deaths), which correspond to convolutions. 
\beq
N_j(t) = \int_0^{\infty} \xi_j(t-a) \ell_j(a) da,\quad \text{and}\quad d_j(t) = \int_0^{\infty} \xi_j(t-a) {d\ell_j(a) \over dt} da \simeq {N_i \over \tau_i},
\eeq
where $d_i$ denotes deaths, $a$ vehicle age and $\ell_j(a)$ the measured survival function for technology $j$. In a scheme where computational power minimisation is sought, deaths can be conveniently and safely approximated with the total population divided by the life expectancy, $N_j/\tau_j$.\footnote{Note that this implicitly assumes an exponential survival function with argument (half-life) $\tau_j$. However changing the shape of the survival function for the same life expectancy changes this result very little, and thus results are not strongly dependent on shape, but rather more so on the life expectancy. Other shapes of the survival function, as measured in further sections, predominantly induce a sliding of $N_j$ forwards in time, i.e. $N_j(t-t_0)$ where $t_0$ stems from the deviation from the exponential form. However for all practical purposes the convolution can really be safely approximated for $N_i/\tau_i$.} A question arises here as to whether the frequency at which vehicle choices take place is as slow as the life expectancy implies. We return to this question in section~\ref{sect:decisionFreq}.

The production capacity by technology category changes through sales and re-invested income. We have shown that under particular forms of the survival function of the production capital, this can also be approximated as proportional to the current population, by category, divided by an industry specific growth rate $t_i$ itself determined by the re-investment rate, the production efficiency and the survival rate of the production capital. This is a reminder that the production capacity established in an industry is built out of income made on selling units in the past, some of which may still be in use. Furthermore, a growing/declining production capacity is inseparable from growing/declining sales, such that a growing/declining population is associated with a growing/declining industry \citep[see][ for a detailed demonstration of this effect]{Mercure2015}. Thus equations~\ref{eq:Words} and \ref{eq:Words2} are rewritten as 
\beq
\Delta N_{j\rightarrow i} = { N_i/t_i \over \sum_k N_k / t_k} F_{ij}{ N_j/\tau_j \over \sum_k N_k / \tau_k}\Delta N_{tot} =  S_i {\overline{t}\overline{\tau} \over t_i \tau_j }F_{ij} S_j {N_{tot} \over \overline{\tau}} \Delta t,
\label{eq:replacements}
\eeq
where $\overline{t}$ and $\overline{\tau}$ are the average industry growth rate and life expectancy, while the $S_i$ are technology category shares of the total fleet. For convenience we define the matrix of time constants $A_{ij} = \overline{t}\overline{\tau} / t_i \tau_j $. For all flow $\Delta N_{j\rightarrow i}$ of substitutions between $i$ and $j$ exists a reverse flow $\Delta N_{i\rightarrow j}$, and thus a net trend
\beq
\Delta N_{ij} = N_i \left(A_{ij}F_{ij} - A_{ji}F_{ji}\right) N_j {N_{tot}^\downarrow \over \overline{\tau}}.
\eeq

The growth of the fleet can also be expressed in a similar way:
\beq
\Delta N_i^{\uparrow} = {1\over n}\sum_j^n { N_i/t_i \over \sum_k N_k / t_k} F_{ij} \Delta N_{tot}^\uparrow,
\label{eq:non-replacements}
\eeq
where $\Delta N_{tot}^\uparrow$ is the time dependent population growth rate, in principle determined by the change in demand and capacity factor. We can combine both equations~\ref{eq:replacements} and~\ref{eq:non-replacements} in a convenient way, by considering expressing it in terms of shares of the total population by technology $S_i = N_i/N_{tot}$, instead of absolute numbers, which must involve a chain derivative: 
\beq
{dN_i \over dt} = N_{tot} {dS_i \over dt} + S_i{dN_{tot} \over dt},
\label{eq:chain}
\eeq
the second term cancels with the equation for the population growth, leaving
\beq
\Delta S_{ij} = S_i \left(A_{ij}F_{ij} - A_{ji}F_{ji}\right) S_j {\Delta t \over \overline{\tau}}.
\eeq

This equation expresses \emph{exchanges of market shares} between technology categories $i$ and $j$ according to preferences and rates of replacement. Cumulating all gains or losses to technology $i$ at the expense or profit of all other categories, we sum over $j$ and obtain the \emph{replicator dynamics} equation, or LVE (eq.~\ref{eq:replicator1})\footnote{Traditionally the LVEs are expressed in absolute numbers $N_i$ while the replicator dynamics equation is expressed in relative numbers $S_i$. These are really equivalent, connected to one-another through the chain derivative eq.~\ref{eq:chain}. See for example \cite{Hofbauer1998}.}:
\beq
\Delta S_i = \sum_j S_i \left(A_{ij}F_{ij} - A_{ji}F_{ji}\right) S_j {\Delta t \over \overline{\tau}},
\label{eq:Replicator2}
\eeq
in which the net flow of shares is regulated by the product of the matrices $A_{ij}F_{ij}$ minus its transpose. While the matrix $A_{ij}$ is interpreted to represent \emph{industrial dynamics} and \emph{reliability}, the matrix $F_{ij}$ is interpreted to represent consumer choices according to our decision-making model, and thus they are completely independent. It is a standard representation of the process of \emph{selection}, identically used in evolutionary biology and economics. This non-linear equation is extremely easy to implement computationally, encapsulating very compactly all relevant population dynamics. In FTT:Transport, $t_i$ is assumed the same for all technologies, and thus we take $A_{ij} = 1/\tau_i$.

\subsubsection{The frequency of decision-making \label{sect:decisionFreq}}

It is not exactly correct to assume that the frequency at which decisions are made matches the frequency of scrappage of vehicles. In fact, what \emph{limits} the rate at which vehicle type decisions arise is tied to the rate at which purchasers of new vehicles go back to the new vehicle dealer. New vehicle buyers do not \emph{always} keep vehicles until the end of their statistical lives. In fact, in many cases, they change vehicle every few years, selling them on to the second-hand market. The rate at which they do this is typically tied to the length of time for which they are contracted to pay for the vehicle, which is much shorter than the lifetime of the vehicle. Contract length are typically between 3 and 5 years while life expectancies are of around 11 years. 

Here, we consider that second-hand markets are `slave' to the new vehicle market, in that its composition is entirely constrained by what has been chosen in the new vehicle market in earlier years. Secondly, vehicles pass through the second-hand vehicle market until scrappage at a variable rate not solely determined by rates scrappage, but also by the rate of acquisition of new vehicles.

We thus use two timescales in the model, one for decision-making and one for vehicle survival. The time scale for vehicle survival, derived from the survival function, is used to calculate the size of the fleet. The purchasing rate is the one used in the replicator equation. This is substantiated by the simple fact that the replicator equation cannot match diffusion rates that we observe in historical data, in many countries and technologies, if we use the survival rate. It is clear from the data that new technologies can diffuse at a rate faster than what would be predicted by the survival rate. Note that these values are upper bounds; the rate of adoption is determined by the product of consumer preferences $F_{ij}$ and the rate of decision making.

These assumptions are supported by our historical database (section~\ref{sect:HistoricalData}), in which we observe rates of diffusion of new technologies that are substantially faster than what would be allowed by standard survival analysis using known scrappage rates. Purchasing timescales (or turnover rate) that enable to fit $\gamma_i$ values for new technologies are of the order of 3-5 years, much shorter than the observed survival time of 11 years. Using purchasing times of 11 years simply does not allow fitting the $\gamma_i$ values of most new technologies (using the method described in section~\ref{sect:gammapractice}). We conclude that this is an important factor to keep in consideration.

\subsubsection{A theory of social influence in vehicle purchases}

In this last part of the theory section, we show here briefly including social influence in a discrete choice model leads to the exact same replicator equation. This is explored in detail in \cite{Mercure2018c}, and only summarised here. 

Discrete choice models define a linear random utility model, in which the utility $U_i^*$ associated to purchasing a particular type of vehicle $i$ is expressed as a function of a number of variables $V$ such as income, gender, distance travelled and so on, and regression parameters $\beta$ and error $\epsilon$,
\beq
U_i^* = \beta^1_i V^1_i + \beta^2_i V^2_i + \beta^3_iV^3_i + ... + \epsilon_i.
\label{eq:randUt}
\eeq
We look for the probability that option $i$ is chosen over other options,
\beq
P \left( U >  \max \left[U_1, U_2, U_3, ... U_n \right] \right) = P \left( U >   U_1 \right) P \left( U >   U_2 \right) ... P \left( U >   U_n \right),
\label{eq:PUg}
\eeq
Following standard theory, this leads to the MNL
\beq
P_i =  { e^{U_i \over \sigma} \over \sum_j e^{U_j \over \sigma}}, \quad \sum_i P_i = 1.
\label{eq:MNL}
\eeq
Taking the probabilistic choice $P_i$ as determining the shares of the market, this determines how the market evolves, in equilibrium, for changes in variables. This can be converted to a CES function, and therefore can be assumed the same as the optimal choice in an equilibrium consumer theory.

If, however, one takes vehicle shares $S_i$ as a simplified proxy for social influence, then one obtains a model in which shares depend on themselves in a recursive way, and the model cannot be solved as an MNL. In fact, we show here compactly that if the random utility is function of vehicle shares, the representative agent cannot exist. This is easily explained: if agents choose according to the shares that they see in the market (their choice likelihood being higher for vehicles with higher shares), this can be interpreted as a group of agents each of which has a different set of knowledge over his/her options. Thus, there cannot be a representative agent.

If an agents $k$ \emph{finds value $V^4_i$ in purchasing vehicle $i$ that other agents $\ell$ also consume}, then a term that links the utility between agents arises \citep[see e.g.][]{Durlauf2010}:
\beq
U_i^*(k) = \beta^1_i V^1_i(k) + \beta^2_i V^2_i(k) + \beta^3_iV^3_i(k) + ... + \alpha f\left( \sum_{\ell} \beta^4_i V^4_i(k,\ell) \right) + \epsilon_i(k),
\label{eq:IntLRUM}
\eeq
The probability of option $i$ being chosen over other options is function of \emph{weighted} sets of choices,
\beq
P \left( U >  \max \left[U_1, U_2, ... U_n \right] \right) = P \left( U >   U_1 \right)^{S_1} P \left( U >   U_2 \right)^{S_2} ... P \left( U >   U_n \right)^{S_n},\nn
\eeq
In this case, the solution is different to the MNL:
\beq
P_i =  {S_i e^{ U_i \over \sigma} \over \sum_k S_k e^{ U_k \over \sigma}},
\label{eq:Sprefs}
\eeq
in which every option is weighted by its own shares. Preferences $P_i$ are instantaneous, but purchases happen at a rate $\tau_i^{-1}$, following consumer needs. We then take preferences as the rate of change of shares (as opposed to be equal to preferences, as it would be in an equilibrium model),
\beq
{dS_i \over dt} =  {1\over \tau_i} {S_i e^{ U_i \over \sigma} \over \sum_k S_k e^{ U_k \over \sigma}}.
\eeq
This is a form of replicator dynamics. Adding to this some notions of survival analysis (see \cite{Mercure2015, Mercure2018c} for details), one can mathematically transform this to something useable for durable goods, which is the particular form of replicator equation that we use in FTT:Transport,
\beq
{d S_i \over dt} = \sum_j S_i S_j \left( A_{ij}F_{ij} - A_{ji}F_{ji} \right), 
\label{eq:Lotka}
\eeq
In a more general model, if we assume the linear random utility function is function of the log of shares, with parameter $\alpha$,
\beq
U_{i}^* = \beta^1_i V^1_i + \beta^2_i V^2_i + \beta^3_iV^3_i + ... + \alpha \log S_i + ... + \epsilon_i,
\label{eq:randUtk}
\eeq
then we obtain the more general replicator equation
\beq
{d S_i \over dt} = \sum_j S_i^{\alpha \over \sigma} S_j^{\alpha \over \sigma} \left( A_{ij}F_{ij} - A_{ji}F_{ji} \right), 
\eeq
in which the particular S-shape of the diffusion is scaled by the ratio of the strength of the social influence $\alpha$ with the heterogeneity $\sigma$. At this moment, due to a lack of data for the strength of social influence in each FTT:Transport region, the parameter is set to 1, and therefore, we use the replicator function given in eq. \ref{eq:Lotka}. Altering $\alpha$ does not radically change the results, but it changes the very particular shape of the diffusion profile. Finally, note that setting the social influence parameter $\alpha$ to zero yields the MNL, and thus, gives back the standard equilibrium consumer theory obtained when optimising at the system level. It follows that optimising at the system level is a result of assuming (1) perfect information, and (2) no social influence.

\subsubsection{Coherence length in FTT}

The coherence length of a function or signal is defined by the length, along an independent variable, over which data points are related to each other. In FTT, we have autocorrelation in time, which means that in time, states of the model are related to states of the model at earlier time. This is due to its path-dependence, as imposed by the replicator dynamics equation. Since the model has no foresight, it does not have forward autocorrelation. For example, in biological systems, populations at a certain time depends on what the population was at earlier times. However, as time goes by, this influence wanes as other more recent events become comparatively more influential. 

The autocorrelation of a model can be measured using a expression of the form $g(t) = \sum_\tau f(t-\tau)f(t)$, where $f(t)$ is the signal. In FTT, the coherence length is of the order of 5-10 years, stemming primarily from purchasing rates (see section~\ref{sect:decisionFreq}). This determines, for instance, the length in time over which the historical data has influence over the trajectory of diffusion. After 10 years, the influence of historical data has declined a factor of around 2. Changes in the historical data would imply changes in populations, which would have an important impact, in comparison to changes in cost data. This is why kick-start policies (defined in the main text) have an important impact on technological trajectories.

\subsection{Integration to E3ME}

The integration of FTT:Transport to E3ME is made through several variables. The vehicle and transport demand econometric equations are technically part of E3ME, not FTT, and interact directly with other variables such as income, prices and output (GDP). These are ways by which interaction takes place with the larger economic model; however these are not the most important. Since transport consumes a large fraction of world production of oil, and since the value of oil is high (oil enables all types of mobility, whether people or goods), any changes in the demand for oil has far reaching consequences for the global economy. FTT:Transport controls the major part of the oil demand in E3ME, which specifies an econometric equation for the demand of `middle distillates' (petrol, diesel, kerosene, etc, excluding heavy oil), by 22 types of fuel users (see \cite{Mercure2018b} for more details on the overall model), including road transport. 

We split road transport between passenger and freight. The freight component is not at this stage developed in a specialised FTT model, unlike the passenger component, as this will be part of future work. We make the split based on estimates of fuel demand per tonne-km (tkm) using \cite{Liimatainen2014,Kamakate2009}, which accounts to one third to half of fuel demand in most countries. We control freight emissions using biofuel mandates; as we later develop a dedicated FTT model for freight, we will include all current technologies (diesel, advanced diesel, CNG and electric) in various size classes using the same method. The passenger component is the main focus here, as we are interested in policies affecting road passenger. Total fuel used calculated in FTT:Transport originates from assumed emissions factors. However, actual fuel use is often more than what is rated by manufacturers, due to driver behaviour not exactly matching the standardised driving profile used by manufacturers for estimating emission factors. We find that total FTT fuel use in all countries accounts for around two thirds or more of fuel use obtained from IEA statistics. Assuming that the error originates from these emissions factors, we scale FTT fuel use to match IEA values at the start of the simulation, and keep these scaling factors constant (one per region) until the end of the simulation.

Changes in fuel use have profound repercussions in E3ME, which we summarise here (more details can be found in \cite{Mercure2018b}).  Changes in the demand for oil for transport affect the price of oil through our fossil resources depletion algorithm \citep{MercureSalas2013}. This module determines the marginal cost of oil production based on a database of types of oil extraction and their cost (conventional oil, offshore, heavy oil, shale oil, tar sands etc). If the demand increases, more costly resources are developed and the marginal cost goes up, while if the demand declines, costly resources are abandoned and the marginal cost goes down.  Oil and gas prices changes in E3ME are set proportional to changes in their marginal cost. When the price of oil changes in E3ME, it affects the oil \& gas industries, which through multiplier effects (input-output tables) leads to changes in economic conditions that can be quite large. For example, it can leads to stranded fossil fuel assets and unemployment in fossil-fuel producing countries.

The demand for electricity from electric vehicles in FTT:Transport is also accounted for in E3ME, which is fed through to the sister model FTT:Power. It can affect for example the development of renewables or affect policies for decarbonisation if the demand increases as EVs diffuse to larger shares over time.

\subsection{Policy in FTT:Transport \label{sect:Policies}}

In this section, we review the representation of all policy types available in FTT:Transport. Eight types of policies are available to use independently, which we divide into two types: the policies that take the form of a pecuniary incentives that are applied at the time of vehicle purchase, and those that do not. The pecuniary incentives can be described by looking at the equation of the LCOT, while the other policies are not related to the LCOT, but rather, to share values. Note that all policies are defined by zero or non-zero values exogenously given between 2018 and 2050, with the exception of efficiency standards, which are currently set as a fixed target.

\subsubsection{Pecuniary incentives}

We reproduce here the LCOT, with added policy parameters. We use double letters to denote policies, while other symbols remain the same as in eq.~\ref{eq:LCOT}:
\beq
LCOT_i = \sum_t {{I_i \over CF_i(t)} + VT_i + CT(\alpha_i) + {FU_i(t) \over \beta_i}FT(\alpha_i,t) + {MR_i \over FF_i} + RT_i(t) \over (1+r)^t} \bigg/  \sum_t 1/ (1+r)^t, 
\label{eq:LCOTPolicies}
\eeq
where,
\begin{itemize}
\item $VT_i$ is a \emph{registration vehicle tax or rebate}, in \$/veh, per vehicle type/class, paid at purchase time,
\item $CT(\alpha_i)$ is a \emph{registration tax based on the fuel economy} $\alpha_i$, in \$/veh/(gCO$_2$/km), not type- or class-specific, paid at purchase time,
\item $FT(\alpha_i,t)$ is a \emph{tax on fuel consumption}, in \$/litre, paid each year, depending on the fuel economy $\alpha_i$,
\item $RT_i(t)$ is a \emph{road tax}, vehicle type/class-specific, paid once per year.
\end{itemize}

These fall into two types: those that are paid once, and those that are paid yearly. The difference in impact is that the yearly policies are discounted, while the on-off policies are not. $CT_i$ is a form of carbon tax, and so is $FT_i$; the difference however stems from the fact that $CT_i$ is a tax on expected lifetime emissions, while $FT_i$ is a tax on actual emissions. These imply slightly different outcomes, as $CT$ tends to have a higher impact per dollar taxed than $FT$, simply due to time preference. $VT_i$ is typically not used at the same time as $CT$, and their inclusion aims at enabling different types of strategies, where for instance $VT_i$ can be made into a tax-feebate scheme in which tax income is recycled into subsidies for low-carbon vehicles. $RT_i$ offers a similar counterpart to $FT$.

\subsubsection{Regulatory and push policies}

Many policies in the real world do not take the form of pecuniary incentives, and often do not apply at purchase time either. These can be of regulatory nature and apply to manufacturers. In FTT, this implies influencing the flow or value of shares in particular technology categories. For instance, it can involves exogenously changing the values of the preferences $F_{ij}$ (see eq.~\ref{eq:Replicator2}).

Policies included are:
\begin{itemize}
\item \emph{Regulations banning the sale of a technology type/class}. This involves setting $F_{ij} = 1$ and $F_{ji} = 0$. Existing vehicles of these types live to the end of their lifetimes. 
\item \emph{Kick-start programme (public procurement)}. This policy exogenously changes the shares of a vehicle type/class at a specific point in time.
\item \emph{Biofuel mandates}. This exogenously determines the relative content of liquid fuels between fossil and renewables types, for all vehicles. 
\item \emph{Efficiency standards}. This exogenously defines the efficiency of new internal combustion vehicles, per vehicle type/class.
\end{itemize}

We note that there is strong cross-policy interaction in FTT:Transport, as is the case in other FTT models \citep[see e.g.][]{Mercure2014}. In particular, regulatory and pecuniary policies can enable each other's effectiveness. For example, setting up a kick-start programme in tandem with a fuel tax promotes the fast diffusion of low-share low-carbon technologies; acting faster than a tax alone.

\subsubsection{Corresponding real-world policies}

Real-world examples for each FTT policy exist, some of which are relatively common. We give some examples below:
\begin{itemize}
\item \emph{Regulations banning the sale of a technology type/class}. UK Diesel car ban. 
\item \emph{Kick-start programme (public procurement)}. China ten Cities, Thousand Vehicles Program; License plate quota for EV (430000 petrol cars quota and 170000 EV quota); China EV credit point; The Chinese government purchased. The Chinese government has directly purchased around 90000 EV in 2017 \\ (see \burl{https://www.ft.com/content/a55e7d36-db8a-11e5-a72f-1e7744c66818}). 
\item \emph{Biofuel mandates}. Brazil National Alcohol Program and EU biofuel mandates.  
\item \emph{Efficiency standards}. US CAFE; EU fuel economy standards; Japan Top-Runner program; China fuel economy limits.
\item \emph{Registration carbon tax} South Africa's registration carbon tax, equal to R100/gCO$_2$/km for each gCO$_2$/km above 120gCO$_2$/km, charged when a new car is purchased.
\end{itemize}

\newpage
\section{Deriving general information from the UK fleet \label{sect:UK}}

This part describes the procedure with which data that was collected for parameterising FTT:Transport. The data takes predominantly one of two forms: distribution of vehicle properties and vehicle numbers. While it is possible to obtain vehicle numbers regionally by technology type, this is not the case for vehicle properties, and in this case it is done for 18 representative regions, including Europe, USA, China, India, Japan and Brazil. We consider these regions to cover well enough most variations between car markets in the world that are of significance for global emissions, while maintaining our work manageable. Some of this works has been published already as \cite{MercureLam2015}.

The UK dataset for vehicle properties was used as a model against which we designed the data collection for the other regions. It is also the richest dataset of all, enabling to derive properties that cannot be obtained for other regions. Given this advantage, the UK dataset was analysed in great detail in order to derive a large amount of vehicle fleet properties. Where data is not available, whenever it makes empirical sense (e.g. in Europe), UK values are used in FTT:Transport. For example, the only survival function derived from actual data in this work is from the UK dataset, which is unlikely to be possible in other regions due to the lack of time-resolved survival data. We thus assume that the shape of the survival function is similar elsewhere. In sections beyond that for the UK, the methodology is the same unless stated otherwise. Similarly, the UK dataset enables to re-construct properties of the current fleet as opposed to more easy to access data about new vehicles. In particular, this enabled us to determine the distribution of current vehicle emissions factors as well as those for new vehicles entering the fleet, which are quite different, a difference that we extrapolated to other regions. 

\subsection{Procedure and datasets}

Two sets of data for vehicle populations in the UK were used, along with datasets related to vehicle prices. These are, respectively, data from the vehicle registration agency DVLA for vehicles registered for the first time and existing registrations \citep{DVLA}, DVLA observations of the vehicle population using cameras and number plate recognition \citep{DVLASurvey}, and data for new vehicles currently on the market \citep{CarPages} and motorcycles \citep{MCNBikes}. The dataset for new registrations was matched with those of vehicle prices in order to derive price, engine size and emissions distributions for newly registered vehicles. Meanwhile, the dataset from the survey was used along with aggregate numbers of vehicles to derive properties relating to the current stock of vehicles, which differ from those of new vehicles.

The registration dataset provides the numbers of new registration entries for vehicles for years from between 2001 and 2012,\footnote{Document VEH0160 (2012)} per vehicle model. For vehicles, this has a very large number of entries, around 30 000, as some models have numerous slightly different variants, and some filtering of entries was necessary. Filtering the entries with less than 100 new registrations, this narrowed down the list to less than 3000, more manageable, while still keeping the major part of the registrations. A large variation in model entry names resulted in many additional entries for vehicles with very similar features. Most of these correspond to special editions of other existing models. 

The car price data from \burl{www.carpages.co.uk} also had of order 2000 entries, each of which features a price value, a measure of emissions and an engine size. This was carefully matched by name, entry by entry, to those of the DVLA new registrations database. In this way, nearly 1.9 million vehicles were assigned a price, an engine size and rated emissions. The price data relates to new 2012 vehicles. Data was available in the DVLA set for new registrations in 2012, the year over which the data matching was done. Registration data is also available for previous years, however prices are not available historically. It is in principle possible to explore registrations in previous years matched to 2012 prices; however when going back several years, some of the models being sold back then do not exist on the market anymore in 2012, and therefore this produces an incomplete picture with missing models, having no counterpart in the price database. A similar procedure was followed to match motorbike registrations to the price database \cite{MCNBikes}, using a filtering of entries with less than 10 registrations. Note that it took quite some time to develop this database, hence we use 2012 data. We may carry out an update shortly.

Meanwhile, the dataset for the survey of observations \citep{DVLASurvey} provides vehicle features such as age, emissions, mass and engine size, for vehicles observed on the roads at 256 sites around the UK during years between 2007 and 2011. These observations were made using digital cameras on roads with automatic number plate recognition, and then matched with the DVLA registration database. This enabled to assign data from registration certificates to these observations. The aim of this exercise by the authorities was to find unlicensed vehicles on UK roads, but the data was released for research, removing personal information, as a statistical resource. This was performed and released for 5 years from 2007 to 2011. Each year features around 1.2 million observations, a number high enough to generate reliable distributions quite representative of the UK's operating vehicle stock, of about 28 million vehicles. The number of observations is however not proportional to the number of vehicles in the UK but varies over the years, and thus the distributions derived from these observations required to be rescaled to real population numbers to be comparable across years. For this, registration data for existing vehicles from the DVLA was used,\footnote{Document VEH0205 (2012)} and thus all distributions produced for the existing population are scaled in fractions of the total population, not in terms of the number of observations. This was used to explore the evolution of the age distribution of UK vehicles, and to determine current distributions of emission factors, engine sizes and vehicles masses. 

Further useful datasets exist from the DVLA. VEH0211 (2012) provides numbers of licensed vehicles by years since their first registration, i.e. by registration year. This therefore follows the gradual decline of vehicles in the fleet as they age year after year. We used this to evaluate the survival function of UK vehicles. Finally, document VEH0153 provides the longest time series of vehicles licensed for the first time (i.e. sales) in the UK, since 1954.

\subsection{Distributions and heterogeneity \label{sect:DistUK}}
\subsubsection{Overall new car price distribution}
\begin{figure}[h]
	\begin{minipage}[t]{0.5\columnwidth}
		\begin{center}
			\includegraphics[width=1\columnwidth]{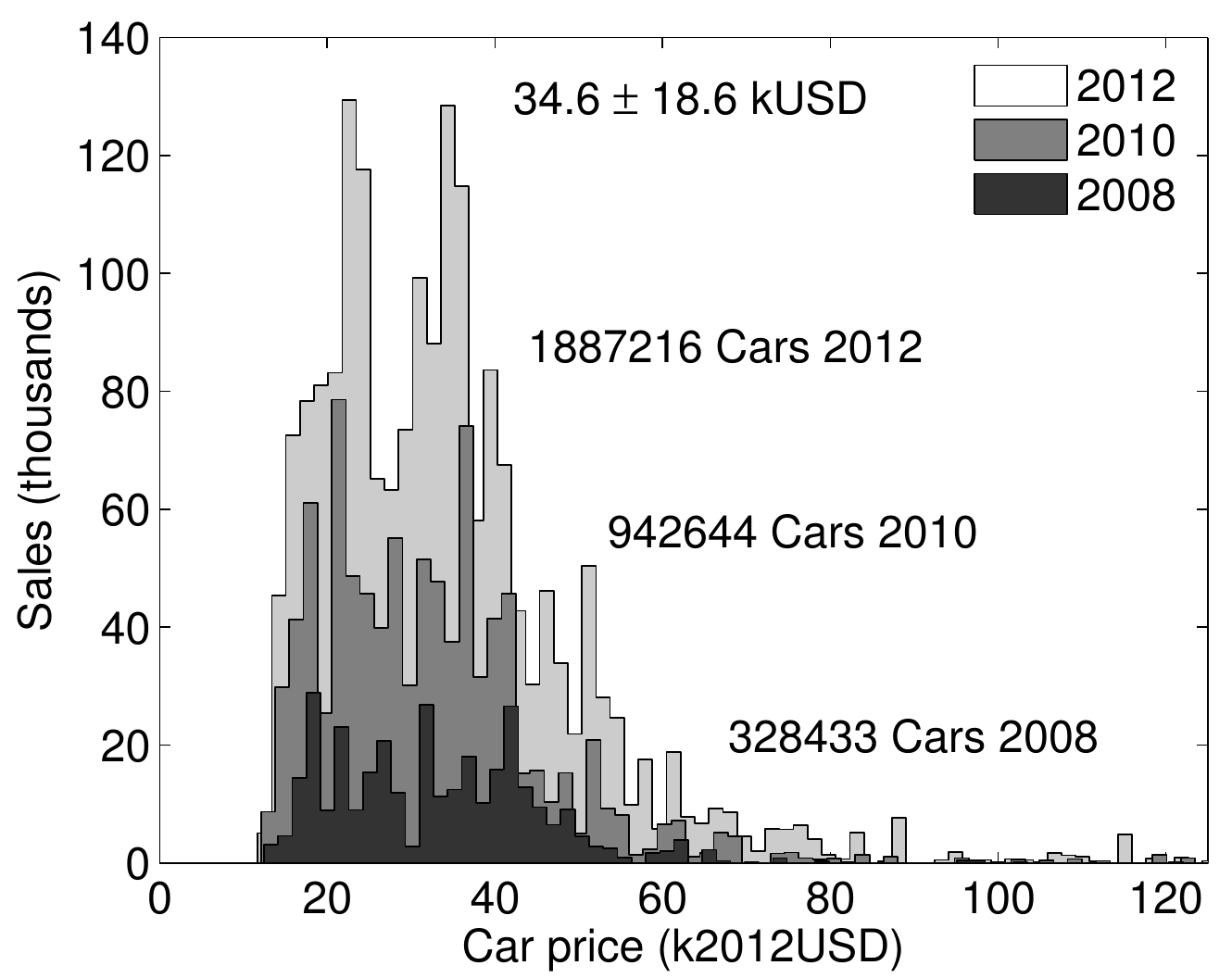}
			\includegraphics[width=1\columnwidth]{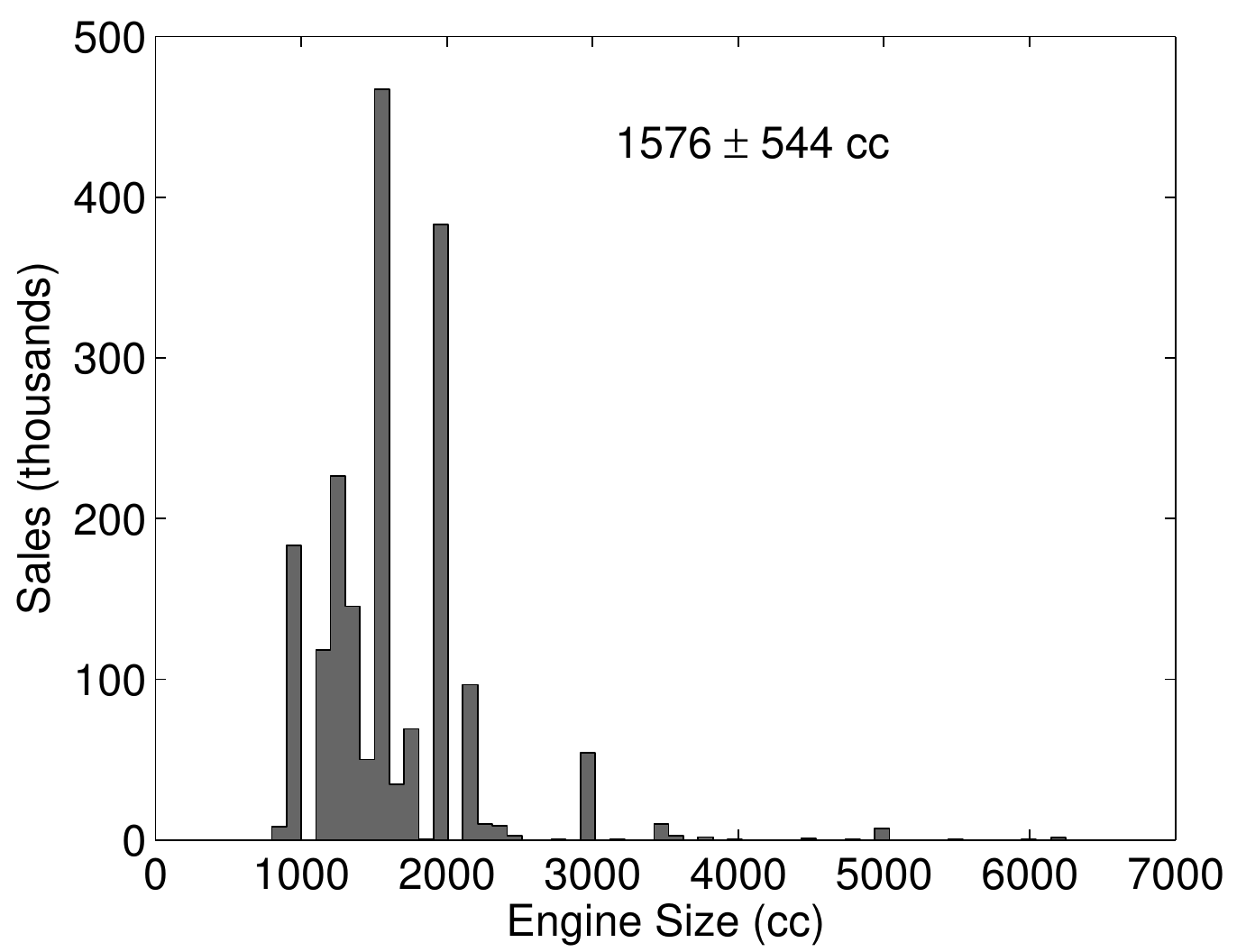}
		\end{center}
	\end{minipage}
	\hfill
	\begin{minipage}[t]{0.5\columnwidth}
		\begin{center}
			\includegraphics[width=1\columnwidth]{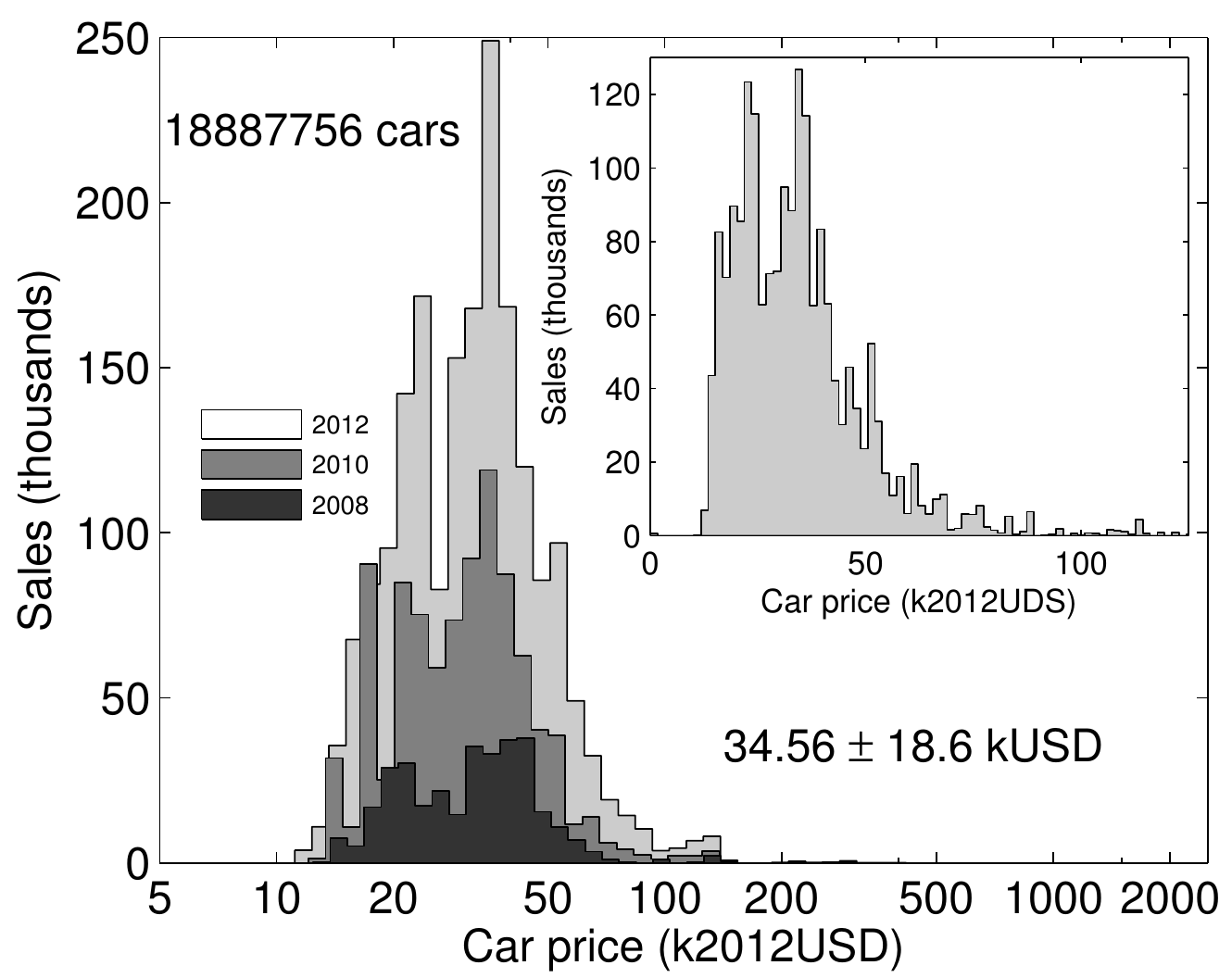}
			\includegraphics[width=1\columnwidth]{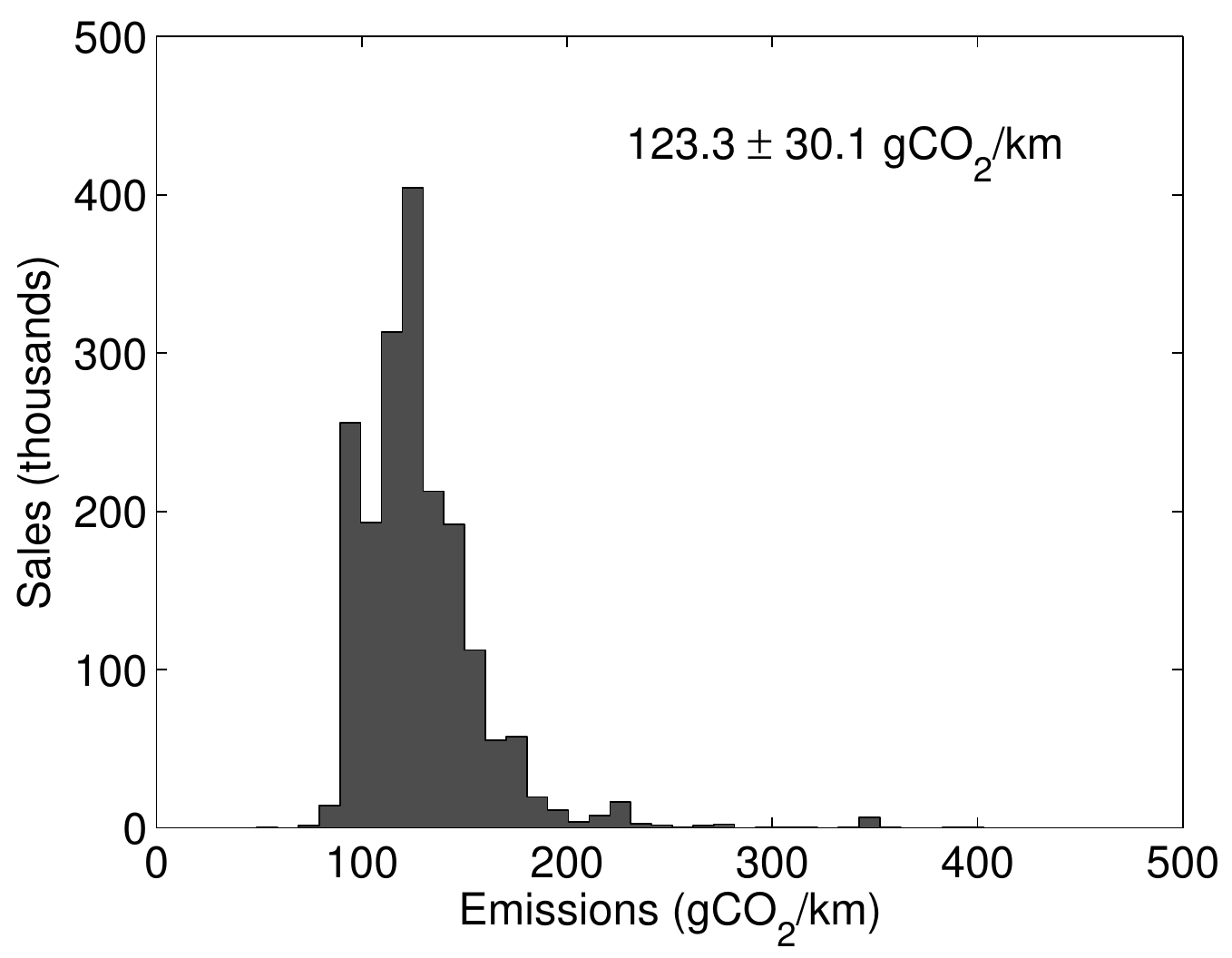}
		\end{center}
	\end{minipage}	
	\caption{\emph{Left} Price distribution of UK vehicles for three different years, based on 2012 prices, scaled linearly in prices, with constant spacing linearly in price. \emph{Right} Same price distribution in logarithmic price space, with constant spacing in the logarithm of the price.}
	\label{fig:UKCarDistributions}
\end{figure}

Figure \ref{fig:UKCarDistributions}, left, shows the overall price distribution of UK car sales in three different years matched to 2012 prices, calculated using ranges of £1000 of constant width in price. The distribution features a clear long tail in the upper price range (note that upper end of the price database ends with a Lamborghini at around £300k, beyond the scaling of this graph). We also observe that the shape of the distribution has not changed across years, which we interpret as evidence that purchasing behaviour has not changed significantly since 2008. In other words, a similar proportion of expensive or economic vehicles was purchased in 2012 compared to 2010 and 2008 (also in 2011 and 2009, not shown).\footnote{Missing models in years earlier than 2012 are not related to their price, as far as we know, and therefore this conclusion can be drawn.} 

The right panel of figure \ref{fig:UKCarDistributions} displays a similar distribution but calculated using constant widths in the natural logarithm (base $e$) of the price of $\Delta \log_e P = 0.1$ (while the plot expresses spacing in base 10). With this we observe that the distributions are symmetric in log space, in other words the distributions are lognormal, as is generally the case for income distribution in most countries, suggesting that these may be related \citep{MercureLam2015}. We conclude from this that price comparisons are better performed in logarithmic space as opposed to linear. In other words, price ratios can be used; if one compares two prices $P_1$ and $P_2$, one evaluates $\log(P_2/P_1)$. The bottom panels show the distribution of engine sizes and emissions, which are equally widely distributed. Averages and standard deviations are given in the charts.

\subsubsection{Price and emissions distributions of new vehicles by category}

\begin{figure}[p]
		\begin{center}
			\includegraphics[width=1\columnwidth]{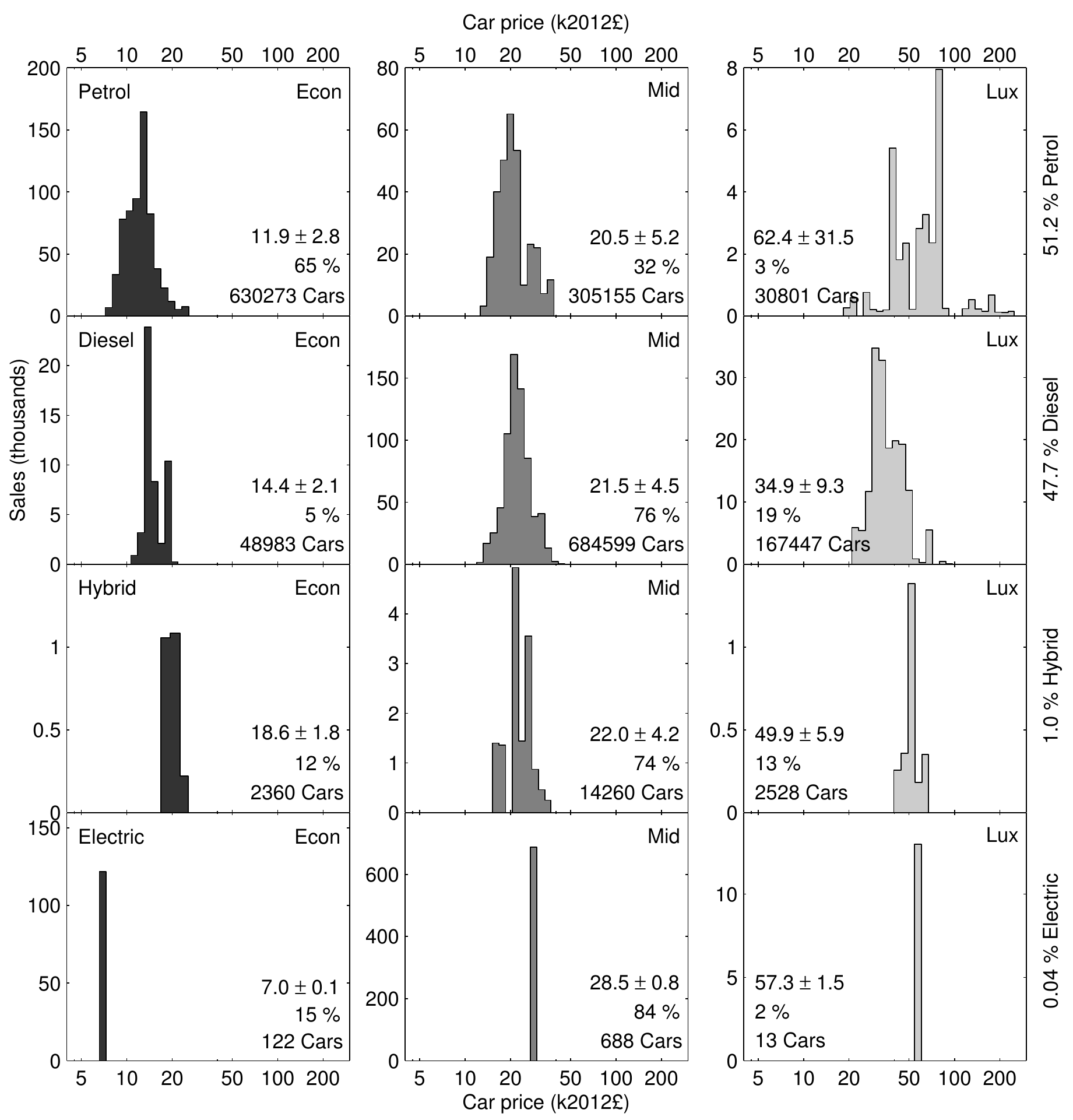}
		\end{center}
	\caption{Price distributions of vehicles matched in the database for three classes (economic, mid-range and luxurious) of engine power by type of engine technology. Percentage values indicate the share of each power class in a technology type. Few instances of electric vehicles were observed in the database, and since the number of available models is restricted, they only have one price value for each class and thus cannot strictly speaking be considered distributions.}
	\label{fig:AllUKDist}
\end{figure}
\begin{figure}[t]
		\begin{center}
			\includegraphics[width=1\columnwidth]{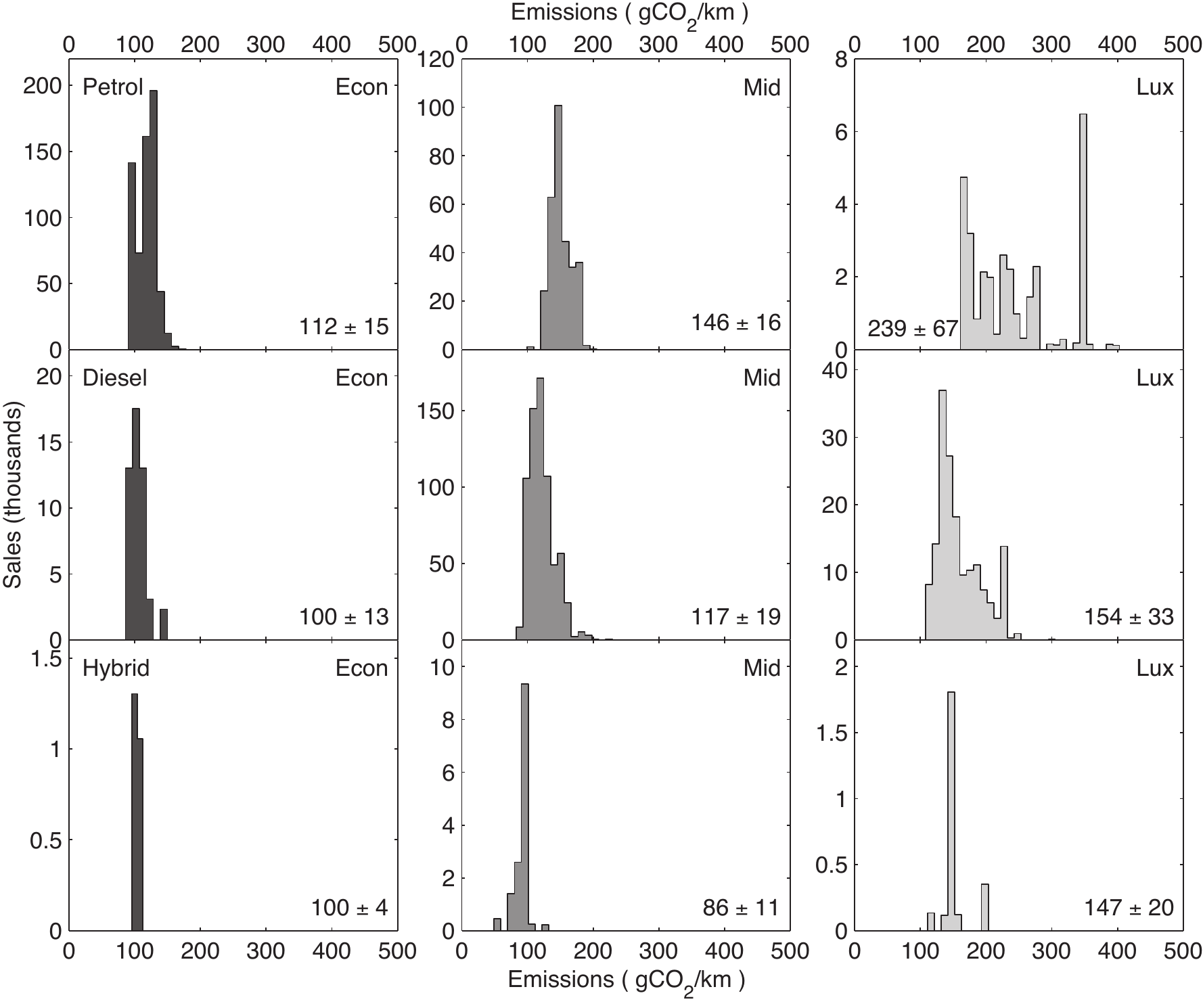}
		\end{center}
	\caption{Emission factors by engine type and class. Numbers in the charts represent averages and standard deviations.}
	\label{fig:AllEmDist}
\end{figure}

The FTT framework requires sub-divisions of engine type categories in order to feature a certain amount of resolution, and because different vehicle classes may evolve differently (i.e. the overall price distribution may change). As seen in the main text, this division is done in terms of engine size class. Vehicles were classified into three classes of engine power, which we term \emph{economic} (Econ), \emph{mid-range} (Mid) and \emph{luxurious} (Lux), for each type of engine technology (petrol, diesel, hybrid and electric). These classes were defined for liquid fuel based technologies according to engine size ranges : $<$ 1400cc, 1400cc to 2000cc, and $\geq$2000cc, based on the classification of Eurostat.\footnote{This was done for convenience when processing vehicle number data for the EU, saving large amounts of time.} For electric vehicles, only three models were found in both UK databases, each of which conveniently happens to match a particular market segment, economic (Renault Twizzy), mid-range (Nissan Leaf) and luxurious (Tesla), and were thus assigned their own category. As can be observed in the chart, price distributions in each class are roughly similar between engine technology type. This supports defining FTT categories as such. 

Price distributions by category are given in figure~\ref{fig:AllUKDist}, and are observed to be unimodal, suggesting that this is an appropriate disaggregation. These can be used to evaluate consumer choices in the FTT binary logit decision framework, as distributed investment terms in the LCOT. The standard deviations are also used and represent consumer and technology diversity. Note that for the UK, CNG and LPG engines are rather rare and were left out of the analysis, even though they are be present in other nations. 

Emission factors vary significantly between engine class and technology, and therefore must be determined accurately in order to track changes in in total emissions as the technology composition changes. This is given in figure~\ref{fig:AllEmDist}. Electric vehicles are considered zero emissions at the tailpipe, electricity emissions being counted elsewhere.

\begin{figure}[p]
		\begin{center}
			\includegraphics[width=1\columnwidth]{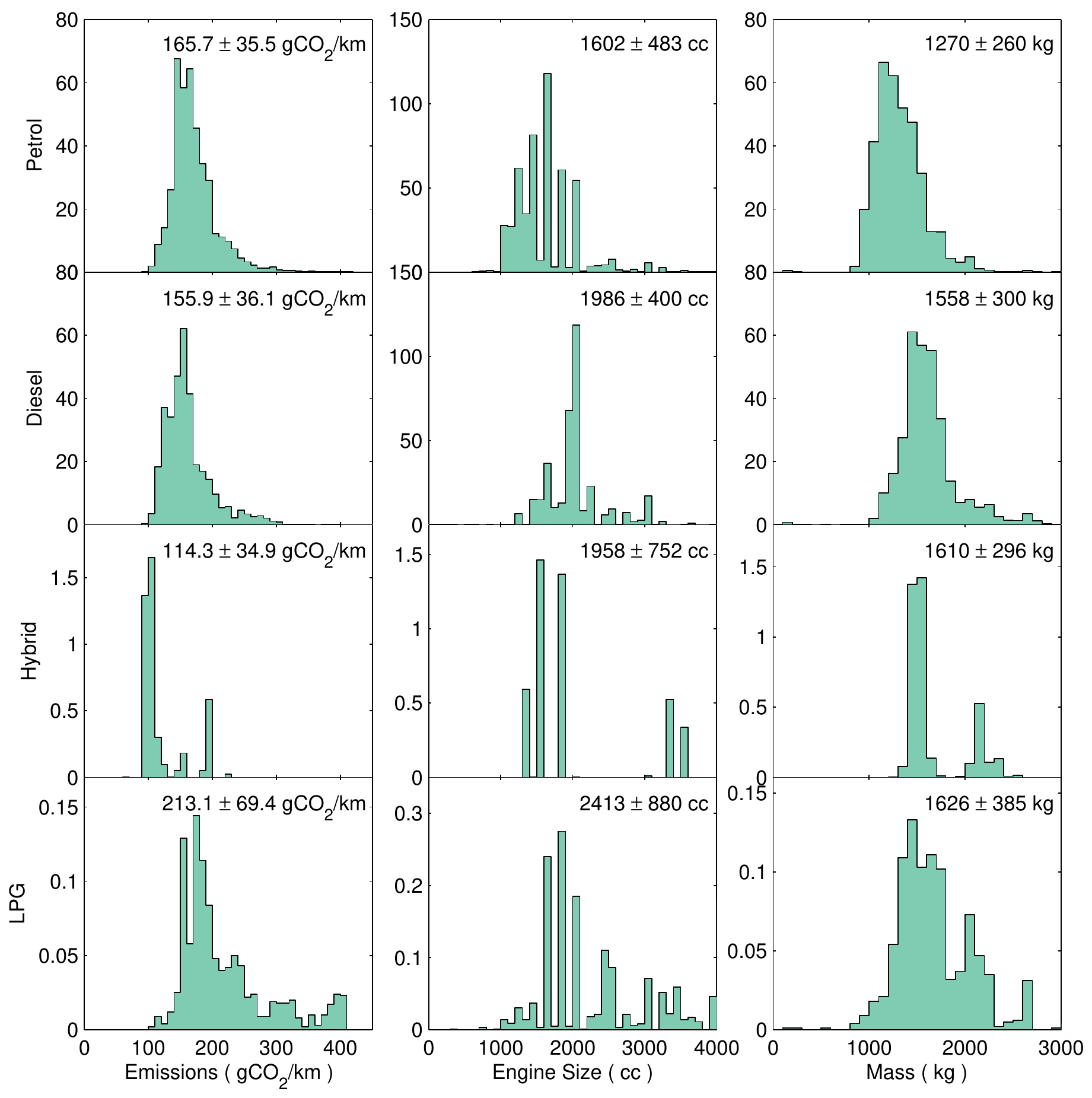}
		\end{center}
	\caption{General properties of the UK car fleet by technology type with size classes combined.}
	\label{fig:AllCarsSEMDist}
\end{figure}
\begin{figure}[p]
		\begin{center}
			\includegraphics[width=1\columnwidth]{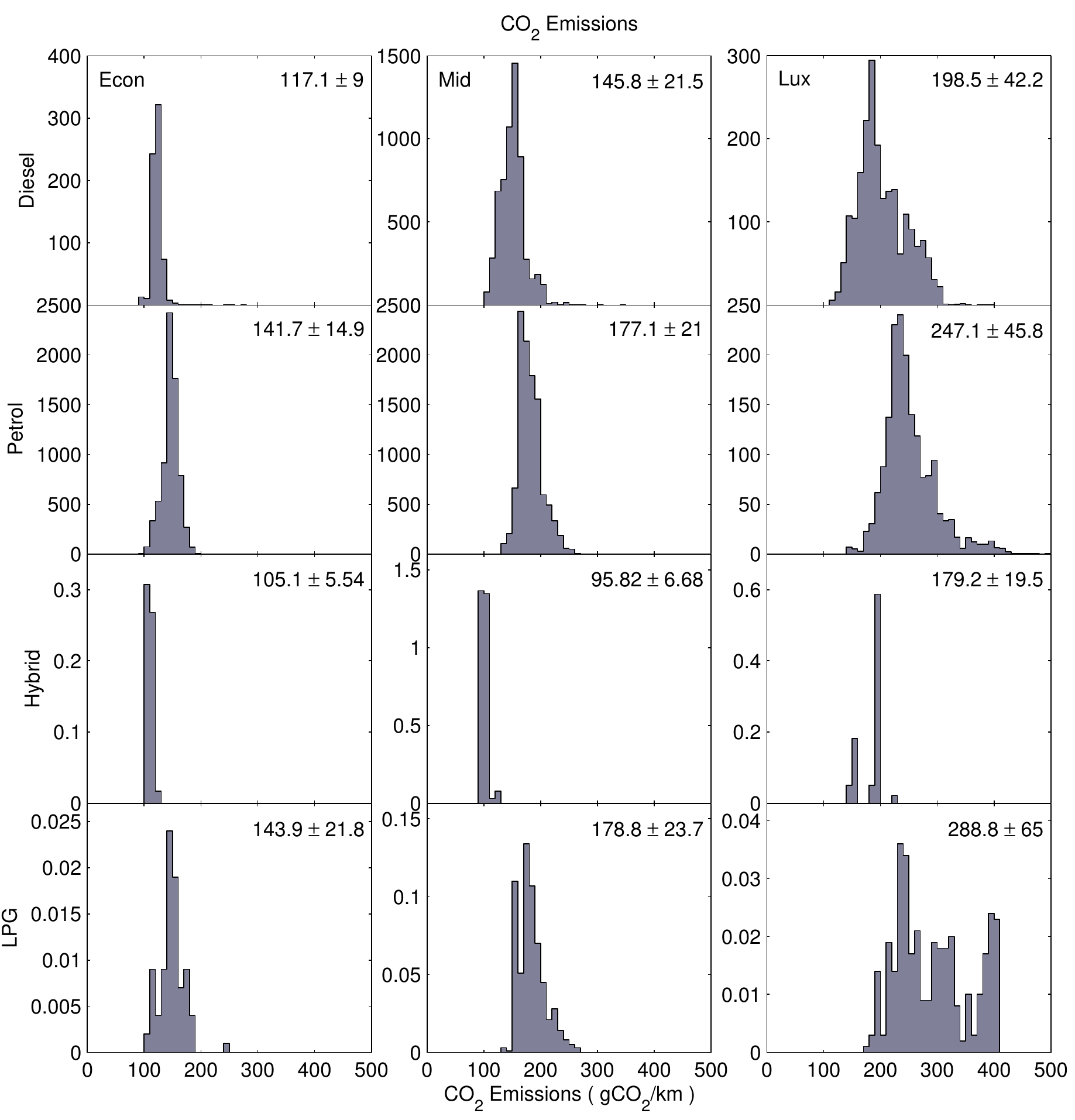}
		\end{center}
	\caption{Emissions by engine type and size class.}
	\label{fig:EmissionsDistTimeD}
\end{figure}
\begin{figure}[p]
		\begin{center}
			\includegraphics[width=1\columnwidth]{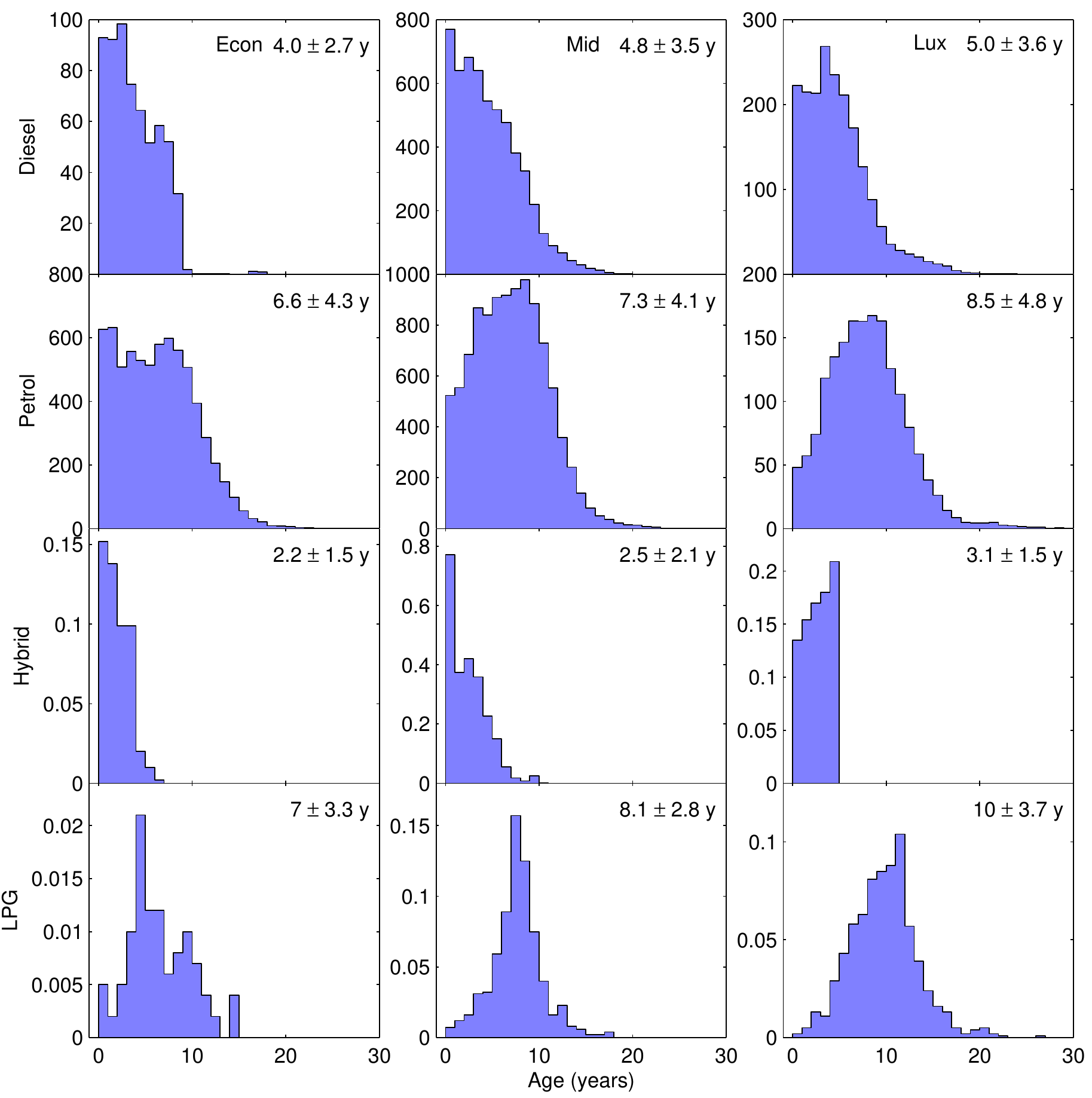}
		\end{center}
	\caption{Age distributions by engine type and size class.}
	\label{fig:AgeDistAll}
\end{figure}
\clearpage
\subsubsection{Distribution of emissions, engine size and vehicle mass  for the current fleet \label{sect:UKsurvey}}

Figure~\ref{fig:AllCarsSEMDist} shows a summary of the profile of the current car fleet in the UK. Using the survey dataset of vehicle observations with number plate recognition, we reconstructed a current snapshot picture of emission factors, engine sizes, age and mass composition of the current UK car fleet. This is extremely important for our model, since the previous dataset on car sales only provides information on new vehicles, and the properties of the fleet can differ significantly from those of new vehicles. This is particularly crucial for emissions factors, and this model's goal is to project future transport-related emissions. 

Figure \ref{fig:AllCarsSEMDist} shows distributions of engine sizes, emissions factors and vehicle masses, with mean and standard deviation values given in each panel, for both petrol and diesel vehicles. Mean values as well as standard deviation values are given in each panel, calculated from the data shown in the panel. This helps understand how the current fleet is composed. This is further subdivided in our usual three classes in the following section for more details. Note that LPG vehicles (converted petrol engines) have come in vogue for some years but their sales have declined again in recent years and thus do not appear in sales data.

\subsubsection{Distributions by size class for the current fleet}

Figure \ref{fig:EmissionsDistTimeD} gives emission factor distributions for all classes and technologies (excluding electric engines) for existing vehicles in the UK. Averages and standard deviations are given in each panel, and these are values used in the FTT model. The distributions in this figure are much better defined and narrow than when looking at the whole fleet, reflecting again that the chosen subdivision between size classes is very important. In particular, the emission factors of the Econ and Mid classes are very narrowly defined. The distribution of emissions for the Lux range is broader, reflecting that this range caters for a range of people with widely differing but high levels of wealth. Emission factors of current vehicles are approximately 20\% larger than those of new vehicles. We model and explain this in section \ref{sect:UKEmTime} using age distributed data.

\subsection{Vehicle lifetimes and survival}

\subsubsection{Age distribution of vehicles by category and survival functions}

Figures \ref{fig:AgeDistAll} shows age distributions by technology type and size class. Their shapes vary significantly between technology types, a representation that some technologies are established since longer than others, hybrid vehicles having been introduced very recently, and diesel vehicles displaying a change in the rate of adoption some years back. Meanwhile petrol vehicles have the expected age distribution decreasing with age following the survival function, and LPG vehicles have seen a past popularity that has waned to the extent that nearly no new LPG vehicles are seen anymore in the fleet. If we can assume that the survival function is a constant of time, these distributions essentially correspond to a product of registrations with the survival function along the age variable. We obtained these age distributions for five consecutive years of observations, from 2007 to 2011. They are clearly not constant, and this is a reflection that sales and registrations have been changing with time. A time series for car registrations can be obtained from DVLA document VEH0153. 

\subsubsection{Survival analysis \label{sect:survival-Data}}

\begin{figure}[h]
	\begin{minipage}[t]{0.5\columnwidth}
		\begin{center}
			\includegraphics[width=1\columnwidth]{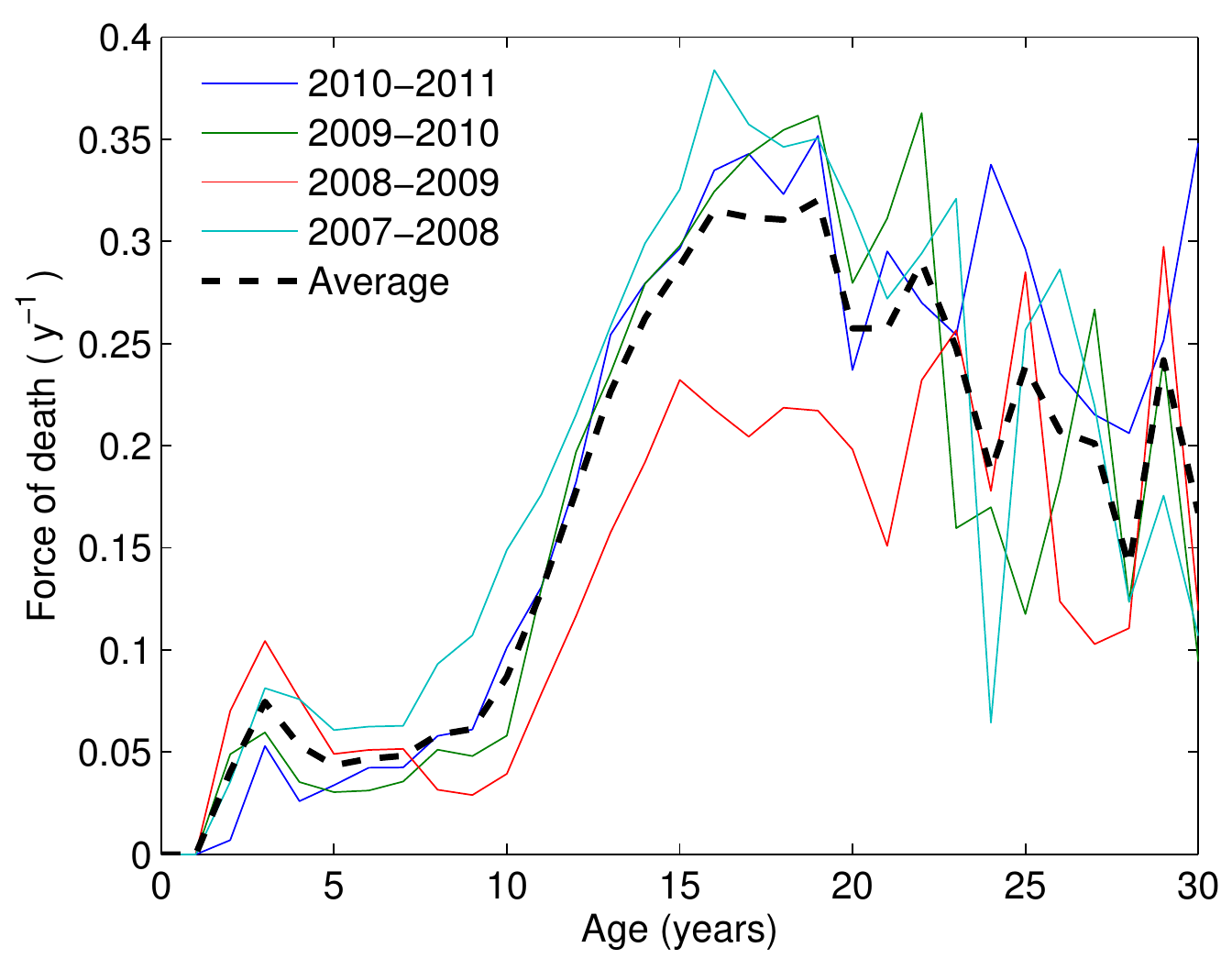}
			\includegraphics[width=1\columnwidth]{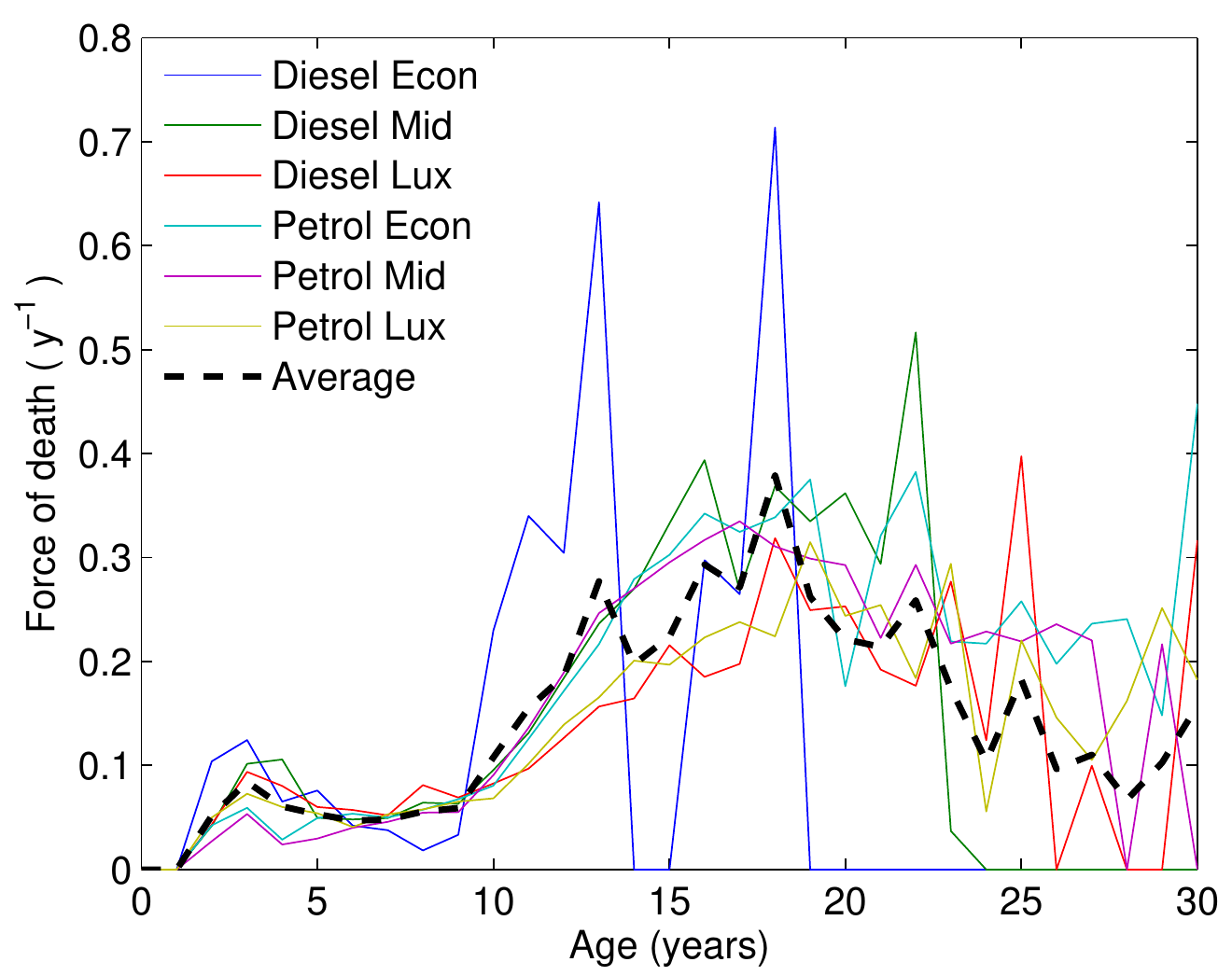}
		\end{center}
	\end{minipage}
	\hfill
	\begin{minipage}[t]{0.5\columnwidth}
		\begin{center}
			\includegraphics[width=1\columnwidth]{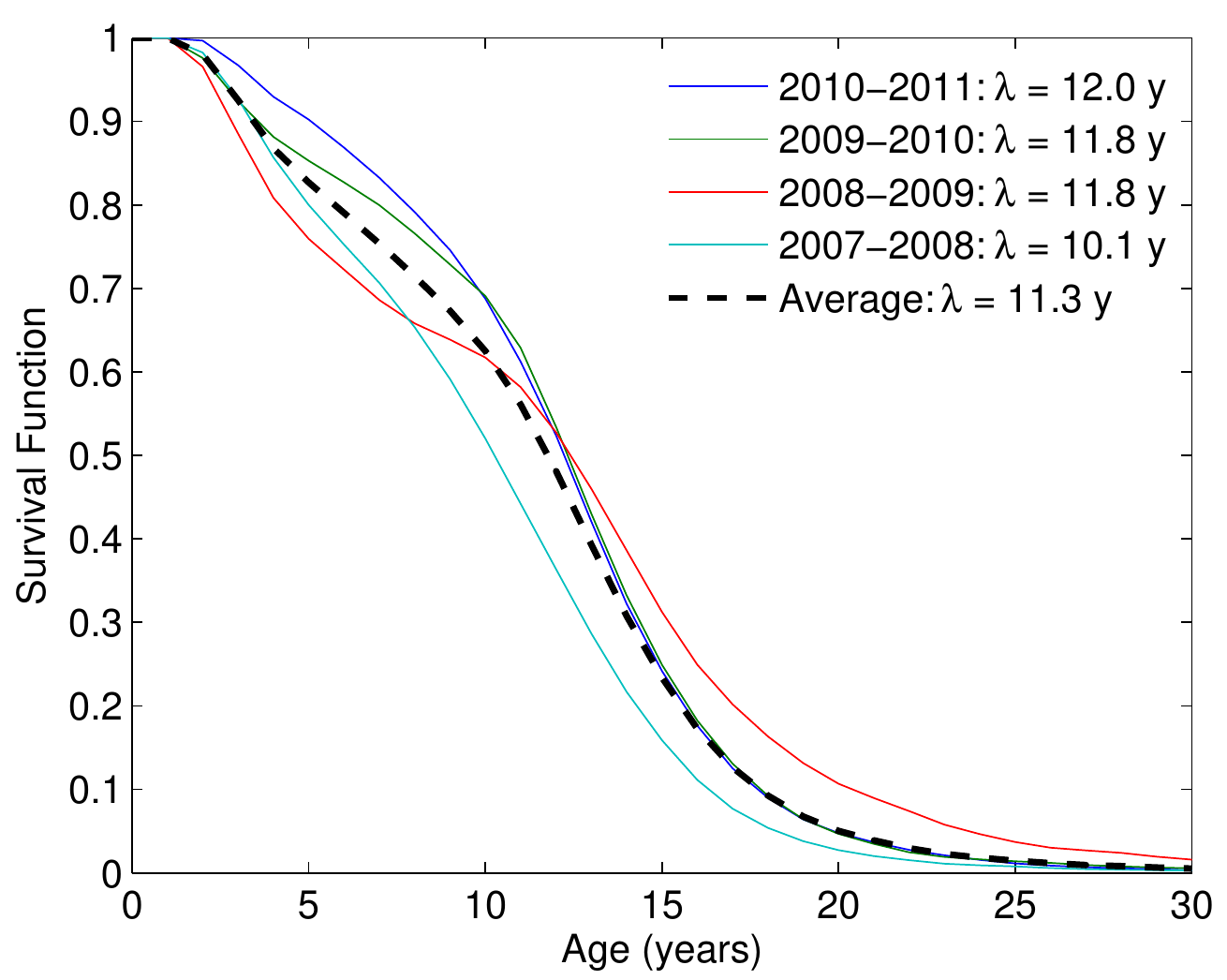}
			\includegraphics[width=1\columnwidth]{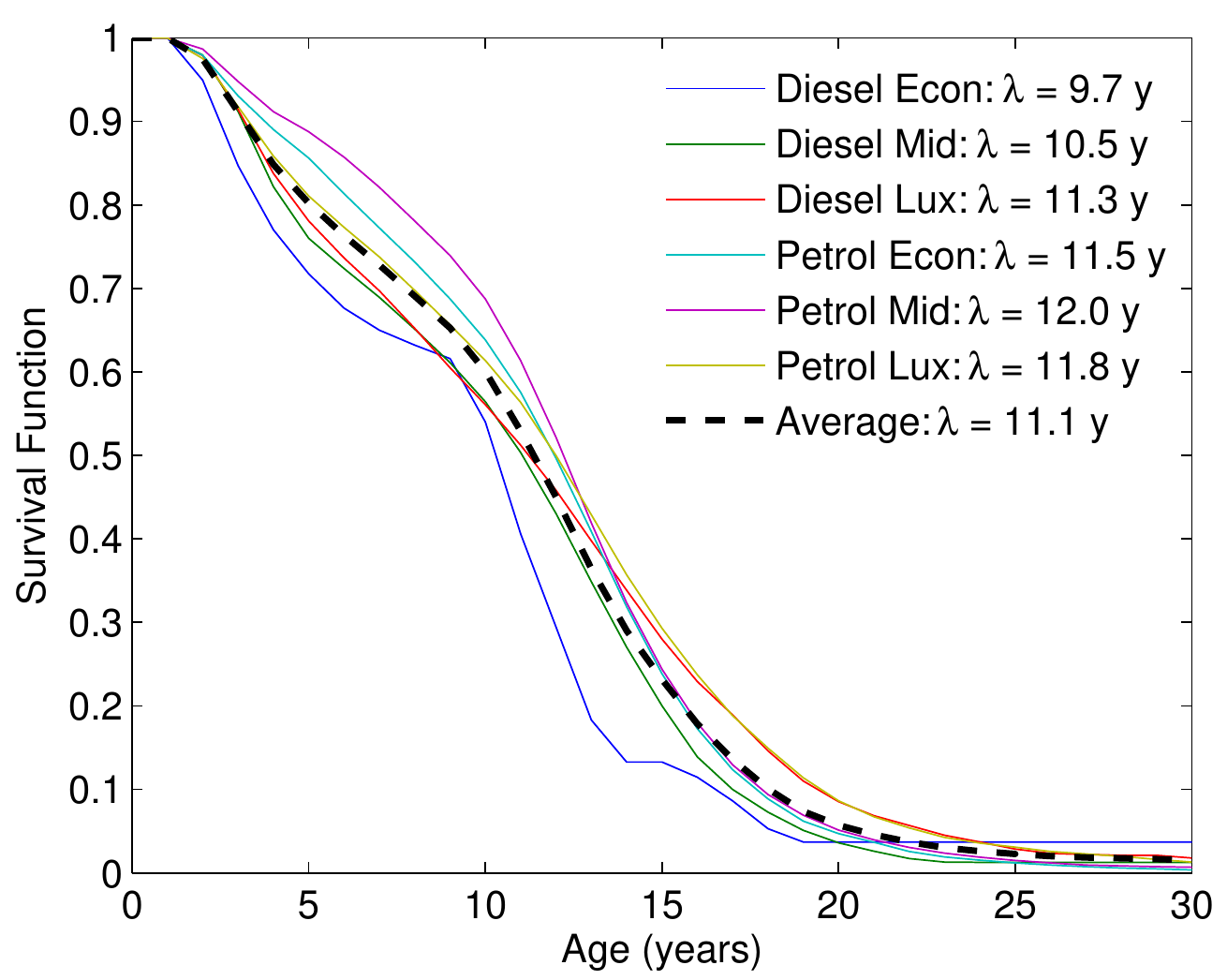}
		\end{center}
	\end{minipage}	
	\caption{Survival functions as calculated using equations of section \ref{sect:survival} with survey plate recognition data.}
	\label{fig:Survival1}
\end{figure}

\begin{figure}[p]
	\begin{minipage}[t]{0.5\columnwidth}
		\begin{center}
			\includegraphics[width=1\columnwidth]{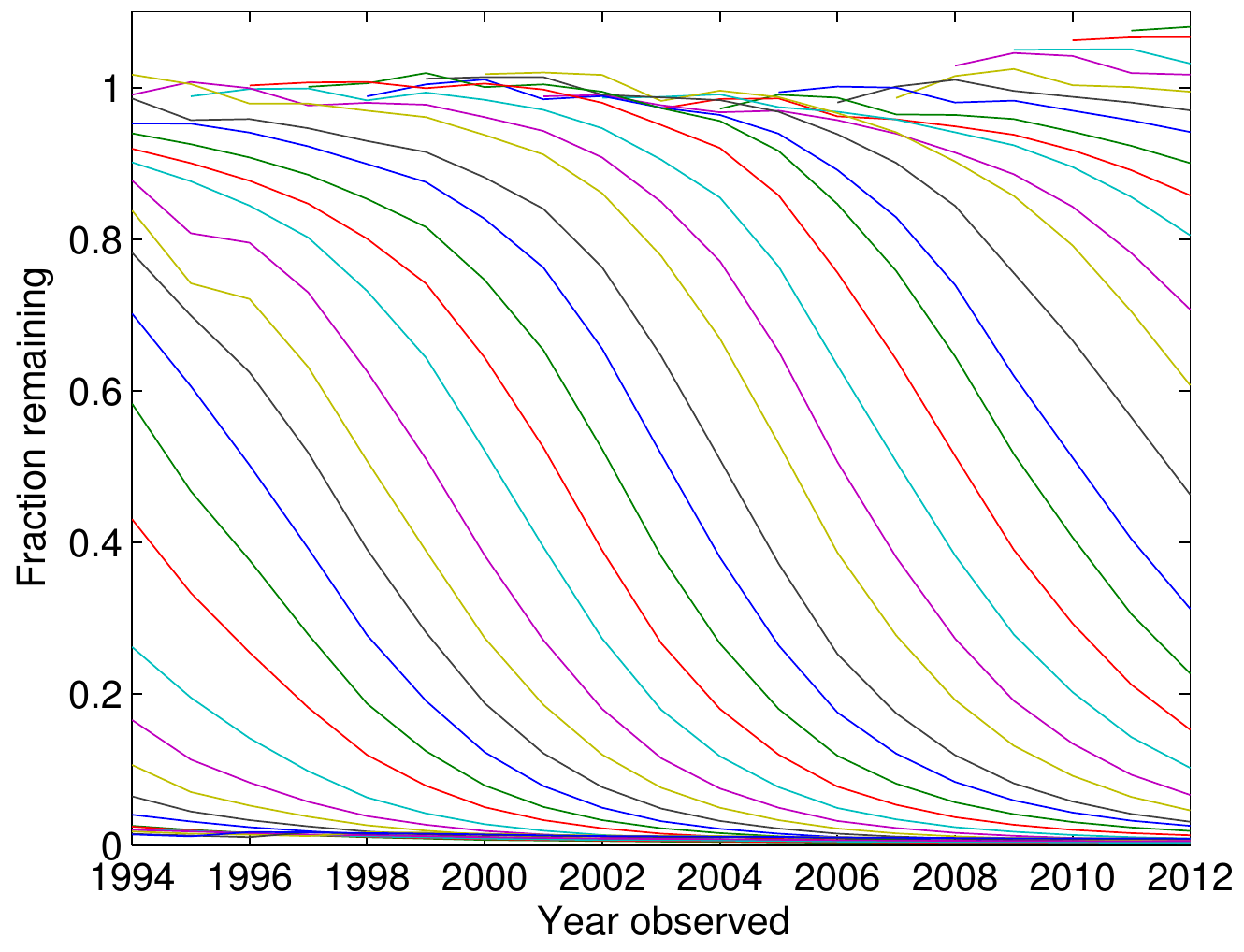}
			\includegraphics[width=1\columnwidth]{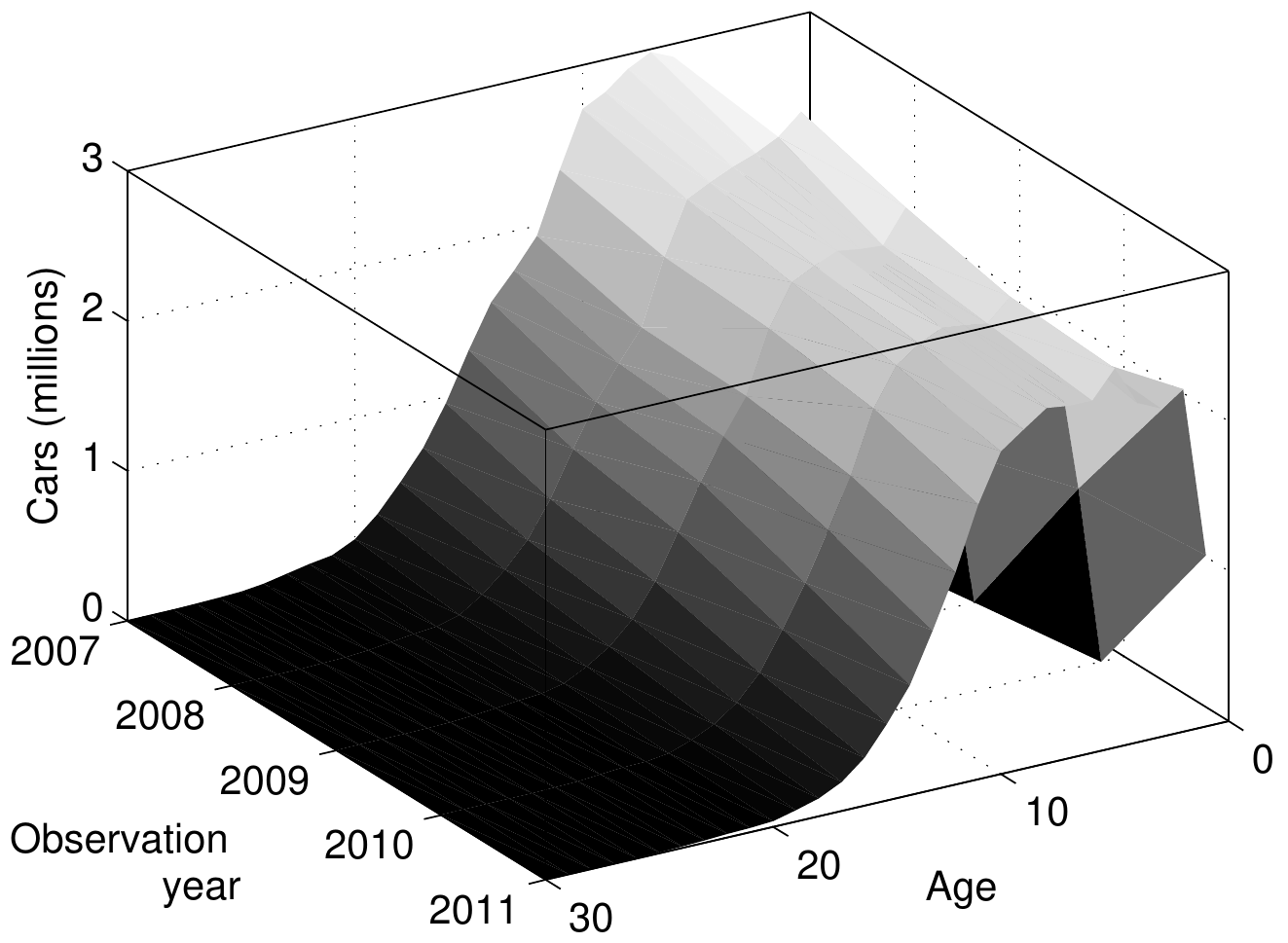}
			\includegraphics[width=1\columnwidth]{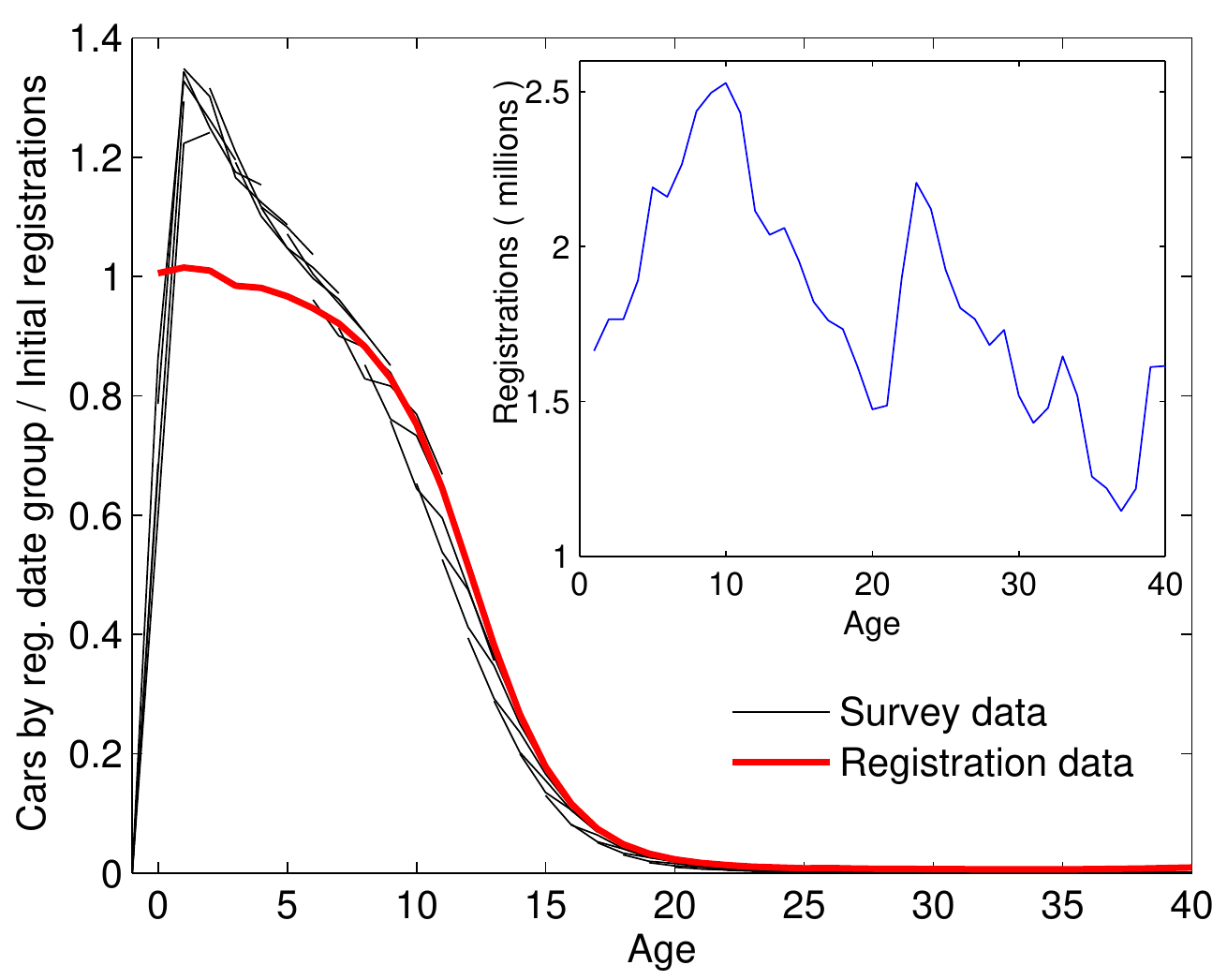}
		\end{center}
	\end{minipage}
	\hfill
	\begin{minipage}[t]{0.5\columnwidth}
		\begin{center}
			\includegraphics[width=1\columnwidth]{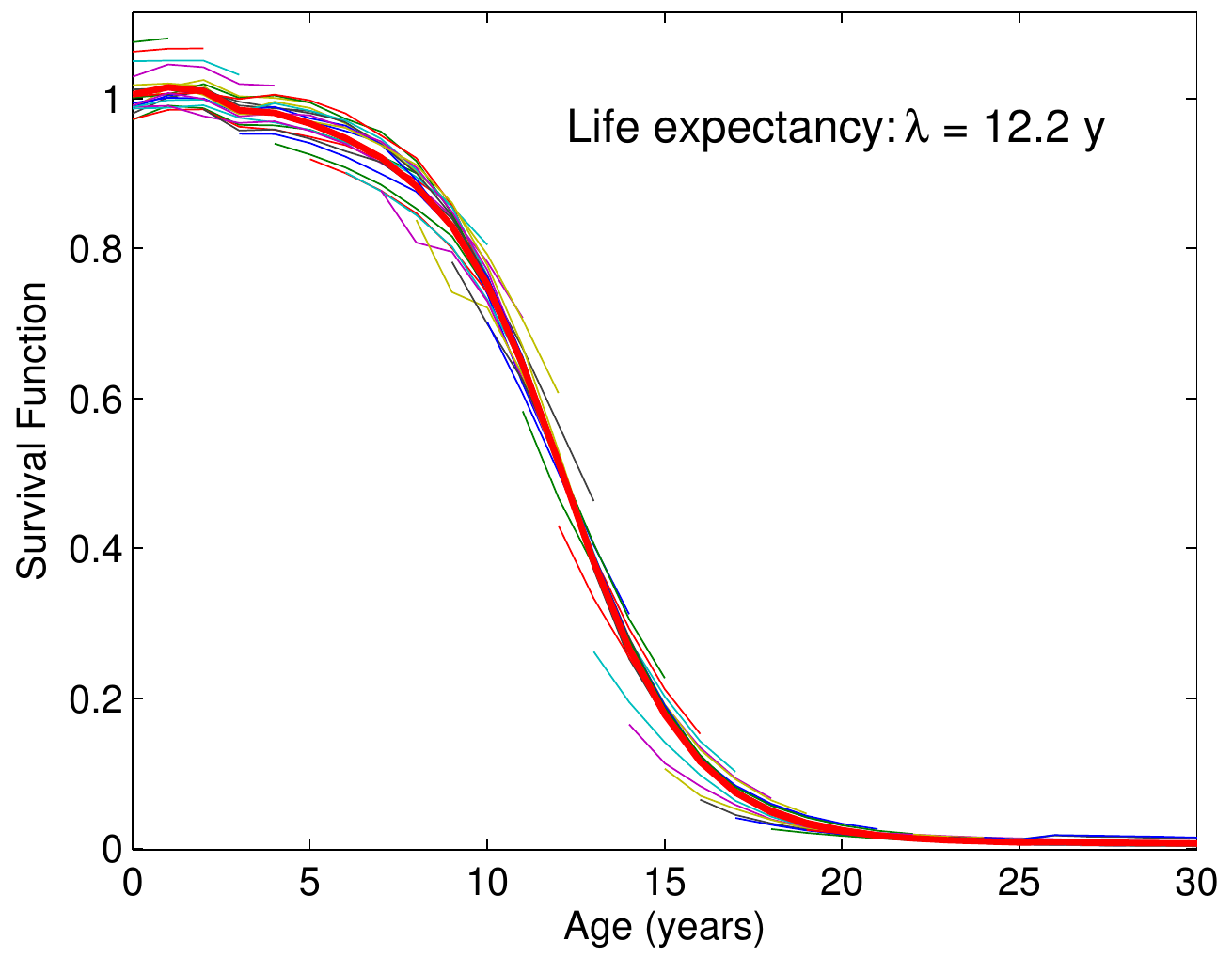}
			\includegraphics[width=1\columnwidth]{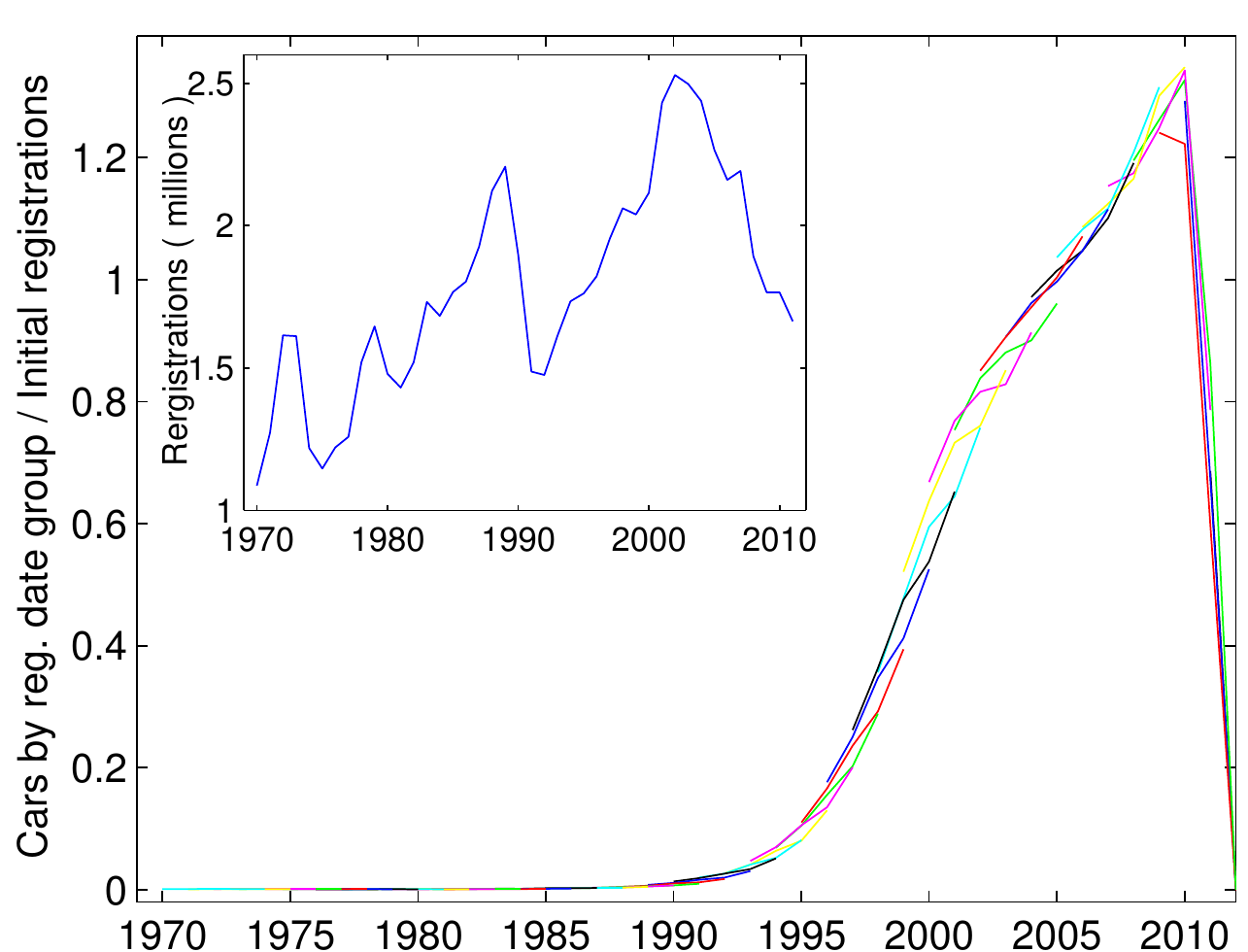}
			\includegraphics[width=1\columnwidth]{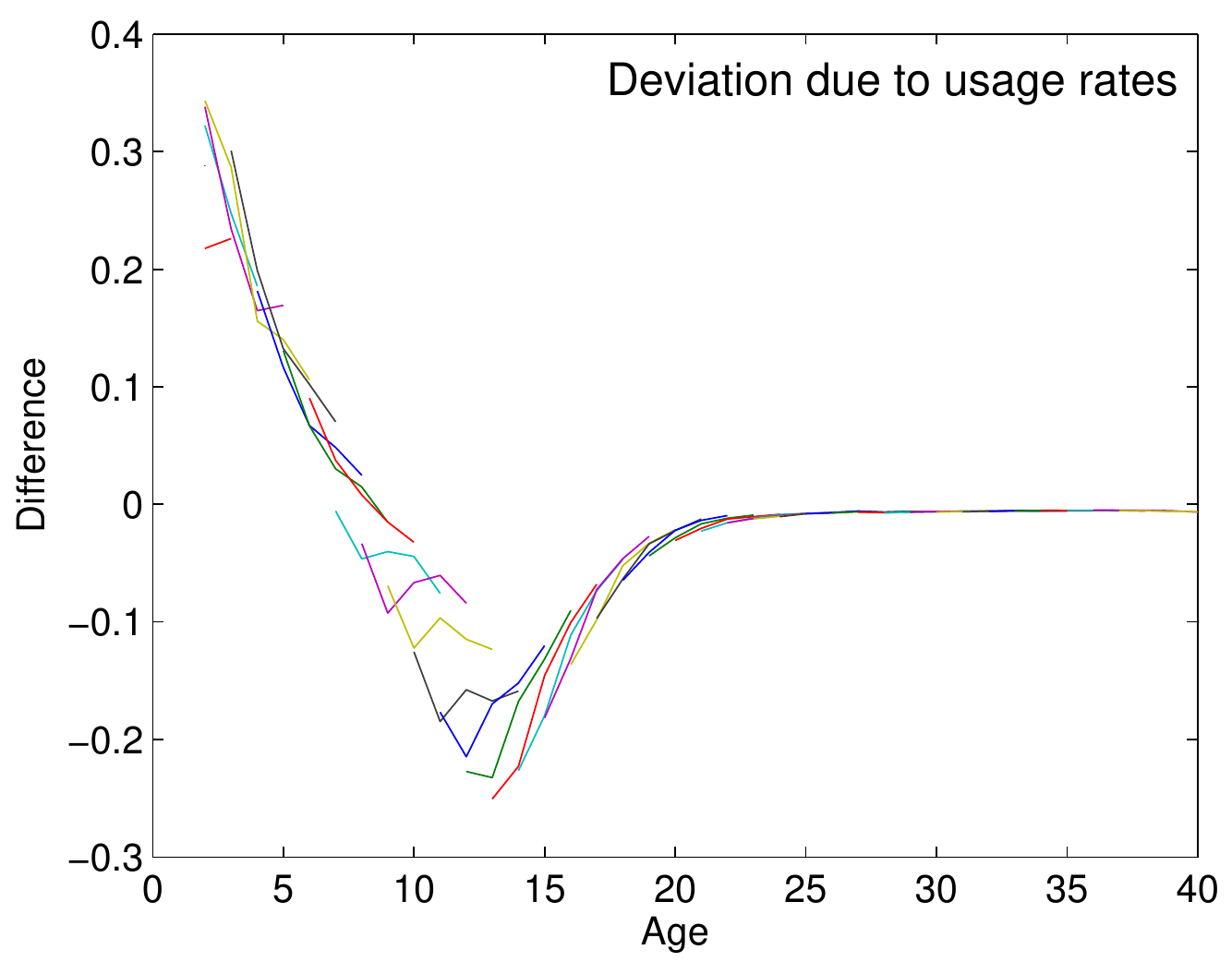}
		\end{center}
	\end{minipage}	
	\caption{\emph{Top two panels} DVLA registration data document VEH0211. \emph{Bottom panels} Survival rates calculated using survey plate recognition data: young vehicles are over represented.}
	\label{fig:Survival2}
\end{figure}

By calculating how many vehicles disappear between years of observation within particular registration year groups, it is in principle to determine the force of death and survival function, as described in the theory given in section \ref{sect:survival}. This was calculated for every engine type and size class, and re-averaged either over observations years or size class. Results are given in fig.~\ref{fig:Survival1}. The resulting survival functions have a peculiar shape with a dip in early years, indicating that vehicle use may be age-dependent.

The DVLA provides a document where the decline of car numbers by year of manufacture are followed. This is shown in fig.~\ref{fig:Survival2}, top left panel, normalised by the registration number. For each year of manufacture, a similar ageing pattern is observed. These can be shown to be very close to logistic functions of one displays them, using $F$ the fraction of initial sales remaining, and plotting $F/(1-F)$ on a logarithmic $y$ axis. This produces linear trends (not shown here). These curves can be overlaid over each other by displaying them according to age. All of these curves trace small sections of a longer trend which represents the survival function, shown in the top right panel.

The deviation of the survey data from this survival function has an explanation. For this we used the data in a different way than what was done in fig.~\ref{fig:Survival1}. Taking the number of observations as a function of year of observation and age, one can draw a three dimensional surface (fig.~\ref{fig:Survival2}, middle left panel). If one follows on this surface trajectories followed by specific age groups, a gradual decline is observed along the time of measurement. If normalised by the initial number of registrations, such trajectories generate sections of the survival function at different ages, 5 data points long, assuming that the survival function has not changed significantly with time. Fractions of vehicles remaining from initial registrations for registration years since 1970 are shown in the middle right panel, with DVLA car registrations in the inset. Note that this fraction increases above one in recent years, apparently contradictory. The same data is shown in terms of age in the bottom left panel, with the same registration data in the inset. Overlaid with this is the survival function taken from the top right panel. We see that above 10 years of age, the survey data agrees very well with DVLA registration data, but that for younger ages the survey data climbs above one. This is interpreted to the fact that vehicles younger than 10 years of age are used significantly more than older vehicles, and approximately linearly so the younger they are. In the survey data, an arbitrary observation frequency arises that could be age-dependent, and this seems to be the case. Thus in the survey data, younger vehicles are over-represented because they are driven more often; meanwhile above 10 years old their frequency of observation in cameras is representative of their registration numbers. The bottom right panel shows the difference between the survey data and the survival curve.

\subsubsection{Emissions factors changing over time \label{sect:UKEmTime}}

\begin{figure}[t]
		\begin{center}
			\includegraphics[width=1\columnwidth]{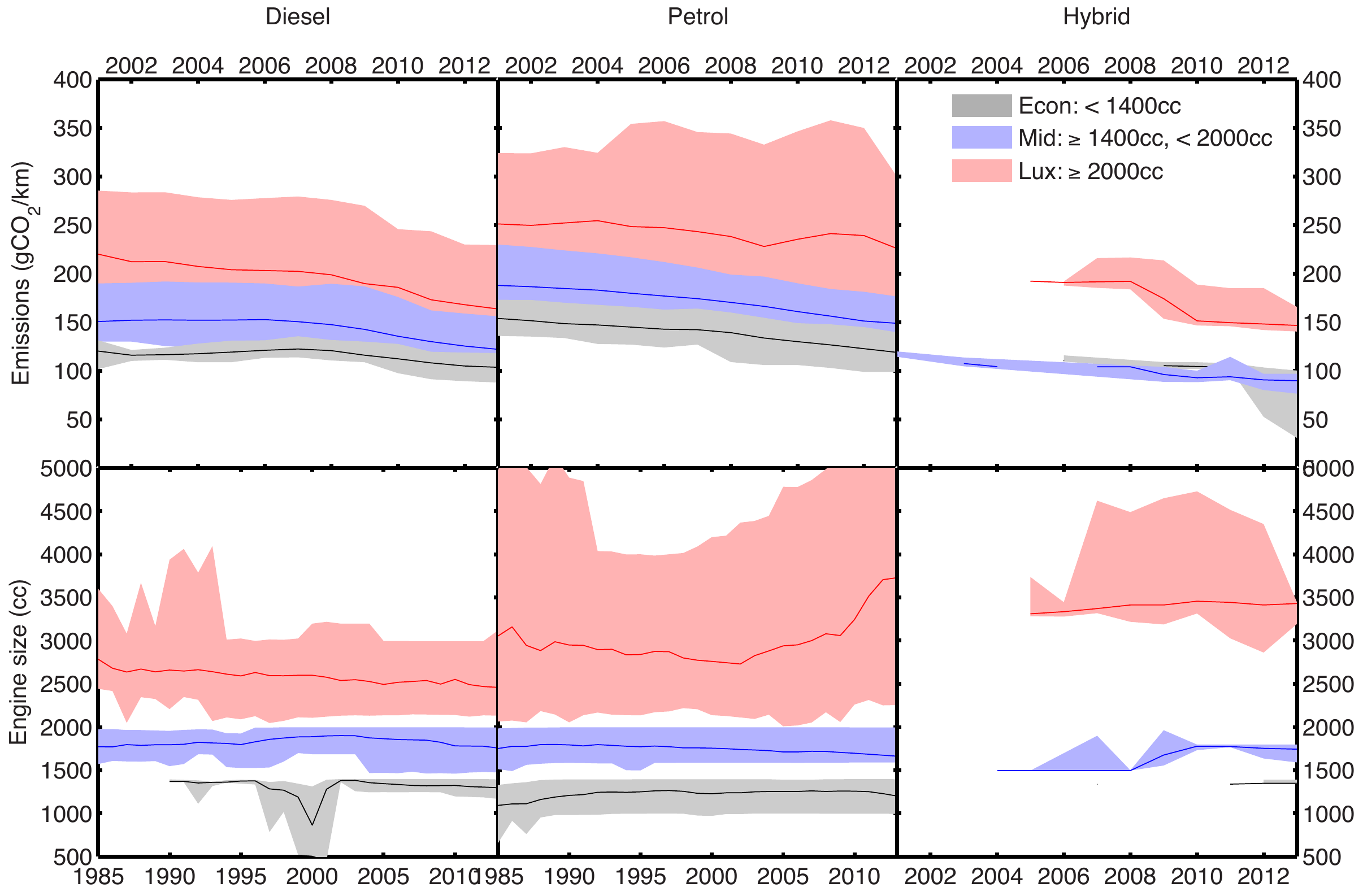}
		\end{center}
	\caption{Changes of engine sizes and emissions factors over time for three engine size classes and three technologies against year of registration. vehicles registered prior to 2001 do not have rated emissions in the database. The shaded areas include 67\% of all vehicles by category (the standard deviation).}
	\label{fig:EmissionsTime}
\end{figure}
\begin{figure}[t]
		\begin{center}
			\includegraphics[width=.65\columnwidth]{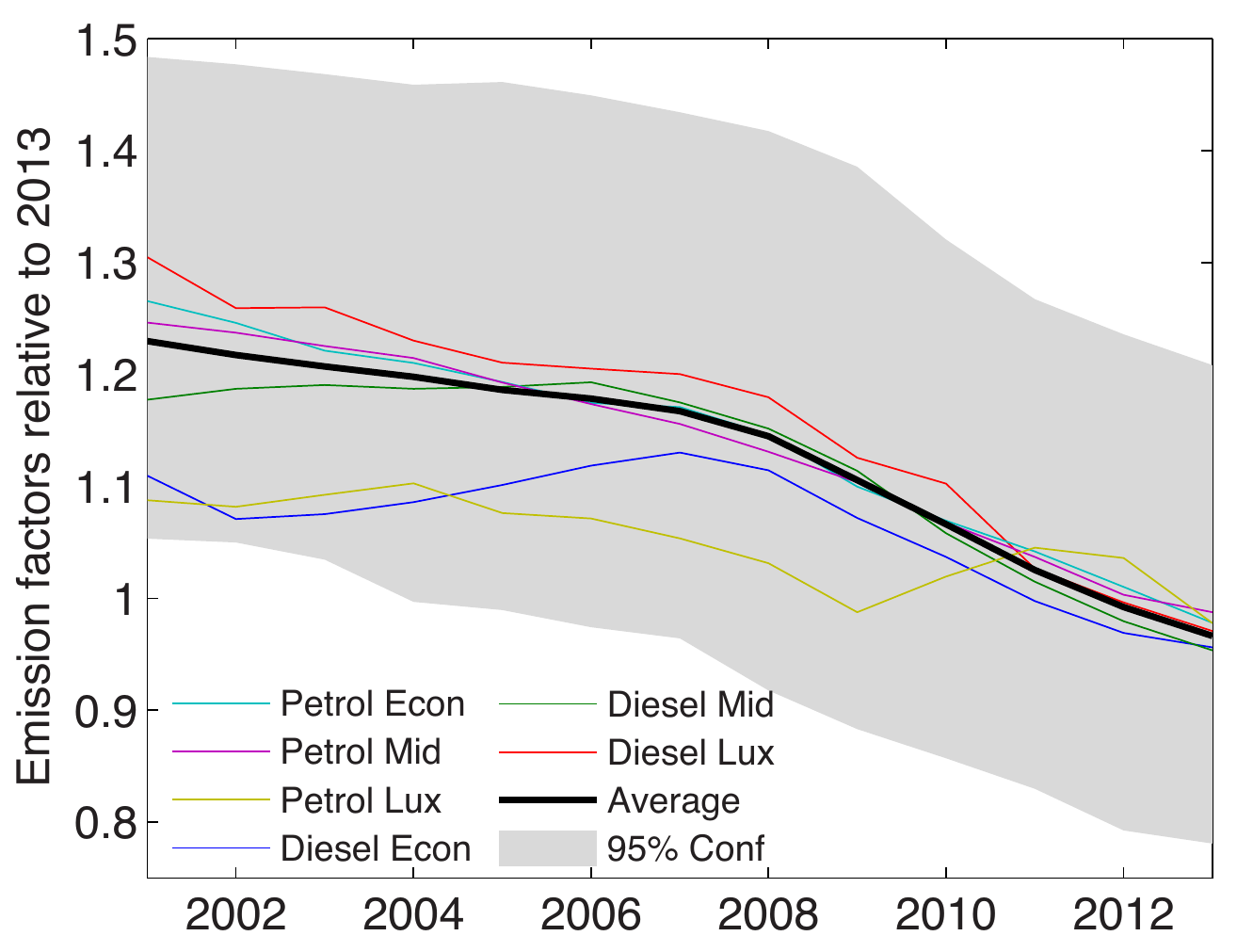}
		\end{center}
		\caption{Trends of emissions factors normalised by their 2011 value (mean curves in the top panel). Solid lines are averages weighted by population numbers. Shaded areas indicate regions of parameter space in which 95\% of the vehicles are located.}
		\label{fig:EmissionsTrend}	
\end{figure}

Section \ref{sect:TimeEmTh} describes how to determine fleet emissions factors from those of new vehicles using sales time series in different regions if the evolution of emission factors can be obtained for at least one country. The UK survey dataset offers this possibility, without which it might not have been possible. By determining the function $f(a)$ of relative changes in emissions in time, one can in principle have an idea how emissions of new vehicles have changed over the years, and use it to evaluate fleet emissions from a time series of sales and a survival function. If we assume that to first order, $f(a)$ is the same worldwide, then we can work back average fleet emissions by category from those of new vehicles. 

In the the UK survey, each entry feature a year of make, an engine size and an emissions factor if the year of make is after 2000. We have analysed how emissions factors and engine sizes are distributed by engine type and size class and year of make. This is shown in figure~\ref{fig:EmissionsTime}, where lines indicate the mean of the distributions and the shaded areas the regions within their standard deviation. One observes that emissions factors have clearly gone down with time in each category, while engine sizes have not changed significantly. This indicates improvements in combustion technologies, not a change in the power of vehicles. We also observe wide variations in these characteristics for petrol and diesel engines, which are established and diverse, while hybrid vehicles feature a small number of available models.

Using these average trends, the function $f(a)$ can be determined for each category, shown in figure~\ref{fig:EmissionsTrend}, by normalising mean emissions by their 2011 value. This thus indicates how emissions factors of new vehicles have changed relatively since 2001. These curves are noisy but a weighted average is given with the thick black curve. The grey shaded area indicates the 95\% confidence level (two standard deviations). Using these numbers and the time series for total registrations of new vehicles given in DVLA document VEH0153 and the survival function calculated above (see figure~\ref{fig:Survival2}, bottom left panel), we evaluated fleet emissions relative to those of new vehicles using the function $f(a)$, which is taken as the average given in figure~\ref{fig:EmissionsTrend}. This can be compared against measured fleet emissions from the survey, given above, in order to check whether this methodology works. Using the sum $\sum f(a) \xi_i(a) \ell_i(a) / \sum \xi_i(a) \ell_i(a)$, we obtained that fleet emissions are 15\% higher than those of new vehicles, with an uncertainty range that spans from 10 to 20\%. Meanwhile, comparing figure~\ref{fig:EmissionsDistTimeD} to figure~\ref{fig:AllEmDist}, we had established that fleet emissions were in general 20\% higher than those of new vehicles entering the fleet. We conclude that this method works but only very approximately. The error stems from working with averages. It is, however, the only method available to us to evaluate fleet emissions in other countries than the UK.

\subsection{Motorcycles}
\subsubsection{Motorcycle age distributions and survival functions}

\begin{figure}[p]
	\begin{minipage}[t]{0.5\columnwidth}
		\begin{center}
			\includegraphics[width=1\columnwidth]{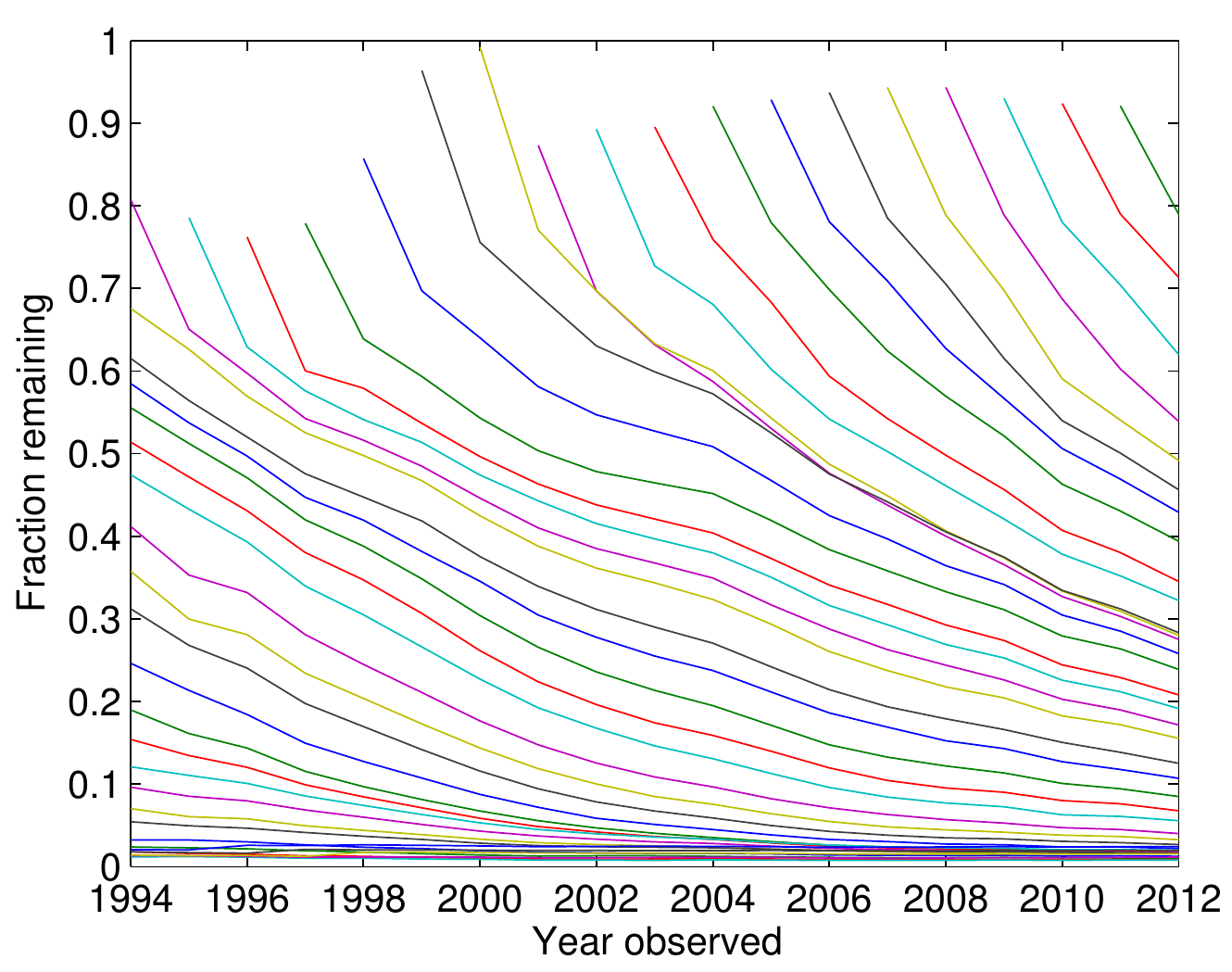}
			\includegraphics[width=1\columnwidth]{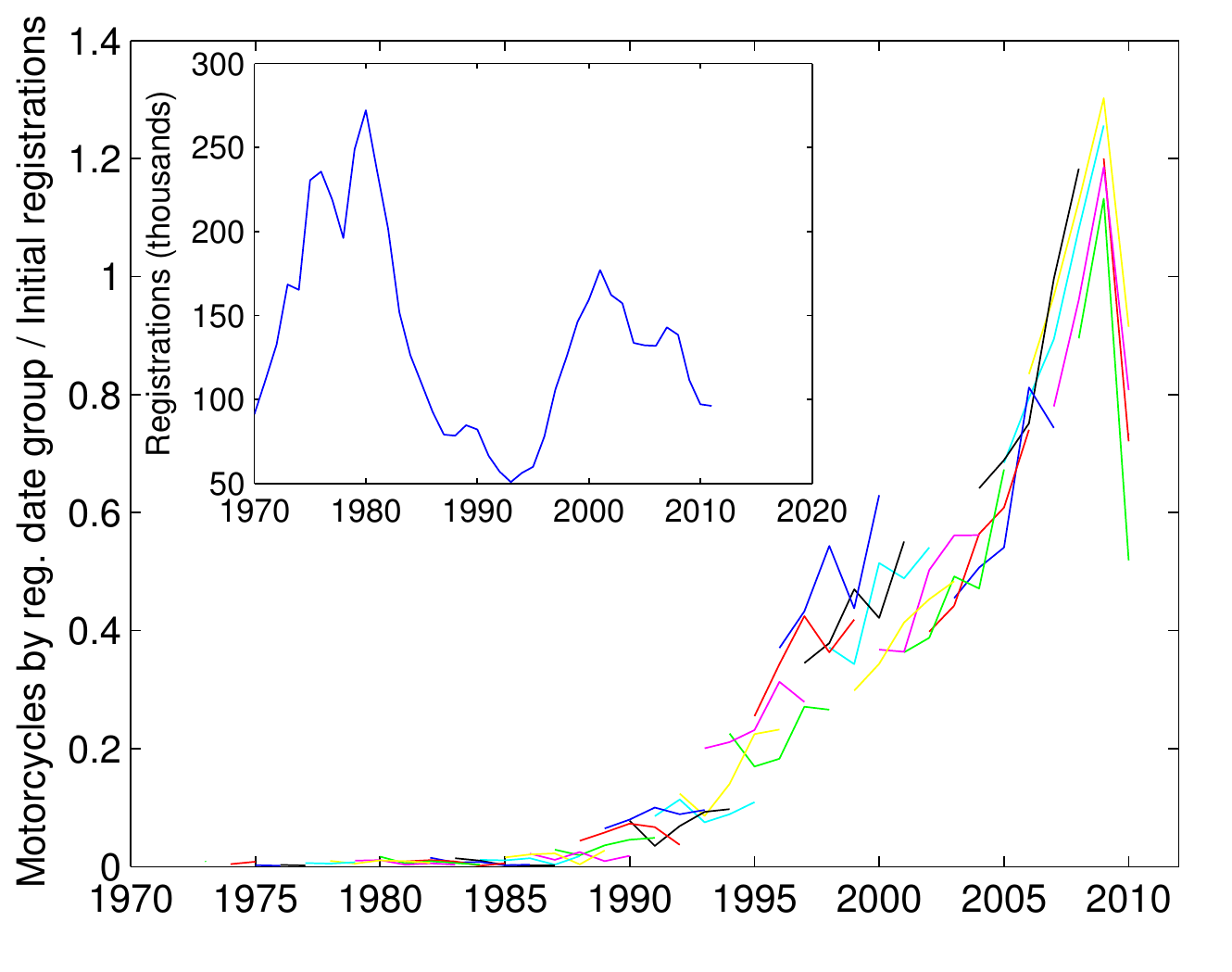}
			\includegraphics[width=1\columnwidth]{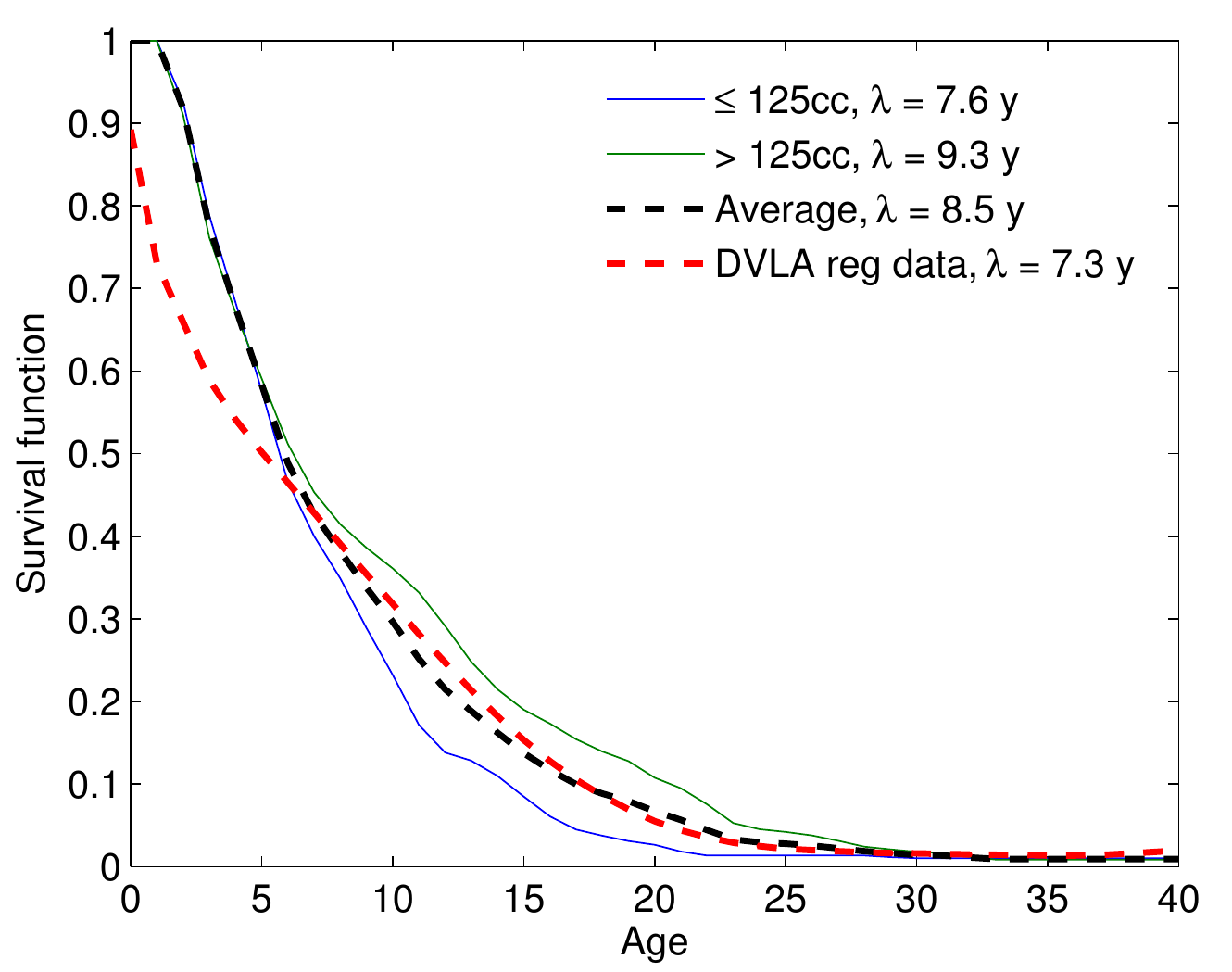}
		\end{center}
	\end{minipage}
	\hfill
	\begin{minipage}[t]{0.5\columnwidth}
		\begin{center}
			\includegraphics[width=1\columnwidth]{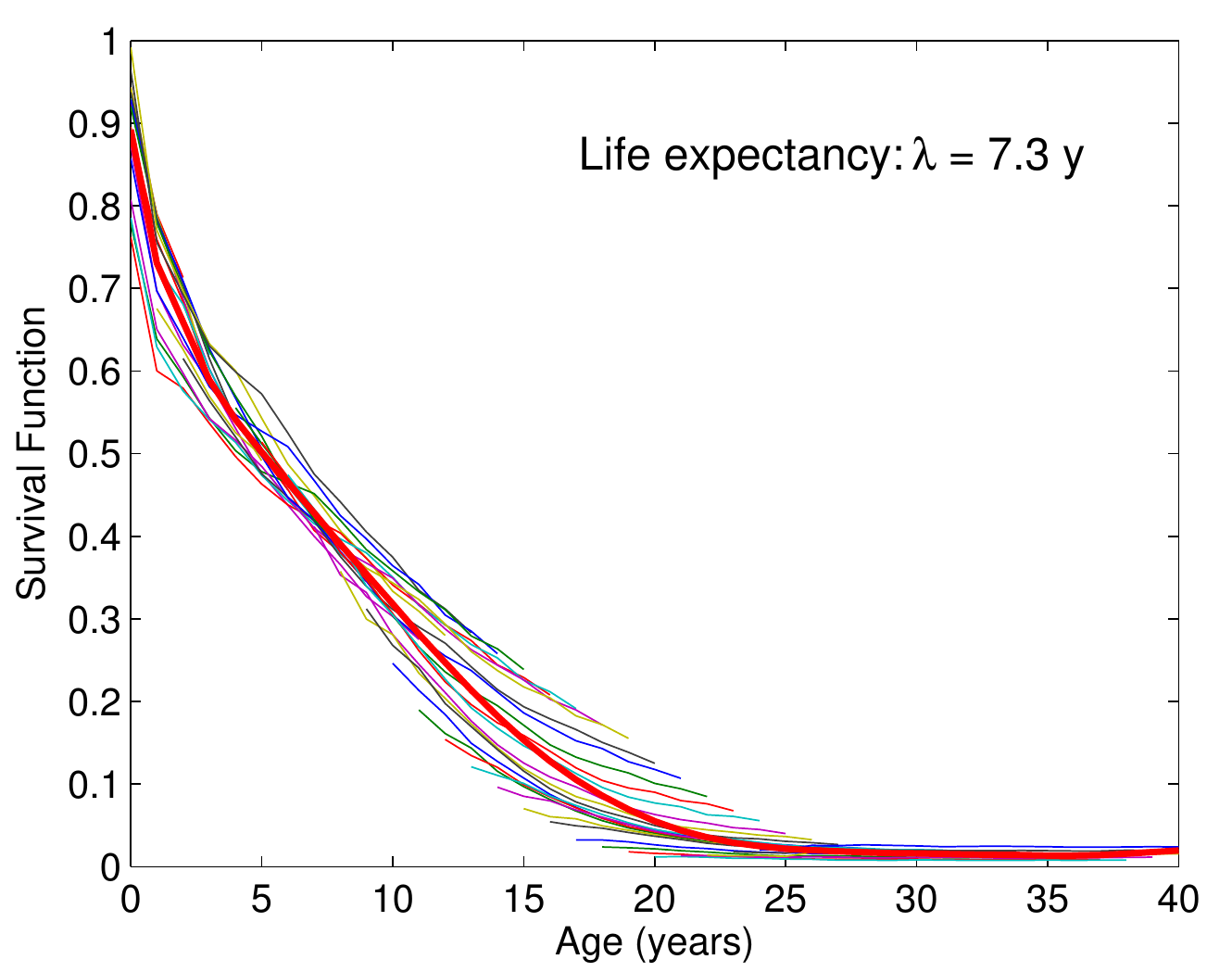}
			\includegraphics[width=1\columnwidth]{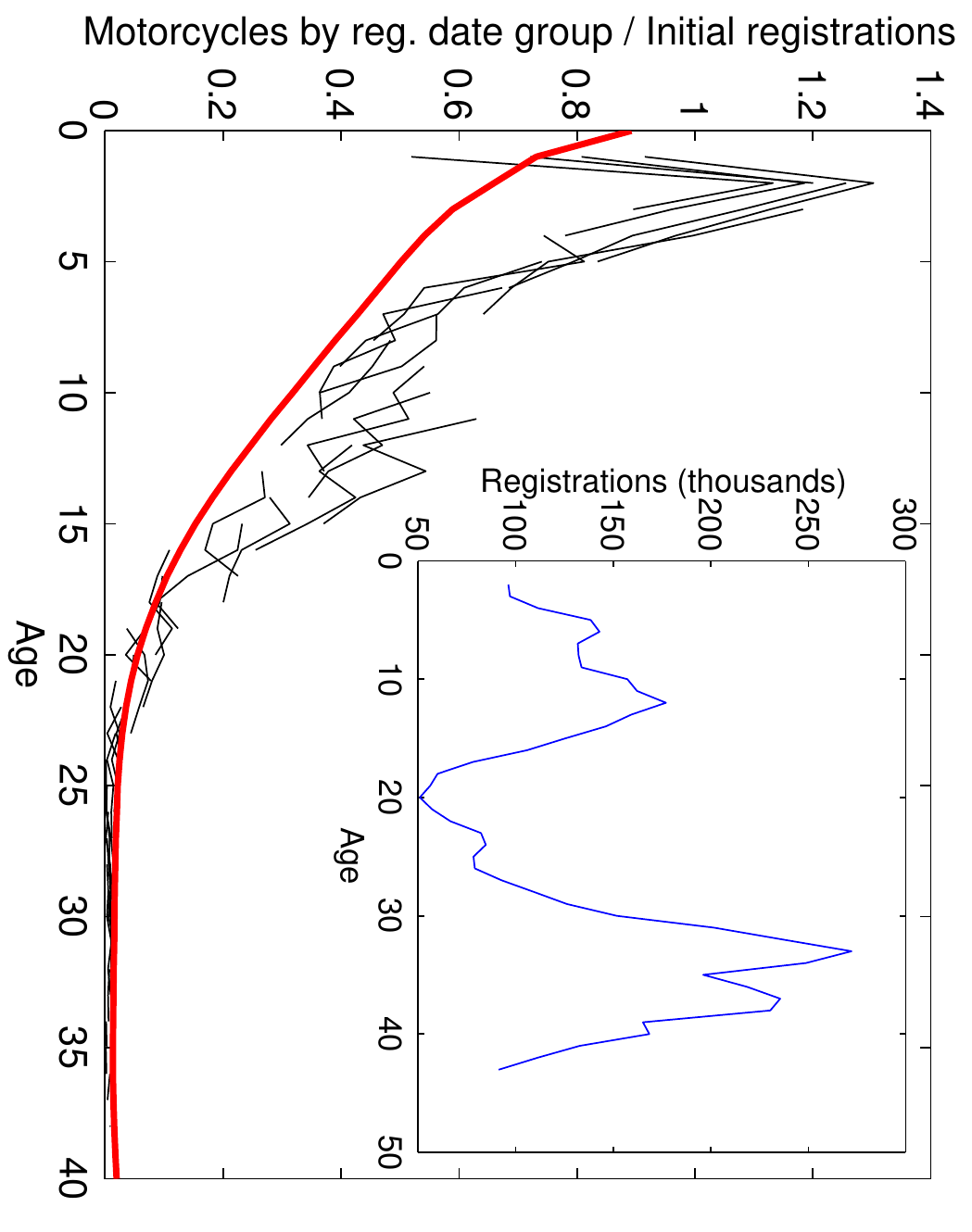}
		\end{center}
	\end{minipage}	
	\caption{\emph{Top two panels} DVLA registration data document VEH0311. \emph{Middle panels} Survival rates calculated using survey plate recognition data. \emph{Bottom panels} Survival functions calculated using survey data and equation \ref{eq:Survival2}, for large and small motorcycles, their average, and the survival function calculated in the top panels.}
	\label{fig:Survival2}
\end{figure}

A survival function was calculated for motorcycles in the UK, which turn out quite different than that for vehicles, and hence it is useful to explore in detail. Following the same methodology as for vehicles, we used the age distribution by years from the DVLA survey data, as well as DVLA registration data from document VEH0311. In exactly the same way as for vehicles, the survey data produces survival functions that include an age dependent level of use of motorcycles, while the registration data does not. 

Obtaining survival functions from registration data using document VEH0311, normalised by first registrations, is as straightforward as for vehicles, and this is shown in the top two panels of fig.~\ref{fig:Survival2}. For every year of registration, the slope of the number of vehicles remaining with time shortly after first registration is very high, and slows down with age (left panel). By displaying the data in terms of age, and averaging, we obtain an average survival function (right panel). This shows that the survival function has not changed significantly with time and is fairly consistent. The average is shown with a thick red line, featuring a life expectancy of 7.3 years. The sharp decline at low ages is ascribed to high rates of accidents for recently purchased motorcycles. This slows down with age, where for old vehicles, breakdowns dominate and those that survive through an accident dominated young age often survive up to 20 years or more. Thus the distribution of ages of death is wider than for vehicles, the latter being dominated by breakdowns.

The survey data was used to determine declines in motorcycle populations by year of registration. This is shown in the middle panels (with historical registrations in the inset). This features an age dependent level of use. The survival function calculated in the top panels is overlaid to this data with a thick red line. We thus observe very rough consistency, and that the age dependent level of use is higher at young ages. Finally, survival functions were calculated from the survey data using eq.~ref{eq:Survival2} as was done for vehicles. This produces survival functions again roughly consistent albeit with higher life expectancy by one year on average, with 1.7 years difference between small and large motorcycles. The difference between calculated survival functions most likely stems from an age dependent level of use, similarly for vehicles.

\newpage

\section{Data parameterisation of FTT:Transport}
For the present study, an original database detailing the technological profile of vehicles and vehicle populations was built. Table~\ref{table:2-1} shows the data that were used to build everything needed to parameterise FTT:Transport. Columns two to four are give the scope and level of detail required, including the selected time periods and the resolution of the data. The following sections explain how each type of data was collected, and shows the data when possible. The rest of the FTT parameterisation can be obtained from the Supplementary Excel File that can be downloaded with this publication.

\begin{table}[h]
    \caption{Data sources for the main variables}
	\small
	\begin{tabular}{p{3cm}p{2cm}p{2cm}p{7.8cm}}
	\hline\hline
         Variable &         Year         &Differentiated in 18 countries?                &Data source                \\
         \hline
         \\
        Vehicle sales by model         &2012            &Yes & Fleet numbers are collected from Marklines  
         \\
         \\
                Vehicle prices      &  2012               &Yes        &  Car manufacturer's website and car sales website (see details in table~\ref{table:4-5},table~\ref{table:4-6} and table~\ref{table:4-7})
         \\
         \\
         Fuel cost &                2013                      &Yes                                                        & Fuel use data are collected from car manufacturer's website and fuel price per litre is collected from the World Bank           
         \\
         \\
         Fuel economy             &2013                   & Yes & Car manufacturer's website and car sales website (see details in table~\ref{table:4-5},table~\ref{table:4-6} and table~\ref{table:4-7})
         \\
         \\
         Discount rate              & 2013                        & No                                                  &    E.g. \cite{inderwildi2012,zhuang2007,harrison2010}
         \\
         \\
         Learning rate               &2012                       & No                                                 & E.g. \cite{AEA2010, weiss2012, IEA2013c, McDowall}                       
         \\
         \\            
            Income data             &  1990-2012.          & Yes                           & E3ME\\
            \\
            Fleet size 		& 1990-2012              & Yes                          & Euromonitor \\
            \\
            Historical fuel prices &1990-2012             & Yes.                           &E3ME (IEA)\\

\hline      
		  \end{tabular}
		\label{table:2-1}
\end{table}

\subsection{Cross-sectional data gathering procedure and distributions for 18 regions and proxies}

This section provides an overview of new vehicle prices and emissions data we have collected. In FTT, all cost data comes in the form of distributions. This required to find data per vehicle model: number of sales, price, emissions, engine size. Such a database either does not exist for most countries, or would be too expensive to purchase. It was possible, however, to obtain vehicle sales per individual model from Marklines in one bundle for many countries. We verified that the totals match national data. The other characteristics of vehicles could be found on various commercial websites, including for instance www.carpages.co.uk, and manufacturer websites. We matched name by name the data for each individual vehicle model. This is quite time-consuming, but provides a very robust dataset. Since this took some time, the data dates back to 2012. We do not expect that the shape of these distributions change substantially over time, since they are mostly related to income distributions. The full methodology is explained in \cite{MercureLam2015}, and more details are given in tables~\ref{table:4-5}, ~\ref{table:4-6} and~\ref{table:4-7} discussed in section~{sect:General}.

\begin{figure}[p]
	\begin{minipage}[t]{1\columnwidth}
		\begin{center}
		        \includegraphics[width=1\columnwidth]{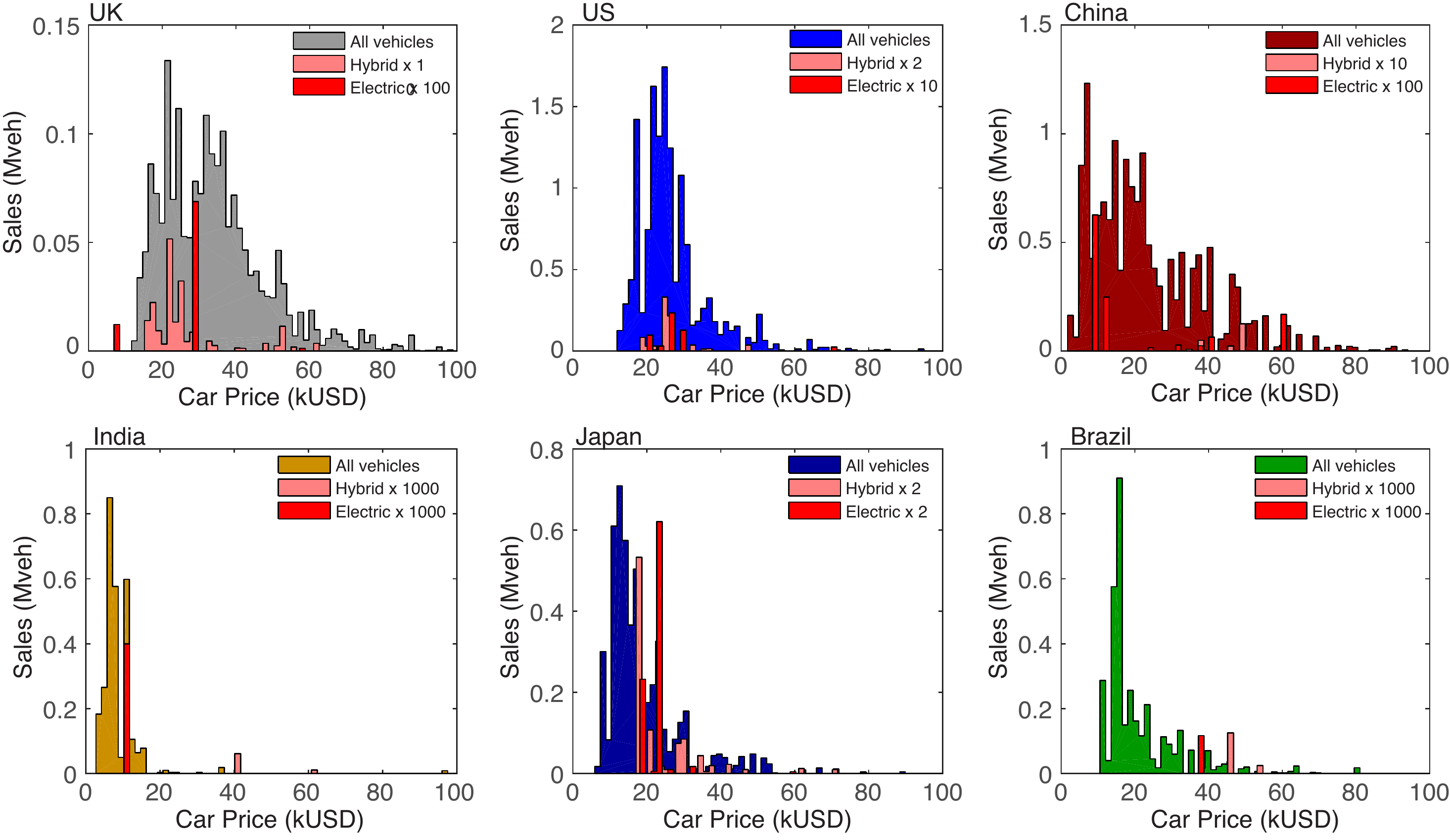}
		        \includegraphics[width=1\columnwidth]{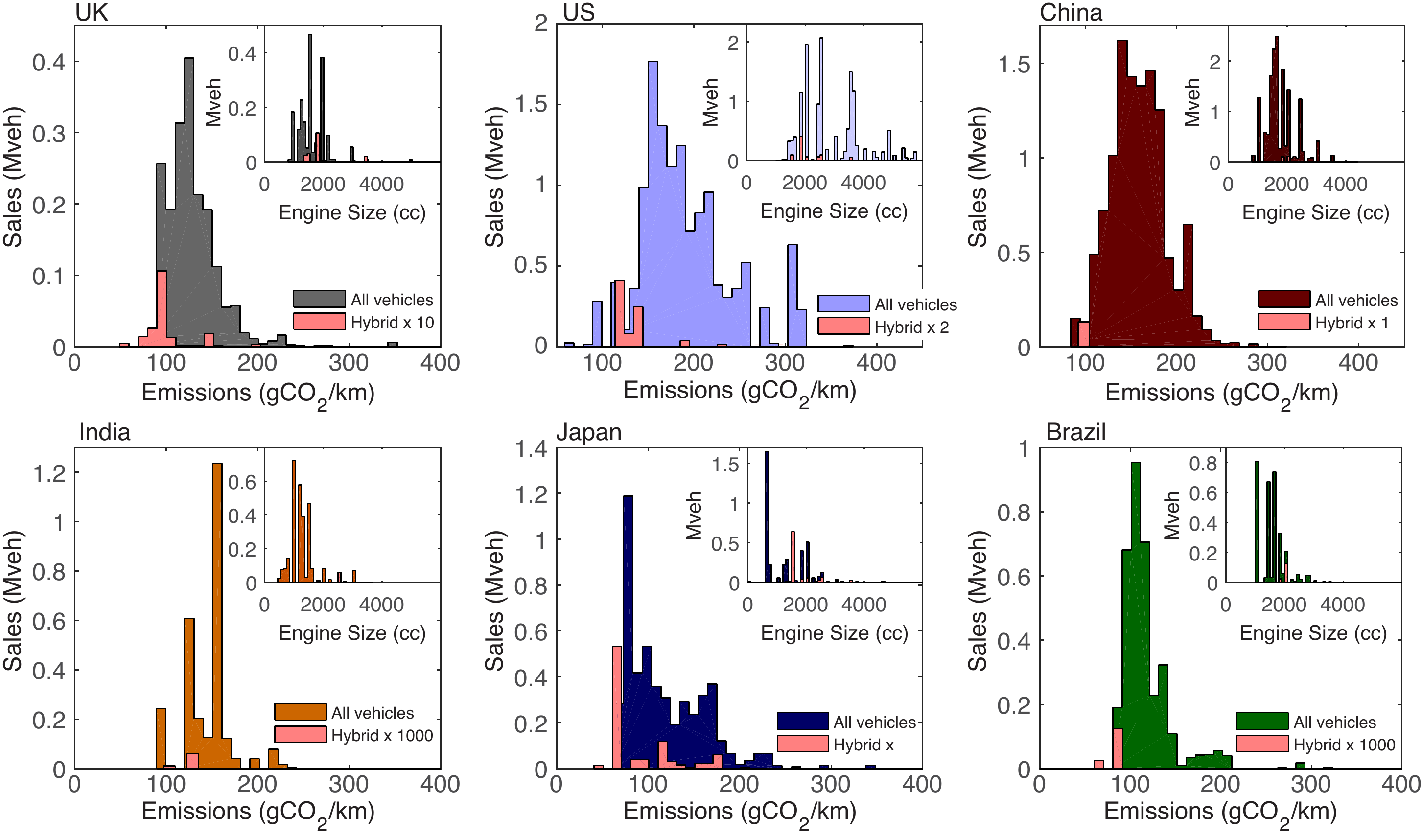}
		\end{center}
	\end{minipage}	
	\caption[Price and emissions distributions for the UK, US, Japan, China, India and Brazil.]{Price and emissions distributions for the UK, US, Japan, China, India and Brazil. Top two rows: price distributions for the six countries with identical price scaling.The price distributions for alternative vehicles are shown in pink. Bottom tow rows: emissions distributions (main graphs) and engine size (insets) of 2012 vehicles sales for the six countries. The emissions and engine size distributions are shown in pink \citep{MercureLam2015}.}
	\label{fig:2-1}
\end{figure}

\begin{figure}[p]
	\begin{center}
		\includegraphics[width=1\columnwidth]{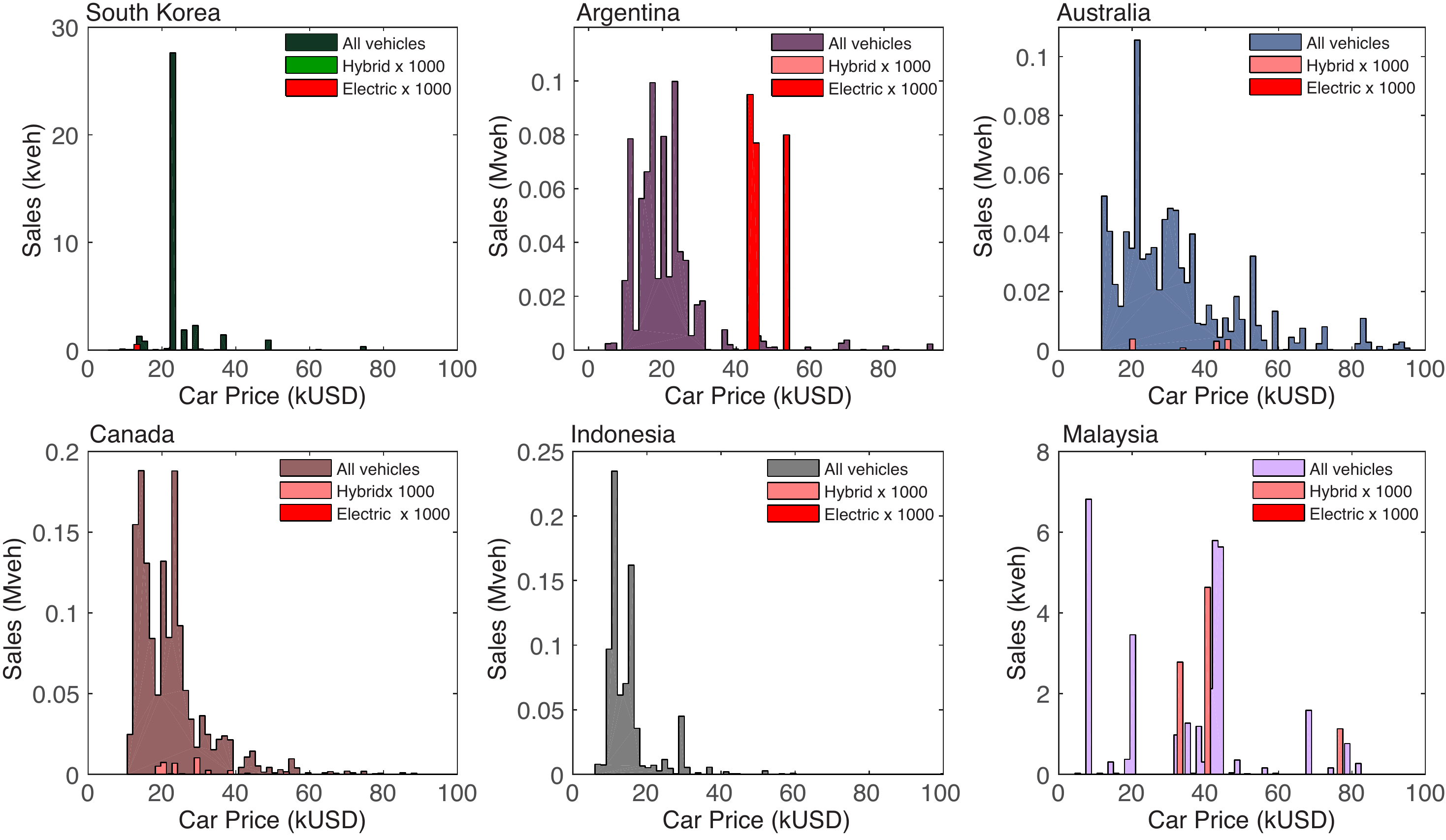}
		\includegraphics[width=1\columnwidth]{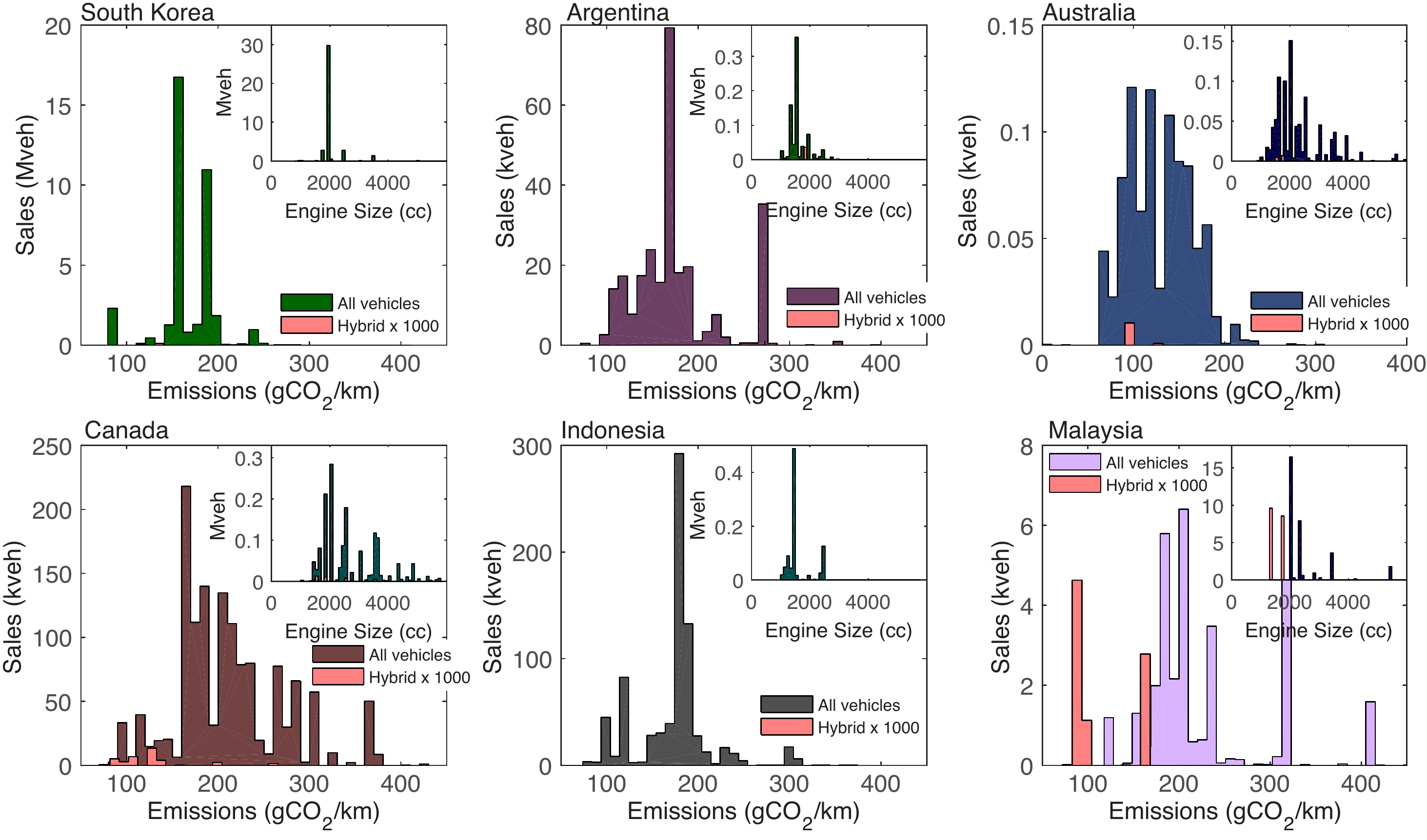}
	       	       	\end{center}
	\caption[Price and emissions distributions for the Korea, Argentina, Australia, Canada, Indonesia and Malaysia.]{Price and emissions distributions for the Korea, Argentina, Australia, Canada, Indonesia and Malaysia. Top two rows: price distributions for the six countries with identical price scaling.The price distributions for alternative vehicles are shown in pink. Bottom two rows: emissions distributions (main graphs) and engine size (insets) of 2012 vehicles sales for the six countries. The emissions and engine size distributions are shown in pink.}
	\label{fig:2-3}
\end{figure}

\begin{figure}[p]
	\begin{center}
		 \includegraphics[width=1\columnwidth]{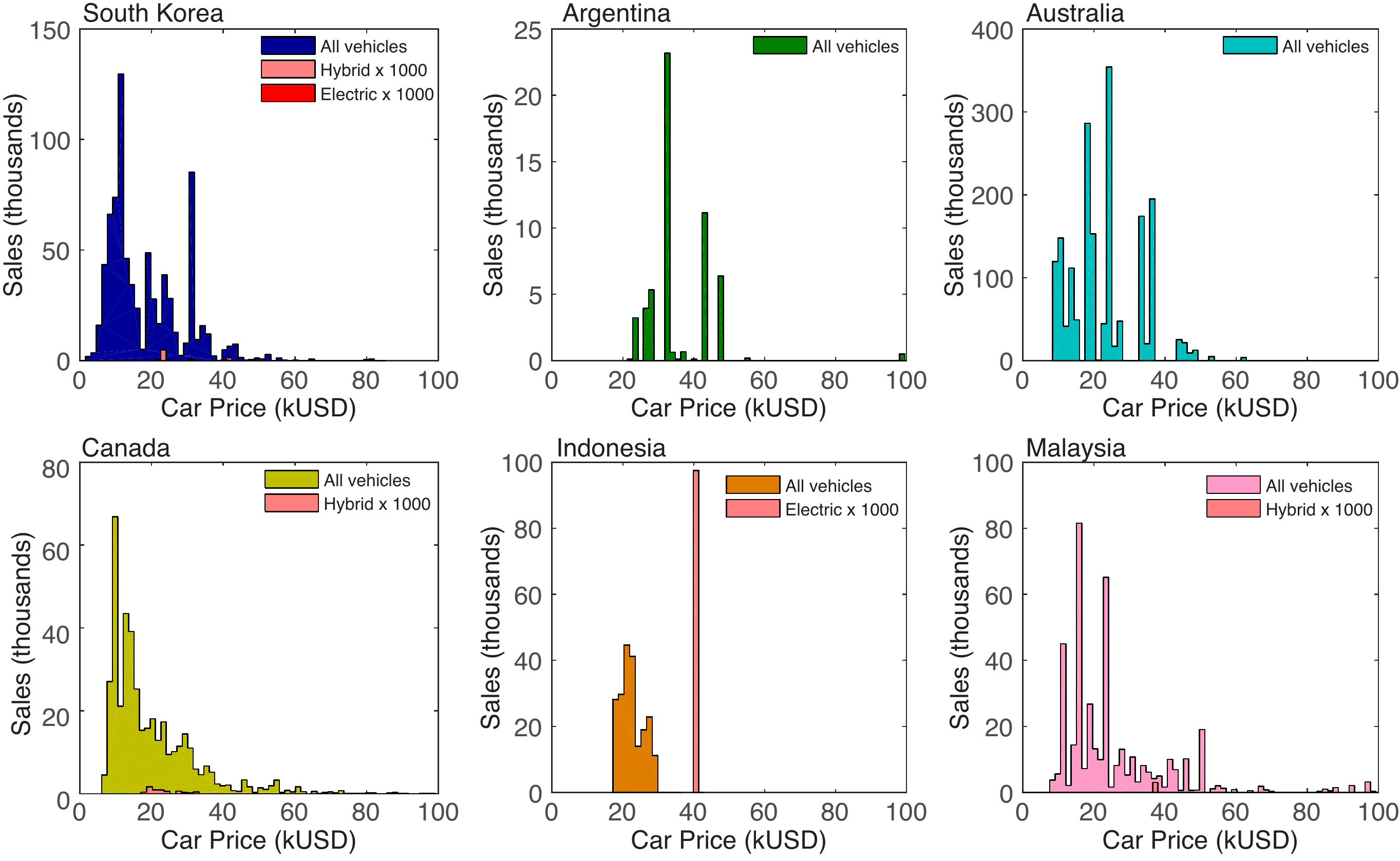}
		\includegraphics[width=1\columnwidth]{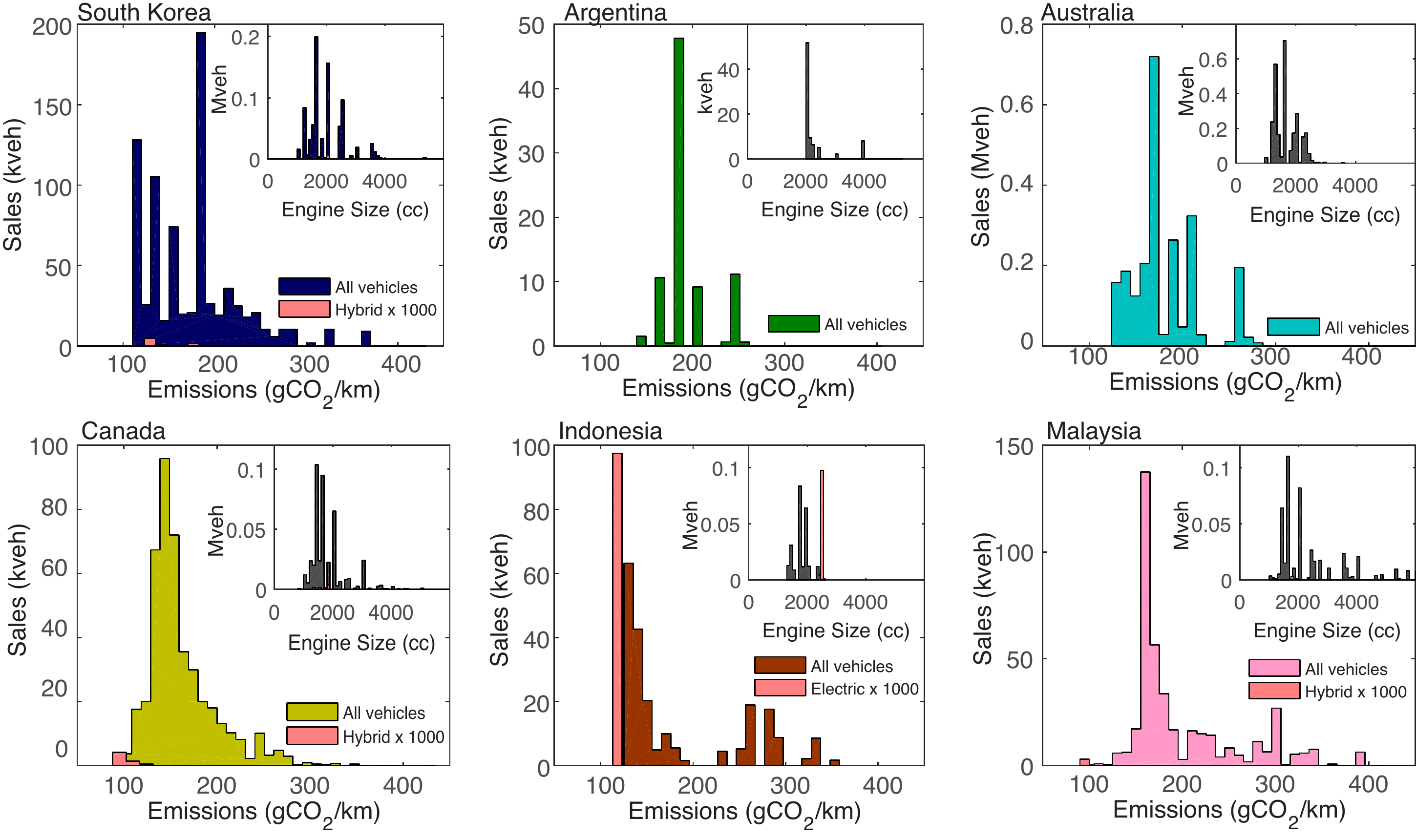}
	      
	\end{center}
	\caption[Price and emissions distributions for Mexico, New Zealand, Russia, South Africa, Taiwan and Saudi Arabia.]{Price and emissions distributions for Mexico, New Zealand, Russia, South Africa, Taiwan and Saudi Arabia. Top two rows: price distributions for the six countries with identical price scaling. The price distributions for alternative vehicles are shown in pink. Bottom two rows: emissions distributions (main graphs) and engine size (insets) of 2012 vehicles sales for the six countries. The emissions and engine size distributions are shown in pink.}
	\label{fig:2-4}
\end{figure}

Figures~\ref{fig:2-1}, \ref{fig:2-3}  and~\ref{fig:2-4} shows our vehicle prices, emissions and engine size distributions for 18 countries. Sales of alternative technologies, hybrid and electric vehicles are shown in pink and red respectively, where available, scaled by factors indicated to make them visible, as their numbers are often orders of magnitude lower. We provide average and median prices, emissions and engine sizes with their standard deviation in Table~\ref{table:2-3}. We observed that the USA has the largest average engine sizes, and hence the largest emissions, while India has the smallest average engine sizes, and hence the smallest average emissions out of all the countries. 

The diagrams give an overview of the diversity of markets in different nations, and are different everywhere. \cite{MercureLam2015} shows that in the UK, where our data is most detailed, this distribution is roughly proportional to the income distribution (see section~\ref{sect:DistUK}). Although we don't demonstrate it, this is probably the case in all countries (down to a cutoff value below which people do not purchase new vehicles). We stress that these exclude second-hand markets. For example, the price distribution in Japan is much narrower compared to China and the USA. Similarly, there is a clear difference between engine sizes and fuel economy between countries, implying that it is essential to consider different fuel economy and engine sizes for individual countries.  For instance, in Saudi Arabia, the distribution of engine sizes covers 1000cc to 6000cc, while in Taiwan engine sizes are concentrated between 1500cc to 2800cc. In terms of emissions distributions, vehicles in the US have the highest emissions, while Japan has the lowest emissions as shown in table~\ref{table:2-3}. 

In terms of alternative technologies, price distributions are very different across countries. For instance, the UK has a wide range of hybrid vehicles and Japan has the highest penetration of hybrid vehicles. Notice that in Japan, Australia and Canada, the market for hybrid vehicles and EV is mostly concentrated in the lower engine class segments, while in the developing nations such as Indonesia, there are very few hybrid vehicles and EVs on the road. Some of this data was published by  \cite{MercureLam2015}.

Following the method shown for the UK in section~\ref{sect:DistUK}, these distributions were segmented into three distributions for each engine size class and technology type. For each, a mean and STD is taken in log scaling, since these distributions are highly assymmetric and roughly lognormal. We do not show this here, but the mean and STD values are given in the Supplementary Excel File.

\begin{table} [t]
\caption{The average, median and standard deviations for car prices, engine sizes and emissions in the six representative regions.}
\small
\begin{tabular}{p{2cm}p{1cm}p{1cm}p{1cm}p{1cm}p{1cm}p{1cm}p{1cm}p{1cm}p{1cm}}
\hline    	
Country & \multicolumn{3}{c}{P(USD)}  & \multicolumn{3}{c}{Engine sizes (cc)}  & \multicolumn{3}{c}{Emissions (gCO$_{2}$/km}  \\ \hline
   & Average   &  Median & S.D.     & Average   &  Median & S.D 

  & Average   &  Median & S.D \\
      \hline             
   USA        & 25959  &   23871  &10570  & 3026  &  2550   &1225  &186  &  176   &50     \\
    \hline
     UK       & 34285      &31520   & 18640      & 1576   &   1498  &  544   & 123.3  &   118.8  &30 \\
    \hline
    Japan      & 18317     &  14968  &    11526   &   1286  & 1252  & 728  &   113  & 102   &44\\
     \hline
     China        & 22826   &   18970   &16633  & 1704  &   1596  &481   & 154   &  153   &31 \\
    \hline
     India       & 8674     &6947      &12418   &1220  &   1170   &445   & 140  &   145   &27 \\
    \hline
    Brazil          &20642     &  16425   &13770   & 1527   &   1558   &458 & 112    &   106  &29  \\
     \hline
 
    Korea       & 19949     	&15432             &13799   	 & 1840	&1998       &652	  & 171  &168  &38	\\
    \hline
     Argentina      & 20850	       &18720        &11578	    & 1646   &1598   &350	         &172 	&167 	&50 \\
    \hline
    Australia          & 31948  	&28415         &19029	   & 2284 	 &1998  &863   & 127  &123	&34	 \\
     \hline
    Canada        & 21407 	&20914     &10680 & 2857   &2500  &1285  	 & 210  	&206	    &60  \\
    \hline
     Indonesia       & 11805	  &15654     & 10021	& 1630	&1495    &440 & 171	&176  &43	\\
    \hline
    Malaysia         & 29976   &14970	&39277 &1714	 &1586    &628& 183  & 181     &47\\
    \hline
    
    Mexico       & 16630  	&13138  &9253 &2002    &1800   &779     & 168		&171   & 57  \\
    \hline
     New Zealand      & 52481  	&29611 &72191& 2280     	&2000   &592    &193 	&186        &26      	 \\
    \hline
    Russia          & 23560    &23325 	&10690 	  & 1656  	&385   &1600     	&188		&190   &54 \\
     \hline
    South Africa        & 19976	&1598	 &601  & 1869   &1598	&601  &173   &161	&84  \\
    \hline
     Taiwan       & 22922   &21318      & 3250  & 1818   &   1798   &247 & 146  &   160   &64 \\
    \hline
    Saudi Arabia       & 27027	&22086 	&21600   & 2378 	&1180 	&2000  &200  &   174   &  59 \\
 
   \hline
\label{table:2-3}
\end{tabular}
\end{table}

\subsection{Construction of historical time-series \label{sect:HistoricalData}}

\subsubsection{Market shares time-series}

To derive the $\gamma_i$ values, one requires historical market shares for different vehicle technologies. To do this, we have either obtained the shares directly from Eurostat or derived the shares using historical sales of new vehicles and their survival rates. The market shares for petrol and diesel vehicles of different engine sizes for the EU countries are available from the Eurostat website \footnote{http://ec.europa.eu/eurostat/statistics-explained/index.php/} from 2002 onwards. 

Outside the EU countries, market shares are not readily available. In particular, it is challenging to find market shares data for engine sizes that match the Eurostat engine size classification for petrol and diesel vehicles. The shares for these vehicles were calculated based on historical car sales from Marklines convolved with the survival function with the following equation:

\beq
N(t) = \sum_a \xi(t-a) \ell(a),
\eeq
where $N(t)$ is the number of vehicles, $\xi(t)$ is the number of sales at year $t$, and $\ell(a)$ is the survival ratio of vehicles at age $a$. The survival function gives the fraction of vehicles that survive up to a certain age. It is typically represented as a monotonically decreasing function that declines from 1 to 0 as age increases. For example, in \cite{zachariadis1995dynamic} the survival rates were simulated using a Weibull distribution,
\beq
f(x) = e^{-(\frac{x+b}{T})^b},
\eeq
where $T$ parameterises the vehicle lifetime and $b$ is the parameter that affects the shape of the survival function. Here, we did not parameterise a function but instead did the calculation directly with data (see Fig.~\ref{fig:SurvivalFn}).

We have purchased annual vehicle sales data from the Marklines website in the year 2013 and 2014, which is an automotive industry portal that consists of motor vehicles market data. Marklines provides for the total sales by model and brand for 63 countries from 2004 onwards. Hence, it is possible to know the sales for each model name for individual countries. Marklines sales numbers were checked for reliability against total sales given by a number of data sources (including official data published by the transport departments of various nations). We concluded that the total sales number in the Marklines dataset is consistent with the official data. This implies that the Marklines data cover all the models available on sale for each country, making our historical shares data reliable. We obtained in this way shares between 2004 and 2012, sufficient to determine $\gamma_i$ values wherever shares are not zero. Since we cover every individual vehicle model, our data is very accurate even when numbers are very small. For example, orders of magnitude differences in sales are observed between petrol and EV vehicles, but that does not make $\gamma_i$ values necessarily less accurate for EVs, unless numbers are very small or the non-zero time series is shorter than 5 years.

\subsubsection{Survival function} 

\begin{figure}[p]
		\begin{center}
			\includegraphics[width=1\columnwidth]{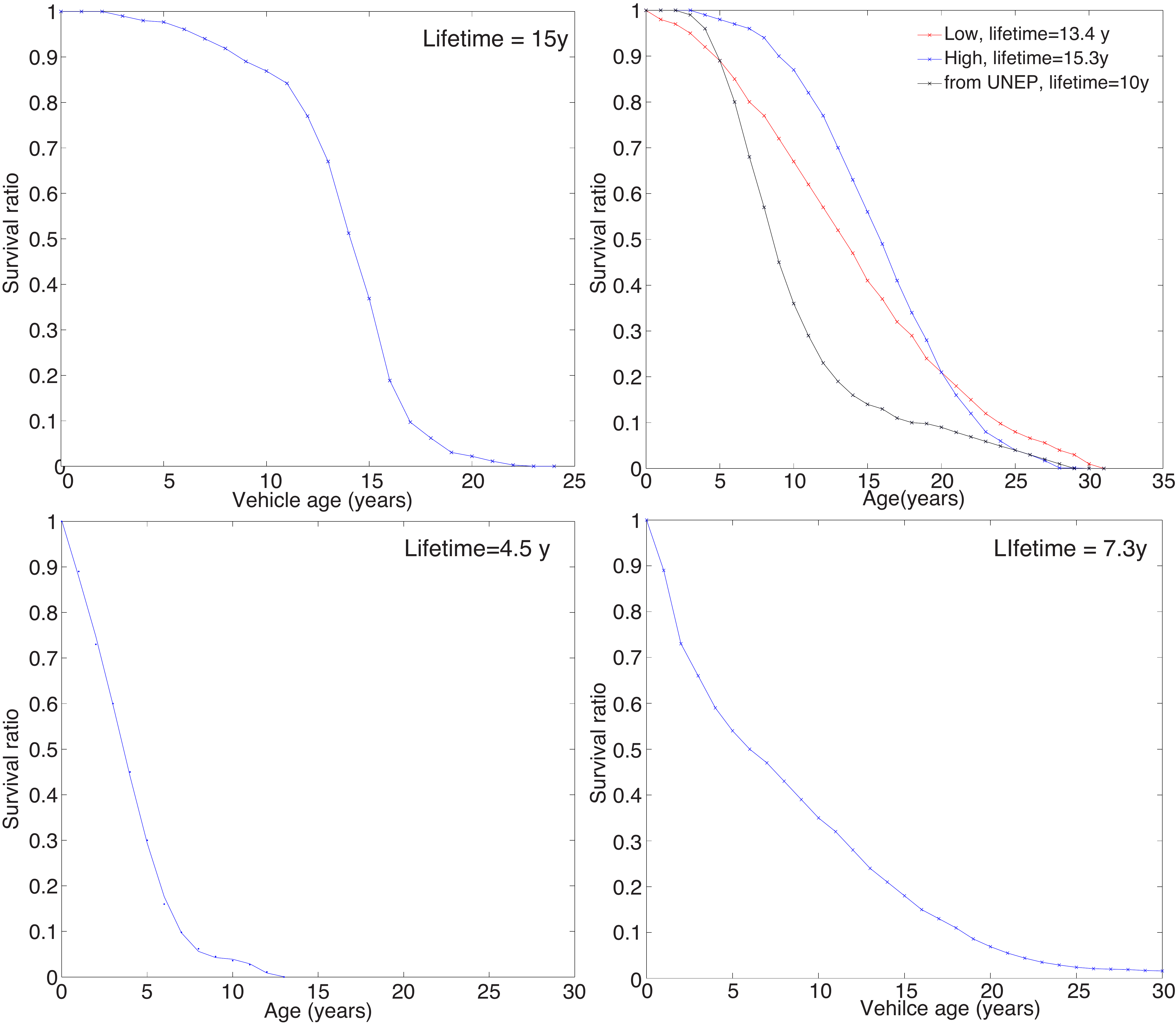}
		\end{center}
	\caption[Survival function for vehicles in China.]{\emph{Top Left} Survival function for vehicles in China. Data from \cite{hao2011} \emph{Top Right} Survival functions for vehicles in India. Data from \cite{goel2013} \emph{Bottom Left} Survival functions for motorcycles, adjusted for China. \emph{Bottom right} UK survival function for motorcycles derived from the \cite{DVLASurvey}. }
	\label{fig:SurvivalFn}
\end{figure}

The determination of vehicle survival rates requires substantial historical information on the stocks and scrappage. Three approaches could be used to find the survival function. Firstly, survival functions for some countries (e.g. China, Japan) have been found directly from existing literature (figure~\ref{fig:SurvivalFn}, see \cite{hao2011,goel2013}). Secondly, when the survival function is not readily available, we could derive a survival function by generating a survival profile based on the survival function derived for the UK (using data obtained from \cite{DVLASurvey}). This is based on the assumption that the reliability functions of mechanical systems for vehicles are similar between countries for the first few years of their lifetime. This is because reliability of vehicles is not necessarily related to political borders, since most firms sell internationally. The difference in survival function between countries is related to weather and traffic context. This assumption is largely consistent with existing empirical evidence \citep{huo2012}. Then, we constantly adjusted the survival values until the difference between total sales and total stocks (available from \cite{Euromonitor} for many countries) became approximately equal (figure~\ref{fig:SurvivalFn}, bottom left). 

When neither total sales nor stocks were available (i.e. Indian motorcycles), we have to borrow survival functions derived from other countries. The variation in survival patterns of vehicles between countries can be attributed mainly to the difference in scrappage policies in different countries, vehicle management and improved technologies. For instance, China has mandatory scrappage standards for vehicles and motorcycles (e.g. 13 years for motorcycles). The scrappage policy causes a shorter average life expectancy for Chinese motorcycles than those in the UK (figure~\ref{fig:SurvivalFn}). Survival patterns should be more similar between countries sharing similar scrappage policies. For motorcycles in India, as for those in the UK, there is no scrappage scheme. While it could be argued that technology is more advanced in the UK than in India, vehicle makers are largely multinational, so technology spillovers almost certainly occur. Thus, the UK motorcycle survival patterns should give a fairly good approximation for Indian motorcycles (see figure~\ref{fig:SurvivalFn}, bottom right). A sensitivity analysis will be carried out to examine the effect of the uncertainties introduced by survival function approximation. 

\subsection{General parameters \label{sect:General}}

As shown in tables~\ref{table:4-5}, ~\ref{table:4-6} and~\ref{table:4-7}, car prices, engine sizes and fuel economy data for each car model listed in Marklines are collected from various sources, including car manufacturers, car sales websites, car industry market reports and government institutions, matched to the car models listed in the Marklines data. Note that the prices obtained are the listed price in the year 2013 when the data were collected. Car fuel economy data were collected from the manufacturers' websites when available. In some cases (such as Taiwan), it was much faster to obtain the car specification and prices from one single car research website where these data were readily available. To ensure the reliability of the data outside the manufacturer's website, we checked the price, engine sizes and fuel economy data from these car sales, research websites and government institutions against the data obtained from the manufacturers. 

In cases where the fuel economy data were not available (e.g. Argentina), we used other countries as a proxy for the fuel economy data in similar markets. This is a reasonable assumption since similar car models are shipped/supplied to a number of Latin American countries. We also borrowed the price data across similar countries in cases where the price data were not available. For instance, in the case of South Korea, the prices for certain car models cannot be obtained from the official manufacturer's website in South Korea, and hence, we borrowed the Japanese car prices. Since we have matched the sales of the outdated models where data are not available readily from the car manufacturers, it is possible to look up the specification from the second-hand market without obtaining the car prices. Note that we have taken the car prices before the addition of any taxation. 

In many cases, each car model has several car price and fuel economy values, depending on vehicle options for a car model. We usually take the mid-value for prices and engine sizes, unless it is known to us that a particular vehicle option/alternative is very popular. The data sources are listed in ~\ref{table:4-5},table~\ref{table:4-6} and table~\ref{table:4-7}. 

Table~\ref{table:4-8} shows the full E3ME-FTT classification. For countries outside the 18 countries above, table~\ref{table:4-9} shows the country we have used as a proxy. For instance, we have used the price, engine sizes and emissions data collected for the UK as a proxy for all European countries. Our approach is based on the fact that it is the relative prices for vehicles within a country that matters for consumers when they are choosing vehicles, instead of the absolute price for vehicles \footnote{We note that for this study it was more important to focus on developing countries rather than on Europe for a global model. We plan to improve the data resolution for Europe in the future.}. Although there could be price diversity between countries, EU countries are regulated by the same emissions standards. While in Latin America, we have taken the Brazilian data for the Latin American countries such as Colombia. This is justified by the fact that similar vehicles in different countries in Latin America have been selling vehicles produced by the same manufacturers. In the case of countries defined as a group (e.g. ASEAN, OPEC), we have picked a major country as representative for the organisation. As shown in table~\ref{table:4-9}, Malaysia is taken as a representative country for ASEAN and Saudi as a representative country for OPEC. Part of the data (the data for US, UK, Japan, China and India) is published in the \cite{MercureLam2015}, with the data collected by the authors. 

\begin{table}[p]
    \caption{Summary of data sources}
	\small
	\begin{tabular}{p{1cm}p{5cm}p{5cm}p{5cm}}
	\hline\hline
         Country  &                 Car price            & Engine size     &fuel economy                \\
     
         \\
         UK     & Car prices are collected from http://www.carpages.co.uk/ for both new models and outdated models. &The car engine sizes are collected from http://www.carpages.co.uk/, along with car prices and fuel economy. &The fuel economy for vehicles is collected from http://www.carpages.co.uk/, along with car prices and car engine sizes. \\
            \\
          USA &  Official websites of car manufacturers in the US for the existing models. For the old/outdated models, the price data were obtained from car dealers such as http://www.autonews.com/section/prices&Official websites of car manufacturers in the US. For the outdated/old models, engine size data were obtained from car dealer such as http://www.autonews.com/section/prices & Official websites of car manufacturers in the US. For the outdated/old models, fuel economy data were excluded from the calculation.
          \\
          Japan    & Official websites of car manufacturers in Japan for the existing models.  The price data for vehicles sold historically were obtained from http://toyota.jp/service/dealer/ &Official websites of car manufacturers in Japan for the existing models. The engine size data for vehicles sold historically were obtained from http://toyota.jp/service/dealer/spt , along with the price data and the fuel economy. & Official websites of car manufacturers in Japan for the existing models.  The engine fuel economy data for vehicles sold historically were obtained from http://toyota.jp/service/, along with the price data and engine size data.\\
          \\
           Canada &Official websites of car manufacturers in Canada for the existing models. For the missing/outdated models, we have used prices from the USA.  &Official websites of car manufacturers in Canada for the existing models. For the missing/outdated models, we have used fuel economy from the USA.     &   Official websites of car manufacturers in Canada for the existing models. For the outdated/old models, fuel economy data were excluded from the calculation.           \\
           \\
           Australia &Official websites of car manufacturers in Australia for the existing car model prices. For old models, we have obtained prices data from http://www.redbook.com.au/&Official websites of car manufacturers in Australia for the existing car model engine sizes.  For old models, engine sizes data can be obtained from  http://www.redbook.com.au/, along with the price data and fuel economy data& official websites of car manufacturers in Japan for the existing models.  The fuel economy data for vehicles sold historically were obtained from http://www.redbook.com.au/, along with the price data and engine size data.\\
           \\
           New Zealand &We have used car prices in Australia for New Zealand & We have used engine sizes obtained for Australia in New Zealand. &We have used fuel economy obtained for Australia in New Zealand.\\
           \\
            Russia &Official websites of car manufacturers in Russia for the existing models.  The prices data for vehicles sold historically/outdated vehicles were obtained from http://auto.mail.ru/ and http://www.avtomarket.ru/&Official websites of car manufacturers in Russia for the existing models.  The engine size data for vehicles sold historically were obtained from http://toyota.jp/service/dealer/spt/search-addr, along with the price data and the fuel economy.&Official websites of car manufacturers in Russia. For the outdated/old models, fuel economy data were excluded from the calculation.\\
           \\
           China & Car prices data (both new vehicles and old models) were obtained from commercial dealer websites, such as China Auto Home (http://www.autohome.com.cn/) and Sohu Auto (http://auto.sohu.com/). &  Engine sizes data were collected from http://www.autohome.com.cn/ and http://auto.sohu.com/, along with price and fuel economy data. &Fuel economy data were collected from http://www.autohome.com.cn/ and http://auto.sohu.com/, along with price and engine sizes data.  \\

                   \hline      
		  \end{tabular}
		\label{table:4-5}
\end{table}

\begin{table}[p]
    \caption{Summary of data sources continued}
	\small

	\begin{tabular}{p{1cm}p{5cm}p{5cm}p{5cm}}
	\hline\hline
         Country  &                 Car price            & Engine size     &fuel economy                \\
         
           \\
           India& Car prices data have been obtained from the official car manufacturers$'$ website for India. For old models, we have obtained prices data from http://www.carwale.com/new/ and http://www.zigwheels.com/newcars &Engine sizes data were obtained from the official car manufacturers$'$ website for India, alongside price data. Similarly, for old models, engine sizes data were obtained from http://www.carwale.com/new/ and http://www.zigwheels.com/newcars&   If available, fuel economy data were obtained from the official manufacturers$'$ website for India. For some new models (where fuel economy data are not available from the manufacturers) and outdated models, fuel economy data were obtained from http://www.carwale.com/new/ and http://www.zigwheels.com/newcars\\
           \\
           Mexico&     Car prices data have been obtained from the official car manufacturers$'$ website for Indonesia.  For old or unavailable models, we have not been able to obtain data so they are not considered in the calculation. &  Car engine sizes data were obtained from the official car manufacturers$'$ website for Mexico. For old or unavailable models, we have used engine sizes obtained for other Latin America countries (e.g. Brazil) if available. &Fuel economy data were obtained from the official car manufacturers$'$ website for Mexico. For old or unavailable models, we have used fuel economy obtained for other North America countries (e.g. USA) if available.   \\
           \\
           Brazil & Car prices data have been obtained from the official car manufacturers$'$ website for Brazil. For old models, we have obtained prices data from http://www.icarros.com.br/catalogo/ & Car engine sizes data were obtained from the official car manufacturers$'$ website for Brazil.  For old models, we have obtained the engine sizes data from http://www.icarros.com.br/catalogo/, along with car prices data. &Car fuel economy data were obtained from the official car manufacturers$'$ website for Brazil. For old or unavailable models, we have used fuel economy obtained for other North America countries (e.g. USA) if available.   \\
           \\
           Argentina& Car prices data have been obtained from the official car manufacturers$'$ website for Argentina.  For old models, we have obtained prices data from http://www.cars.com.ar/& Car engine sizes data were obtained from the official car manufacturers$'$ website for Argentina. For old models, we have obtained the engine sizes data from http://www.cars.com.ar/, along with car prices data.&Car fuel economy data were obtained from the official car manufacturers$'$ website for Brazil. For old or unavailable models, we have used fuel economy obtained for other North America countries (e.g. USA) if available.\\
           \\
            South Korea & Car price data have been obtained from the official car manufacturers$'$ website for South Korea. For old or unavailable models, we have taken the prices from Japan. & Car engine size data were obtained from the official car manufacturers$'$ website for Indonesia. For old or unavailable models, we have taken engine data from Japan. & Car fuel economy data have been obtained from the official car manufacturers$'$ website for South Korea.  For old or unavailable models, we have taken data from Japan.\\
            \\
            Taiwan & Car prices data have been obtained from the official car manufacturers$'$ website for Taiwan. For old models, we have obtained prices data from https://tw.autos.yahoo.com/car-research&Fuel economy data were collected from https://tw.autos.yahoo.com/car-research, along with price and engine sizes data.\\
            \\

                                                         \hline      
		  \end{tabular}
		\label{table:4-6}
\end{table}

\begin{table}[p]
    \caption{Summary of data sources continued}
	\small

	\begin{tabular}{p{1cm}p{5cm}p{5cm}p{5cm}}
	\hline\hline
         Country  &                 Car price            & Engine size     &fuel economy                \\
         
                     \\
            Indonesia & Car prices data have been obtained from the official car manufacturers$'$ website for Indonesia. For old or unavailable models, we have not been able to obtain data so they are not considered in the calculation&   Car engine size data have been obtained from the official car manufacturers website for Indonesia.  For old or unavailable models, we have not obtained data so they are not considered in the calculation. &   Car fuel economy data were obtained from the official car manufacturers website for Indonesia. For old or unavailable models, we have not obtained data so they are not considered in the calculation. \\
            \\
             Malaysia &     Car prices data have been obtained from the official car manufacturers$'$ website for Malaysia and Singapore. For old models, we have obtained prices data from http://www.sgcarmart.com/newcars/ &Car engine sizes data were obtained from the official car manufacturers$'$ website for Malaysia and Singapore. For old models, we have obtained the engine sizes data from http://www.sgcarmart.com/new\_cars/ &        Car fuel economy data have been obtained from the official car manufacturers$'$ website for Malaysia and Singapore. For old or unavailable models, we have not obtained data so they are not considered in the calculation.\\
             \\
             Saudi Arabia &Car prices data have been obtained from the official car manufacturers$'$ website for Saudi Arabia. For old models, we have obtained prices data from http://saudi.dubizzle.com/en/home/,  http://www.carsemsar.com/en/saudi-arabia and http://www.drivearabia.com/  & Car engine sizes data were obtained from the official car manufacturers$'$ website for Saudi Arabia. For old models, we have obtained the engine sizes data from http://www.drivearabia.com/ , along with car prices data.& Car fuel economy data were obtained from the official car manufacturers$'$ website for Malaysia and Singapore. For old or unavailable models, we have not obtained data so they are not considered in the calculation.\\
             \\
             South Africa &Car prices data have been obtained from the official car manufacturers$'$ website for South Africa. For old models, we have obtained prices data from http://www.cars.co.za/. &Car engine sizes data were obtained from the official car manufacturers$'$ website for Saudi Arabia. For old models, we have obtained the engine sizes data from http://www.cars.co.za/. , along with car prices data. &  Fuel economy data were obtained from the official manufacturers$'$ website for South Africa.  For some new models (where fuel economy data is not available from the manufacturers) and outdated models, fuel economy data have been obtained from http://www.cars.co.za/.\\
             \\
                        
                                                         \hline      
		  \end{tabular}
		\label{table:4-7}
\end{table}

\begin{table}[p]
    \caption{List of E3ME world regions}
	\small
	\begin{tabular}{ l l l l l l }
	\hline\hline
        E3ME region			&Countries		&E3ME region	&Countries	&E3ME region	&Countries	\\
        \hline
        1-28: EU-28 			&EU-28 			&39: Russian Fed. 		&Russian Fed.				&50: Indonesia		&Indonesia						\\
        29: Switzerland			&Switzerland		&40: Rest Annex I		&Belarus					&51: ASEAN		&Thailand, Cambodia,  			\\
        						&				&					&						&				&Lao, Malaysia,  				\\
						&				&					&						&				&Myanmar, Philippines,		\\
						&				&					&						&				&Singapore, Vietnam		\\
        30: Iceland			&Iceland			&41: China			&China					&52: OPEC		&Iran, Iraq, Kuwait, 		\\
						&				&					&						&				&Qatar, UAE			\\
        31: Croatia			&Croatia			&42: India				&India					&53: Rest of World	&All other countries					\\
        32: Turkey				&Turkey			&43: 	Mexico			&Mexico					&54: Ukraine*		&Ukraine							\\
        33: Macedonia*			&Macedonia		&44: Brazil			&Brazil					&55: Saudi Arabia*	&Saudi Arabia						\\
        34: USA				&USA			&45: Argentina*			&Argentina				&56:	Nigeria*		&Nigeria							\\
        35: Japan				&Japan			&46: Colombia			&Colombia				&57:	South Africa*	&South Africa						\\
        36: Canada			&Canada			&47: Rest of 	 		& Bolivia, Chile, 			&58:	Rest of Africa*	&Rest of Africa			\\
        						&				&	Latin America		&Ecuador, Peru, 	  		&59:	Africa OPEC*	&Algeria, Angola, 					\\
						&				&					&Central America 			&				&Libya							\\
        37: Australia			&Australia			&48: South Korea		&South Korea				&				&								\\
        38: New Zealand		&New Zealand		&49: Taiwan*			&Taiwan					&				&								\\
        	\hline
	Notes:	&\multicolumn{5}{ l }{\footnotesize Note: stars denote countries not included in FTT:Transport, where fuel used is assumed proportional to the } \\
			&\multicolumn{5}{ l }{\footnotesize World average. In these countries, either some data is missing, or it has to do with that the E3ME classification } \\
			&\multicolumn{5}{ l }{\footnotesize recently changed from 53 to 59 countries and FTT:Transport has not yet been updated accordingly. Thus, the } \\
			&\multicolumn{5}{ l }{\footnotesize FTT:Transport classification is bound to change in the near future. Some regions, however, may never be modelled, }\\
			&\multicolumn{5}{ l }{\footnotesize due to lack of data.}\\
	\end{tabular}
	\label{table:4-8}

    \caption{Parameterisation of the model outside the 18 nations where direct data were available.}
	\small
	\begin{tabular}{p{5cm}p{10cm}}
	\hline\hline
         Country  &                 Proxy country              \\
        
         \\
         EU 28, Norway, Switzerland, Iceland, Croatia, Turkey and Macedonia    &  UK  \\
            \\
          Rest of Annex 1                                   &   We have taken Ukraine as the representative country in Rest of Annex 1.We have used car prices in Russia for Ukraine.  
          \\           
          \\
          Colombia  & We have used car prices collected for Brazil to represent prices in Colombia.  \\
          \\
           Rest of Latin America  & We have taken Brazil as the representative country in Rest of Latin America.    \\
           \\
          ASEAN &  We have taken Malaysia as the representative country in ASEAN. \\
           \\
           OPEC  & We have taken Saudi Arabia as the representative country in OPEC. \\
           \\
            Rest of world  &  We have taken South Africa as the representative country in this category. \\
           \\
                           
                   \hline      
		  \end{tabular}
		\label{table:4-9}
\end{table}

\subsection{Econometrics for transport demand and vehicle sales \label{sect:econometrics}}

Transport demand depends on the needs of the economy and preferences of people in society. Passenger transport demand can be measured in different ways, including the number of trips, vehicle kilometres and passenger kilometres. Two factors are particularly important in projecting emissions, namely the number of vehicles registered to be on the road and the number of passenger kilometres, which measure the quantity and volume of travel required by future passenger vehicles.

The future demand for transport is related to economic activity (GDP), fuel costs and demand elasticities. In the present model, by regressing the vehicle travel demand (pkm) with respect to income and fuel prices, the model predicts the vehicle travel demand up to 2050. Notice that the limitation for the regression is that it omits variables that are potentially useful in predicting vehicle travel demand, such as the public transport network, congestion, geographic area etc., due to the difficulty of finding sufficient data for 59 regions over a 20-year period. However, this would not be picked-up substantially well in the econometric specifications since many of these omitted parameters are not likely to change rapidly over time, while the econometric equations model changes. Equation \ref{eqPM} and equation \ref{eqVS} show the demand for passenger kilometres and the number of new vehicle sales, used by the E3ME-FTT model: 
\beq
PM_j = \alpha_{1,j}*RRPD_j + \alpha_{2,j} * PFRM_j + \epsilon_j
\label{eqPM}
\eeq
and
\beq
VS_j =  \beta_{1,j} *RRPD_j + \beta_{2,j}*PV_j + \beta_{3,j}*ST_j + \epsilon_j
\label{eqVS}
\eeq
where $PM$ is the demand for transport in passenger kilometres (pkm), $RRPD$ is the real income, $PFRM$ is the price of middle distillates for road transport, $VS$ is the number of new vehicle sales and $PV$ is the real price of vehicles and $ST$ is the number of vehicles on the road, all specified for each region $j$.

In the FTT:Transport integration to E3ME, $RRPD$ and $PFRM$ are modelled endogenously, while $ST$ is obtained from $VS$ by integrating with the survival function. Thus E3ME supplies endogenous values of $PM$ and $VS$ to FTT:Transport, which determines fleet compositions, and feeds $PV$ and $ST$ back to E3ME. FTT:Transport furthermore feeds back to E3ME its estimation of fuel use for road transport. It is to be noted that in this formulation of feedbacks, changes in the use of liquid fossil fuels from FTT:Transport has large impacts on economic activity in oil producing countries.

For the projections of future car sales, we have obtained the number of car fleets (numbers of vehicles registered) for the global 59 countries (1990-2012) from the Euromonitor website.  

\subsection{Determining the non-pecuniary $\gamma$ values in practice \label{sect:gammapractice}}

\begin{figure}[p]
		\begin{center}
			\includegraphics[width=1\columnwidth]{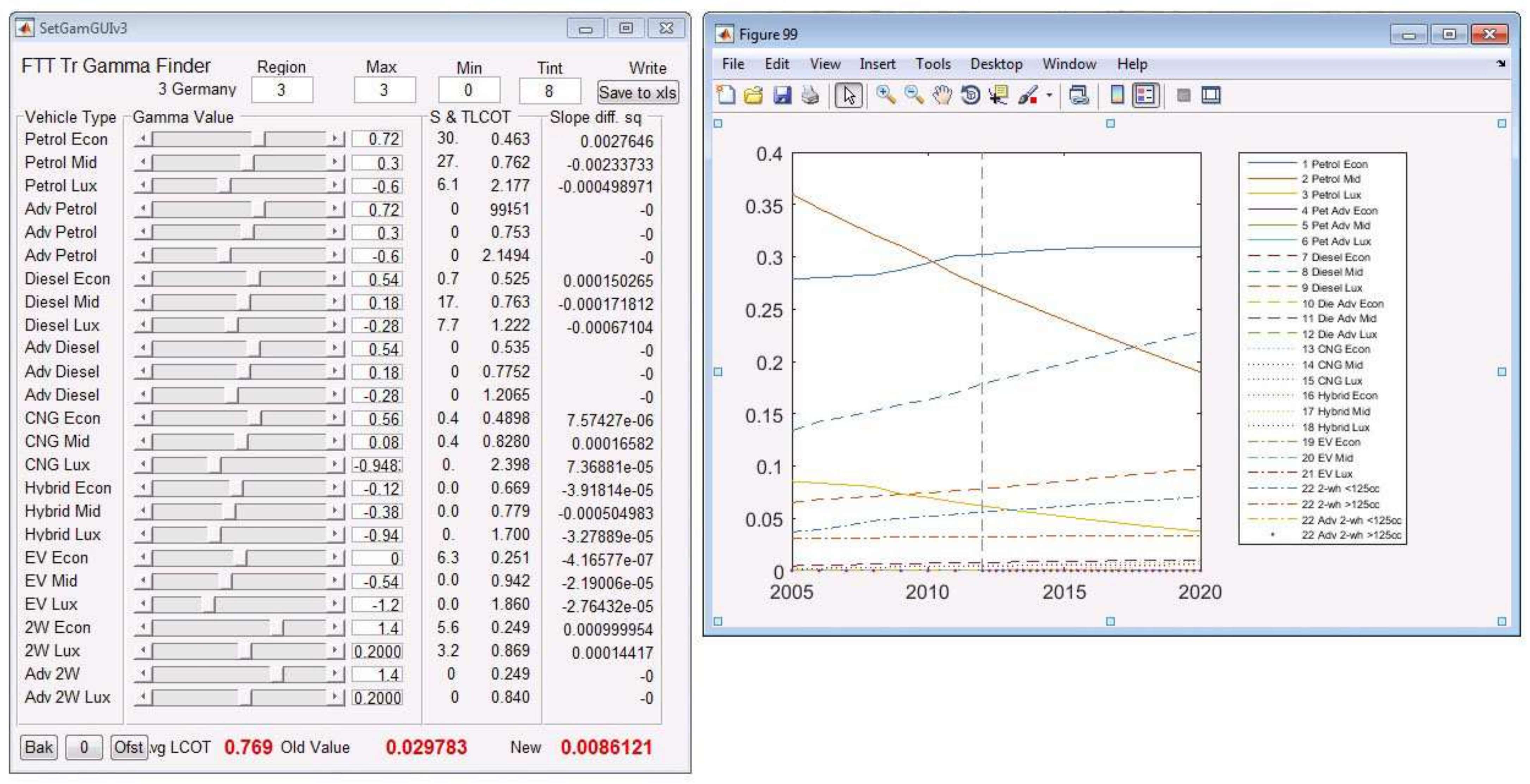}
		\end{center}
		\begin{center}
			\includegraphics[width=.85\columnwidth]{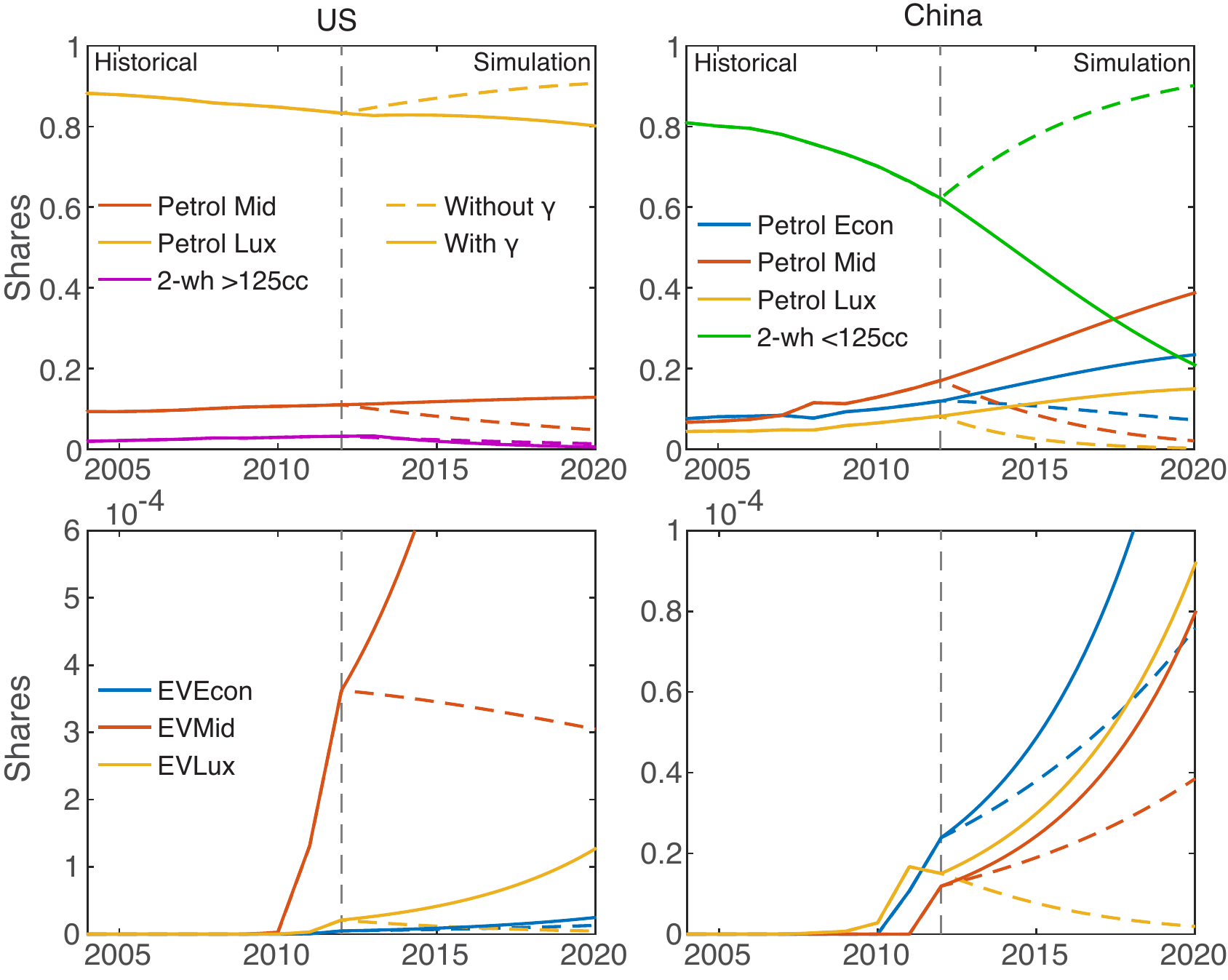}
		\end{center}
	\caption{Graphical user interface used to determine $\gamma$ parameters. Sliders or value inputs are used in order to change the diffusion trajectory of the model (to the right of the dashed line), until it is consistent with historical data (to the left of the dashed line). This is done for every technology in every country, meaning that 59$\times$25 = 1475 parameters are adjusted. The fact that this is done by hand ensures that every single diffusion profile has been inspected. Technologies with lower shares are parameterised in the same way by zooming in.}
	\label{fig:SetGamGuiv3}
\end{figure}

When $\gamma_i = 0$, we obtain a rate of diffusion that does not normally match historical diffusion (see Fig. 4 of the main paper, reproduced here in fig.\ref{fig:SetGamGuiv3}). One, and only one, set of $\gamma_i$ leads to the diffusion of technology in the simulation to have the same rate as the historical rate at the starting point of the simulation. Since decisions are based on cost differences, the number of $\gamma_i$ parameters equals that of technologies minus one (or one of the $\gamma_i$ equals zero). As found by experience, the unique set of $\gamma_i$ cannot be obtained by simple optimisation, as too many spurious local solutions arise. We thus designed a dedicated graphical user interface software that enables to robustly determine these parameters by hand. This is done for each technology in every region, thus a time-consuming procedure, but visual inspection ensures that the parameters are not spurious.\footnote{Determining the set takes approximately one hour.} We find that $\gamma_i$ values follow what is expected: luxury models have large negative values (large benefits). We typically keep the value for the small economic petrol vehicles near zero; however any arbitrary constant can be added to all $\gamma_i$ without consequence to the results. This is a reflection that FTT represent flows of shares according to relative generalised cost differences between categories. Note that since generalised cost differences already exist in the baseline, \emph{diffusion trends exist in the baseline}, a fact that is observed in the data, and the determination of the $\gamma_i$ parameters is of primary importance. Although it does not provide significant information on the non-pecuniary benefits themselves, it is in this way a robust methodology. 

Figure \ref{fig:SetGamGuiv3} shows an example of visual adjustment of the $\gamma_i$ parameters for the UK, with the graphical user interface used. Note that it is critically important to obtain good values for technology categories with large share values. Meanwhile, it is sometimes \emph{difficult} or \emph{ambiguous} to determine $\gamma_i$ values for some new technologies, when they have small and/or noisy historical data, and sometimes no data is available at all. However, the $\gamma_i$ values are \emph{strongly tied to vehicle class categories} (Econ, Mid, Lux), as we observe across countries, but less so to engine technologies types (Petrol, Diesel, EVs). However, they are not particularly tied to particular countries. For example, a luxury diesel vehicle is desirable in a roughly similar way to a luxury petrol vehicle, but not to an economic petrol vehicle. The relative difference in $\gamma_i$ between technology categories is always much smaller than across engine size classes. It is thus feasible to use proxies for missing values, either across similar technology types (Petrol and Diesel) or across countries for the same technologies (e.g. EV Econ). Thus in instances where $\gamma_i$ values could not be obtained using the visual interface, for instance if they have zero shares, they were inferred from other technology categories within the same engine size class. This is because it is not realistic to assume that these technologies would have zero shares forever, and indeed, we use in some cases kick-start policies for these categories to start diffusing (e.g. EVs in India).

Finally, it is to be noted that changing one $\gamma_i$ value in a set for one region requires to re-determine all the others, as it changes the relative value of all technologies. Furthermore, if the definition of the LCOT is changed for any reason (e.g. adding a pecuniary parameter, or changing the discount rate), the empirical $\gamma_i$ must all be re-determined since their meaning also changes. In this sense the $\gamma_i$ contain everything of relevance that is not explicitly represented in the LCOT; the more parameters are included in the LCOT, the less are implicitly represented in the $\gamma_i$. It takes a few hours to determine all $\gamma$ values.

\section{Sensitivity analyses and model validity \label{sect:Sens}}

In this section, we carry out a sensitivity analysis over most relevant technological parameters of FTT:Transport. We say \emph{most} since we could, in principle, produce pages and pages of tables of numbers for additional different sensitivities, but we consider that this would increase substantially the amount of insight in comparison to what is given here. We chose the parameters that we expected would generate the most changes in emissions and technological shares. For example, we put emphasis on EVs since these have the most impact on transport emissions. However, the technological trajectory itself is important, and therefore we provide both changes on emissions and changes on shares of technologies. All sensitivities are carried out globally, meaning that the parameters are changed in the same way in every country. It is clear that in each country, these changes have a different impact (e.g. changing EV costs in countries with few EVs has low impact), however, we do not consider sufficiently instructive to generate this data for every country.

Results are given in table~ref{table:models}. Numbers shown refer to percent changes in a scenario in which a parameter variation is imposed, against the corresponding scenario without the variation. Changes in shares are in 2050, while changes in emissions are cumulated to 2050. Variations are in percent of the cost if applied to cost values (e.g. 20\%), or in percentage points if applied to a rate (e.g. learning and consumer discount rates). We find that results are not highly sensitive to any particular parameter. 

The highest changes we see are the changes that arise for changes in the $\gamma$ values. In particular, if we amplify the differences between the $\gamma$ (first and second rows, $\gamma \pm 20\%$), we see re-allocations of shares between categories, especially across engine size classes, of at most 30\%. If we change $\gamma$ of vehicle classes, we also see substantial re-allocations of shares across engine size classes. Importantly, however, changing the $\gamma$ of EVs does not produce large changes in the allocation of shares. This is because shares of EVs are low to start with, so changing their parameters has a comparatively low influence on the overall trajectory, until EVs take a substantial market share in around 2050. It is to be noted that changing values of $\gamma$ implies diffusion trajectories that are inconsistent with historical data for all vehicle types (i.e. broken at the start year of the simulation), and therefore can in themselves be quite unrealistic. For example, changing the $\gamma$ of mid-range petrol vehicles implies changes in diffusion trajectory of all other vehicle types, larger for larger share categories. $\gamma$ values for low-share low-carbon vehicles are more uncertain than those for large-share categories, and correspondingly, have less impact on the overall trajectory (i.e. larger changes are needed).

We provide in table~ref{table:models} combinations of uncertainties across all variations carried out. We do not know what the uncertainty over these parameters is in reality. However, experience with data tells us that these variations are reasonable. The interpretation of this is therefore that if the real uncertainty over these parameters matched what has been used here, and that all of these were introduced simultaneously (e.g. in a Monte-Carlo analysis), we would obtain the variations given in the rows labelled `Root sum square', where the root of the sum of the squares was calculated. In the bottom row of the table, combined uncertainty is given for a scenario comparison between the baseline and the 2$^\circ$C scenario. We find that at most 50\% uncertainty is generated (in this case, for EV shares in 2050). This may seem relatively large; however this corresponds only to 30\% change in transport emissions, and thus does not have a major impact on any climate scenario studied. 

\newpage
\subsection{Sensitivity of technological parameters}

\begin{table}[p] 
	\small
		\begin{tabular}{ l l r r r r r r r r r}
			\hline
					& Parameter				&	\multicolumn{9}{ l }{\% change in CO$_2$ and technology shares over the same scenario without changes}\\
					&						& CO$_2$	& Econ	& Mid	& Lux	& Hybrid	& CNG	& EV		& ADV	& FF \\
			\hline
			\hline
\parbox[t]{2mm}{\multirow{27}{*}{\rotatebox[origin=c]{90}{\bf 2$^\circ$C Scenario}}}
					&1- All $\gamma$ +20\%			&2.31	& -11.58	& 6.86	& 12.04	& 17.47	& -1.75	& -2.35	& 1.07	& 9.30	\\
					&2- All $\gamma$ -20\%			&-0.05	& 7.47	& -6.14	& -10.30	& -16.09	& -14.44	& 12.12	& -8.00	& -16.16	\\
					&3- Learning rates +5\%			&3.38	& -2.97	& 0.52	& 7.62	& -1.06	& 0.49	& 1.37	& -0.02	& -6.71	\\
					&4- Learning rates -5\%			&-13.05	& -8.22	& 3.09	& -5.80	& 2.53	& -20.24	& 4.53	& -7.98	& 7.14	\\
					&5- Discount rates +10\%			&-16.14	& -13.80	& 10.09	& -5.60	& 14.18	& -28.07	& 12.89	& -12.12	& -3.75	\\
					&6- Discount rates -10\%			&14.30	& 8.35	& -7.17	& 6.76	& -11.26	& 20.55	& -11.79	& 9.89	& 5.34	\\
					&7- EV prices +10\%				&1.67	& -0.06	& 0.84	& -1.93	& 1.76	& 0.88	& -2.46	& 1.13	& 0.09	\\
					&8- EV prices -10\%				&-1.38	& 0.37	& -0.95	& 1.78	& -1.44	& -0.84	& 2.85	& -1.24	& -0.14	\\
					&9- Fuel efficiency +20\%			&7.90	& -0.09	& -1.12	& 2.45	& -5.78	& -0.14	& -0.59	& 0.41	& -4.03	\\
					&10- Fuel efficiency -20\%			&-6.99	& 0.33	& 1.46	& -3.54	& 6.82	& 0.14	& 0.81	& -0.52	& 5.43	\\
					&11- Turnover Rate +50\%		&55.82	& -15.83	& -3.02	& 9.65	& -27.07	& 24.00	& -79.83	& 28.33	& 1304.87	\\
					&12- Turnover Rate +25\%		&37.00	& -4.11	& -4.88	& 0.86	& -10.43	& 30.00	& -57.26	& 23.68	& 393.97	\\
					&13- EV $\gamma$ +10\%		&-0.65	& -1.19	& -0.39	& 3.31	& -0.55	& -0.62	& 0.83	& -0.42	& -0.01	\\
					&14- EV $\gamma$ -10\%			&0.16	& 1.27	& 0.42	& -3.52	& 0.83	& 0.70	& -1.04	& 0.53	& 0.01	\\
					&15- Hybrid $\gamma$ +10\%		&1.66	& -0.88	& 3.13	& -3.26	& 11.90	& -0.17	& -0.80	& 0.57	& -0.06	\\
					&16- Hybrid $\gamma$ -10\%		&-0.98	& 0.82	& -2.92	& 3.12	& -10.43	& 0.12	& 0.57	& -0.41	& 0.06	\\
					&17- CNG $\gamma$ +10\%		&0.34	& -3.25	& 2.52	& 1.04	& 0.25	& -0.44	& 0.09	& -0.49	& 0.14	\\
					&18- CNG $\gamma$ -10\%		&0.38	& 3.13	& -3.04	& 0.10	& -0.22	& 1.17	& -0.14	& 0.50	& -0.12	\\
					&19- ADV $\gamma$ +10\%		&0.13	& -0.88	& -2.32	& 6.33	& -1.57	& -0.30	& -0.25	& 0.22	& -0.51	\\
					&20- ADV $\gamma$ -10\%		&0.40	& 1.32	& 1.20	& -5.14	& 1.26	& 0.44	& 0.15	& -0.13	& 0.46	\\
					&21- 2W $\gamma$ +10\%		&0.75	& 0.51	& -0.02	& -0.52	& -0.16	& 2.26	& -2.03	& 1.22	& 5.58	\\
					&22- 2W $\gamma$ -10\%		&-2.65	& -2.14	& 0.48	& 1.24	& 0.18	& -6.82	& 4.76	& -3.26	& -7.65	\\
					&23- Econ $\gamma$ +10\%		&-1.15	& 2.59	& -3.49	& -0.79	& 1.26	& -2.30	& 0.78	& -0.87	& -0.77	\\
					&24- Econ $\gamma$ -10\%		&1.26	& -4.84	& 6.29	& 0.48	& -1.10	& 2.17	& -0.80	& 0.85	& 0.89	\\
					&25- Mid $\gamma$ +10\%		&-0.54	& -9.31	& 14.89	& -6.32	& 9.05	& 0.37	& -0.52	& 0.41	& 0.03	\\
					&26- Mid $\gamma$ -10\%		&1.32	& 6.56	& -11.68	& 6.57	& -7.85	& -0.17	& 0.30	& -0.26	& -0.03	\\
					&27- Lux $\gamma$ +10\%		&3.38	& -2.09	& -5.83	& 15.41	& -0.30	& 0.19	& -0.22	& 0.27	& 0.71	\\
					&28- Lux $\gamma$ -10\%		&-1.38	& 1.95	& 4.13	& -11.89	& 0.69	& 0.47	& -0.09	& -0.05	& -0.58	\\

			\hline
					&\bf Root sum sq. 1-10		&27.71	& 22.95	& 15.90	& 21.10	& 31.37	& 42.81	& 22.25	& 19.41	& 23.07	\\
			\hline
					&\bf Root sum sq. $\gamma$	&5.52	& 14.15	& 22.41	& 24.00	& 20.06	& 8.05	& 5.59	& 3.93	& 9.61	\\
			\hline
			&Table continued... &&&&&&&&&\\
  		 \end{tabular}
 \end{table}
\begin{table}[p] 
	\small
		\begin{tabular}{ l l r r r r r r r r r}
			\hline
					& Parameter				&	\multicolumn{9}{ l }{\% change in CO$_2$ and technology shares over the same scenario without changes}\\
					&						& CO$_2$	& Econ	& Mid	& Lux	& Hybrid	& CNG	& EV		& ADV	& FF \\
			\hline
\parbox[t]{2mm}{\multirow{27}{*}{\rotatebox[origin=c]{90}{\bf Current Trajectory Scenario}}}
					&1- All $\gamma$ +20\%			&4.34	& -9.96	& 4.74	& 33.76	& 5.44	& -1.07	& -5.36	& 1.40	& -0.19	\\
					&2- All $\gamma$ -20\%			&-4.91	& 11.48	& -14.12	& -24.74	& -4.77	& -3.00	& 14.59	& -2.78	& -1.19	\\
					&3- Learning rates +5\%			&-1.86	& -4.30	& 7.62	& 5.36	& 7.53	& -1.41	& 2.16	& 11.45	& -27.28	\\
					&4- Learning rates -5\%			&6.70	& 3.64	& -6.44	& -3.88	& -6.78	& -0.72	& -9.23	& -7.85	& 22.29	\\
					&5- Discount rates +10\%			&-2.19	& 2.78	& -4.77	& -5.89	& 2.63	& -0.86	& 6.93	& 0.25	& -4.15	\\
					&6- Discount rates -10\%			&2.46	& -4.00	& 7.16	& 8.01	& -2.28	& 1.02	& -8.98	& 0.10	& 4.42	\\
					&7- EV prices +10\% 			&3.13	& 0.84	& 0.38	& -3.18	& 3.94	& 1.09	& -9.67	& 1.66	& 0.99	\\
					&8- EV prices -10\% 				&-6.66	& -9.01	& 14.45	& 6.68	& -9.69	& -1.87	& -46.47	& -1.53	& 26.37	\\
					&9- Fuel efficiency +20\%			&1.86	& 0.07	& 0.11	& -0.32	& -0.36	& -0.02	& -0.03	& 2.33	& -5.31	\\
					&10- Fuel efficiency -20\%			&-1.49	& -0.04	& -0.03	& 0.13	& 0.29	& 0.02	& 0.02	& -1.90	& 4.33	\\
					&11- Turnover Rate +50\%		&30.18	& -21.85	& 26.73	& 25.20	& -45.51	& -17.40	& -88.55	& -15.83	& 79.62	\\
					&12- Turnover Rate +25\%		&16.39	& -9.15	& 14.83	& 5.85	& -21.99	& 2.13	& -70.07	& -0.14	& 34.80	\\
					&13- EV $\gamma$ +10\%		&-1.36	& -0.84	& -0.39	& 3.21	& -1.52	& -0.48	& 4.08	& -0.69	& -0.45	\\
					&14- EV $\gamma$ -10\%			&1.62	& 0.91	& 0.44	& -3.49	& 1.95	& 0.51	& -4.74	& 0.80	& 0.51	\\
					&15- Hybrid $\gamma$ +10\%		&0.16	& -0.49	& 1.87	& -0.90	& 8.62	& -0.25	& -1.30	& 1.34	& -2.41	\\
					&16- Hybrid $\gamma$ -10\%		&-0.03	& 0.29	& -1.26	& 0.70	& -6.82	& 0.13	& 1.17	& -1.07	& 1.88	\\
					&17- CNG $\gamma$ +10\%		&-0.03	& -1.64	& 4.34	& -0.81	& -0.19	& 0.95	& -1.59	& 0.52	& -0.40	\\
					&18- CNG $\gamma$ -10\%		&0.12	& 1.42	& -3.98	& 0.99	& 0.22	& -0.98	& 1.42	& -0.52	& 0.48	\\
					&19- ADV $\gamma$ +10\%		&0.49	& -0.58	& -1.96	& 5.25	& -1.08	& -0.33	& -0.37	& 2.47	& -5.44	\\
					&20- ADV $\gamma$ -10\%		&-0.21	& 0.68	& -0.02	& -2.52	& 0.43	& 0.11	& 0.19	& -1.44	& 3.18	\\
					&21- 2W $\gamma$ +10\%		&1.63	& -0.86	& 3.52	& 2.51	& -1.19	& -0.31	& -3.30	& 0.22	& 1.32	\\
					&22- 2W $\gamma$ -10\%		&-2.52	& 0.75	& -3.27	& -2.21	& 0.80	& -0.36	& 6.02	& -0.69	& -1.58	\\
					&23- Econ $\gamma$ +10\%		&0.93	& 2.94	& -5.84	& -1.29	& 0.55	& 0.29	& -1.75	& 0.25	& 0.29	\\
					&24- Econ $\gamma$ -10\%		&-0.85	& -3.18	& 6.83	& 0.68	& -0.47	& -0.40	& 1.42	& -0.22	& -0.21	\\
					&25- Mid $\gamma$ +10\%		&-0.74	& -6.15	& 19.44	& -5.70	& 4.63	& 0.25	& -0.76	& 0.82	& -1.44	\\
					&26- Mid $\gamma$ -10\%		&1.09	& 4.77	& -15.75	& 5.41	& -3.28	& -0.23	& 0.24	& -0.39	& 0.73	\\
					&27- Lux $\gamma$ +10\%		&2.02	& -2.70	& -9.71	& 23.86	& -1.46	& -1.26	& 1.87	& -0.76	& 0.88	\\
					&28- Lux $\gamma$ -10\%		&-0.93	& 2.38	& 6.57	& -17.76	& 1.42	& 0.92	& -2.51	& 0.72	& -0.46	\\

			\hline
					&\bf Root sum sq. 1-10		&12.73	& 19.19	& 24.58	& 44.15	& 16.64	& 4.37	& 52.08	& 14.72	& 44.97	\\
			\hline
					&\bf Root sum sq. $\gamma$	&4.69	& 10.05	& 30.19	& 31.92	& 12.94	& 2.36	& 10.47	& 3.91	& 7.61	\\
			\hline
					&\bf Combined error		&31.35	& 34.58	& 47.65	& 63.16	& 42.79	& 43.84	& 57.87	& 24.98	& 52.01	\\

			\hline
			\hline
  		 \end{tabular}
    \caption{Sensitivity analysis on key technological parameters. Each number refers to a percentage change in CO$_2$ emissions or technological shares (share of the total fleet), in a scenario with parameter change, with respect to the same scenario without variation. Variations used are what we consider realistic uncertainty values. Changes in rates (e.g. learning rates) are percentage point changes (e.g. learning rate of 15\% changed to 20\%). Outcome changes on emissions are cumulated to 2050, while for shares the values are in 2050. Turnover rates represent the rate of replacement of the fleet. Note that when changing the turnover rates, the $\gamma$ values lose meaning and the projections \emph{do not match} trends observed in historical data. In other words, slower turnover rates cannot reproduce observed rates of technological change, and are included for reference, but are not realistic. The root of the sum of the squares of the variations are given for the technological parameters (rows 1-10) and all the individual $\gamma$ values (rows 13-28), excluding turnover rates. Root sum square values can be interpreted as combined uncertainty if the variations were normally distributed and known. Here, there are not precisely known or known to follow a particular distribution, and thus these values do not correspond to actual uncertainty, but instead, give an indication of model response to variations. The variations in rows 13-20 are shown graphically in fig.~\ref{fig:Sens_ba}.}
	\label{table:sensitivity}
\end{table}

In this section, we analyse responses of the model to changes in its key parameters. Key parameters of FTT:Transport include 
\begin{enumerate}
\item All cost parameters explicitly specified (vehicle prices, fuel costs, consumer discount rates, learning rates etc),
\item Technical parameters (fuel efficiency, emissions factors, lifetimes),
\item Non-pecuniary costs not explicitly specified (i.e. the $\gamma_i$ values).
\end{enumerate}
Some parameters are equivalent or related, for example emissions factors and the fuel efficiency. Some parameters are derived from data (e.g. $\gamma_i$ values, prices, emissions factors), while others are directly taken from the literature (e.g. learning rates, discount rates). Therefore, it is possible that some choice of parameters are non-sensical or lead to violating the assumptions of the model. This happens for instance if we vary vehicle prices without re-estimating $\gamma_i$ values, which makes diffusion trajectories in the simulation inconsistent with trajectories observed in historical data. Such scenarios are by definition not self-consistent and thus not realistic, but useful for analysing the properties of the model. 

With these issues in mind, it is important to analyse model responses to variations in key parameters, in order to ensure that the model is not `highly sensitive' to very specific values for any particular parameter. As a benchmark, if the outcome variation is less than the input variation, we assume here that the model is not `highly sensitive' to the particular values chosen. 

We varied 28 parameters or sets of parameters, by quantities that we consider either representative of uncertainty, or sensible values for observing model responses, shown in table~\ref{table:sensitivity}. We did this in both our `current trajectory' and our 2$^\circ$C scenarios. The reason for doing so is that policies constrain model evolution in particular directions, and thus one should not expect to find the same response (decarbonisation policies constrain outcomes more). Variations were carried out in all countries simultaneously, for simplicity, as we expect similar results to be observed at the regional level.

However, we did not vary the list of possible technologies, as is often done in cost-optimisation models. The reason for this lies with our modelling philosophy in order not to violate our assumptions. The reason for which we would not remove technologies is that FTT:Transport only models the diffusion of technologies already sold in the market, as evidenced by our historical database. It would not be consistent with reality to remove any of these. The reason for which we do not add technologies (e.g. `disruptive technologies') is due to our modelling time-frame: we already know from diffusion data, turnover rates and literature that technologies with vanishingly small market shares in the present day can only marginally change the market by 2050. Thus, sensitivity analyses that would consist of only adding a radically new technology at near-zero shares will by FTT construction make a difference of a marginal order to model outcomes. This is true unless specific action is taken, for example, with a relatively large government-led purchasing programme. Linking an ambitious programme of that kind with a new technology could indeed change outcomes to an appreciable degree; however we consider this within the realm of technology push policy, not a sensitivity analysis.

The parameters varied here are as follows:
\begin{enumerate}
\item \emph{ All $\gamma_i$ values simultaneously}; (for all vehicle types)
\item \emph{Learning rates of all vehicle types}; 
\\ Note that learning cost reductions for conventional vehicles are marginal even when changing rates, due to their existing market domination.
\item \emph{Consumer discount rates}; (equal for all vehicle types)
\item \emph{The price of electric vehicles}; 
\\ Note that changing vehicle prices without changes of $\gamma_i$ breaks FTT assumptions and leads to broken diffusion trends at the start of the simulation. Changes can made within fitting accuracy are acceptable. 
\item \emph{The fuel efficiency of new liquid fuel engines}; (ADV category). 
\\ This affects both emissions and the attractiveness of vehicles. 
\item \emph{The rate of vehicle purchases (turnover rates)}; 
\\ Note that, as noted in section~\ref{sect:decisionFreq}, these are not the same as the rates of scrappage, and in fact are not directly related. Turnover rates here mean the rate of acquisition of new vehicles, i.e. the rate of decision-making. Note further that changing turnover rates without changing $\gamma_i$ leads to highly inconsistent model definitions, and moreover, often leads to the inability to explain historical data in the case of fast diffusing technologies (such as electric vehicles). 
\item \emph{$\gamma_i$ values for EVs, Hybrids, CNGs, ADV and 2-Wheelers;};
\\ ADV stands for higher efficiency new internal combustion engines, while 2W are motorcycles. 
\item \emph{$\gamma_i$ values for vehicle class categories};
\\Note that these are changed for all vehicle types, for one engine size class at a time.
\end{enumerate}

We conclude with this analysis the following broad findings. (1) Learning and discount rates and vehicle prices have a relatively large impact on results, however, only of a similar order to the variations imposed; (2) Changing turnover rates has a relatively large impact on results (changes in outcomes much larger than changes in the parameters). (3) changing individual $\gamma_i$ values has, in general, little impact on overall results (much smaller than the variations themselves), but leads to some re-allocation of shares across alternate vehicle types or class;

(1) This result is self-explanatory. Outcomes will depend on the particular form chosen for calculating the generalised cost to the consumer. This includes the choice of form for discounting future costs. Indeed, no-one can truly know what thought process takes places in the minds consumers when choosing vehicles, and that furthermore, one cannot ascertain the degree of diversity of explicit or intuitive methods used by consumers when deciding (whether all consumers even use similar decision methods or criteria, see e.g. \cite{Knobloch2016}). Thus, decision-making criteria are not clear; however, it is important to remember that in the FTT formulation, all cost values (excepting $\gamma_i$) are distributed, and that these distributions are combined. This allows for some degree of flexibility of interpretation, as different formulations for decision-making will not lead to radically different outcomes if differences remain within variations already explicitly specified. For example, if consumer discount rates are in reality lower than explicitly specified, or of consumers do not in fact discount following the standard method, if the changes remain within the range specified in vehicle prices data (mean and STD values), little difference will be observed in model outcomes. 

\begin{figure}[p]
		\begin{center}
			\includegraphics[width=1\columnwidth]{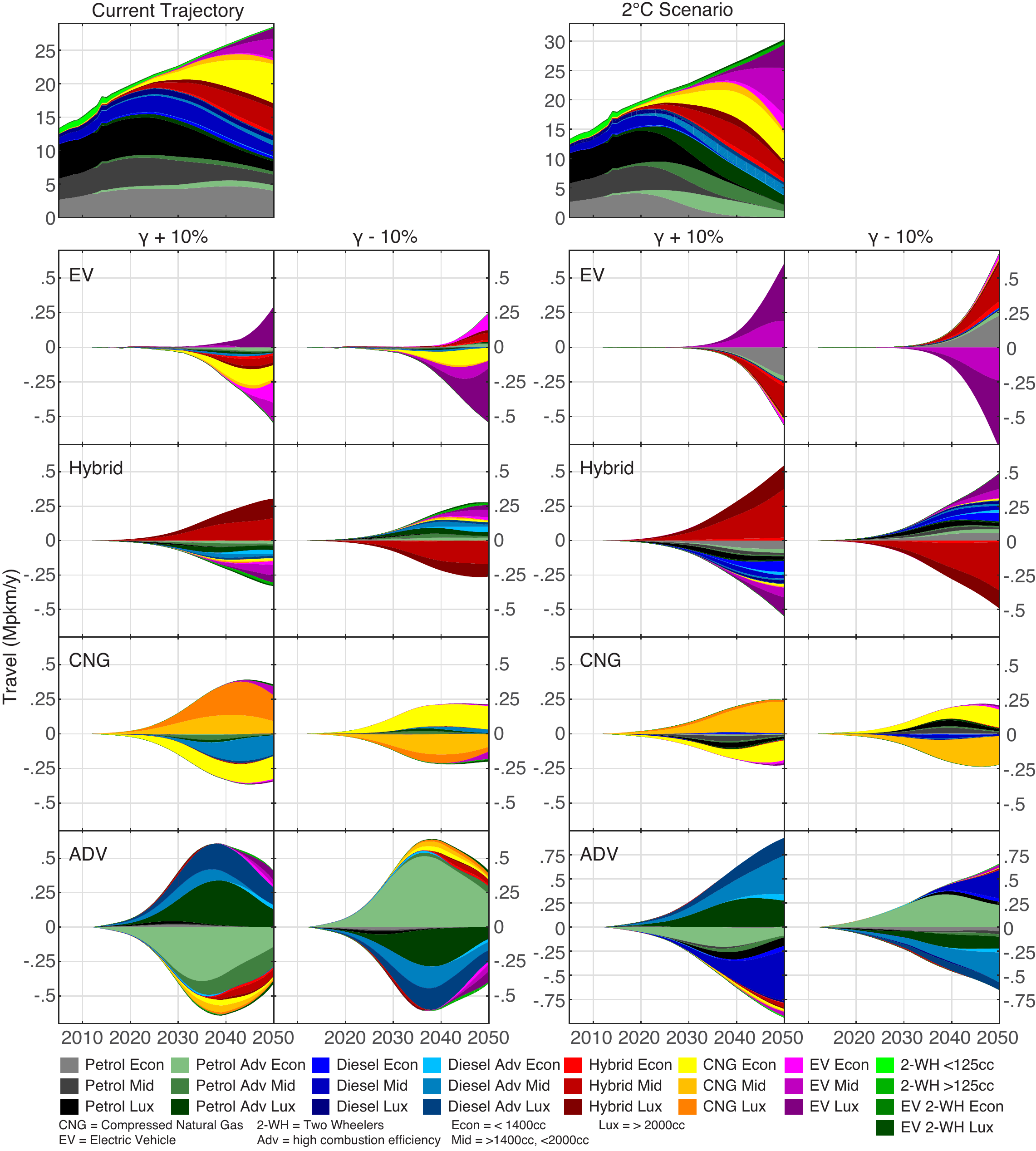}
		\end{center}
	\caption{Sensitivity analyses of rows 13-20 in table~\ref{table:sensitivity}, in which $\gamma$ values are changed by $\pm$10\% individually, for electric  (EV), hybrid, compressed natural gas (CNG) and new higher efficiency internal combustion engine (ADV) vehicles. Changes are expressed in terms of travel demand supplied by different vehicle types. To the left are variations applied to the 'current trajectory' scenario, while to the right are variations applied to a scenario with policies for achieving emissions consistent with a 2$^\circ$C scenario. The actual scenarios without variations are shown in the top row, while the changes resulting from changing $\gamma$ values are given below.}
	\label{fig:Sens_ba}
\end{figure}

(2) Turnover rates in FTT models are chosen in a way that (i) enables to fit the historical data, and (ii) is consistent with typical financial constraints. The reasoning behind this is that once costs are sunk, consumers are free to take new financial contracts. Here, it means that once a consumer has completed payments for a vehicle purchase (which typically takes 3-5 years), he is free, if he wishes to, to sell the vehicle and acquire a new one. However, it doesn't imply that he does so, and therefore, turnover rates in fact are upper limits to the rate of vehicle acquisition. In this, we assume that new vehicle markets are independent from second-hand vehicle markets, but that second-hand vehicle markets are `slave' to new vehicle markets (see section~\ref{sect:decisionFreq}). Changing turnover rates effectively implies changing the duration for which new vehicle consumers are financially constrained before they can replace their vehicle. This is not directly related to the lifetime of vehicles (or scrappage rates, which depend on a combination of failure rates and prices in the second-hand market). It also implies a different definition to $\gamma_i$ values, which are contingent to a turnover rate. Changing the turnover rates changes the pace at which technological change takes place; such changes are partly compensated by corresponding changes in $\gamma_i$.

Here, changes of turnover rates are carried out without changes of $\gamma_i$. This implies highly inconsistent diffusion trajectories between projections and historical data. However, these changes also imply large changes in model outcomes. We found that slower rates makes fitting $\gamma_i$ values often impossible in the cases of new technologies, in particular electric vehicles, as observed in historical data. It is clear that longer time-series would enable to better constrain turnover rates empirically. We therefore urge caution when assessing the impact of changing turnover rates in FTT:Transport. Here, the changes correspond to changing financial schedules length from 4 to 6 and 8 years. 

(3) Changing one value of $\gamma$ implies changing the attractiveness of one vehicle type, to the benefit or expense of all others. This leads, by construction, to violating the premise that diffusion trajectories are inferred from historical data. However, fitting $\gamma_i$ values is accurate only to a certain extent, which we estimate at between 5\% (established technologies) to 20\% (technologies with short time-series). As noted in section~\ref{sect:gammapractice}, estimating can be reliably be made with time-series as short as 5 years. Here, we vary the $\gamma_i$ by 10\% for each technology type. We find that outcomes vary by between 0 and 10\%. As intuitively expected, changes in the $\gamma$ for one technology mostly affects its own pace of diffusion. The relatively low impact of varying these parameters is  explained by the fact that these are done for individual technologies one at a time, which has relatively low impact overall in a multi-technology system: each alternative gains or loses a relatively small amount of market share. Changing several or all $\gamma_i$ simultaneously has a higher impact, however. Note that additively changing all $\gamma_i$ by a constant has no impact by construction, but multiplying them by a factor does, since it creates relative changes. 

\begin{figure}[h]
		\begin{center}
			\includegraphics[width=1\columnwidth]{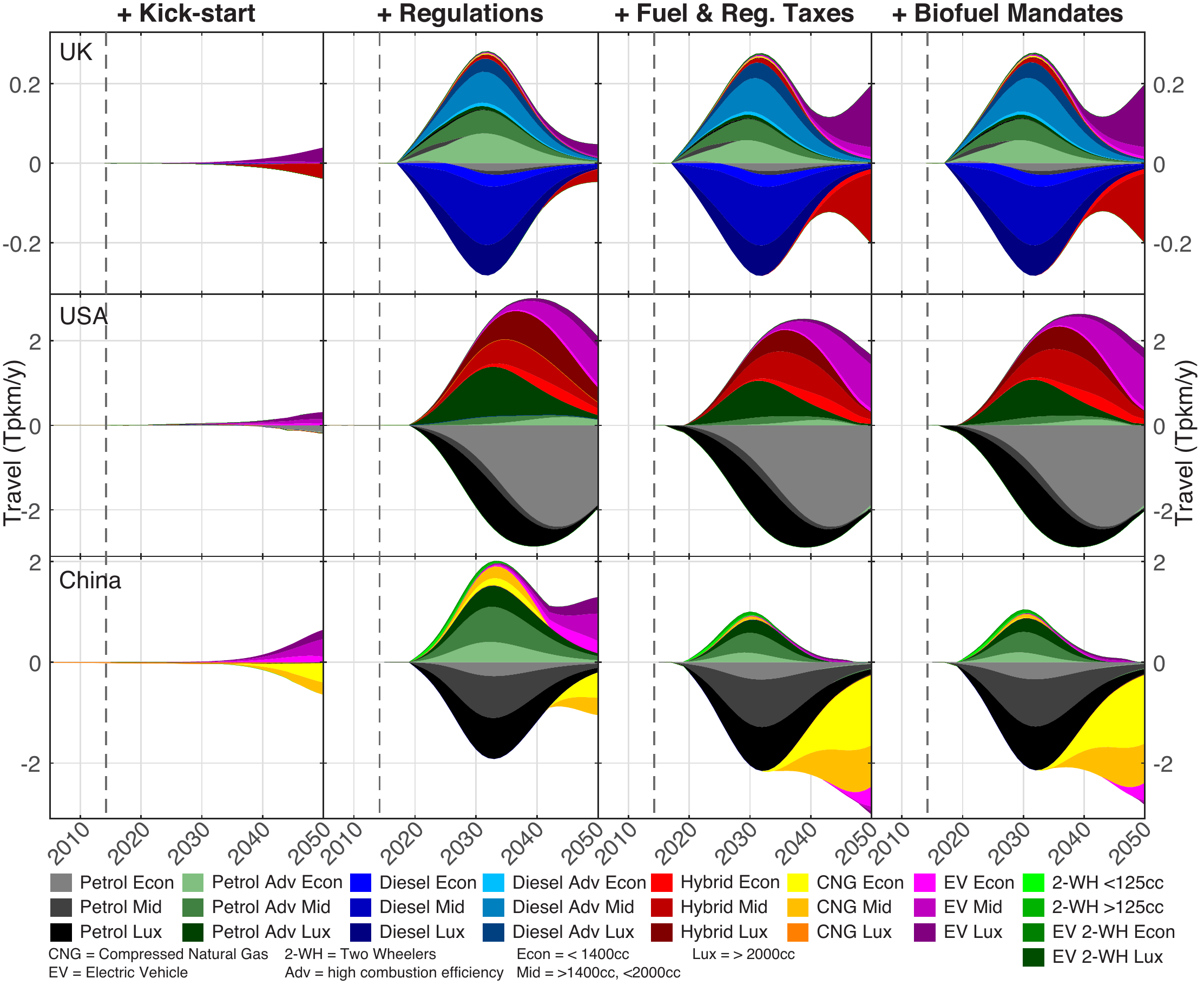}
		\end{center}
	\caption{Figure showing the differences from the between the panels of figure 2 of the main text. Differences are taken from the baseline. }
	\label{fig:Sens_policy}
\end{figure}

Figure~\ref{fig:Sens_ba} demonstrates this in more detail, where changes are shown over modelling time-span. It can be observed that variations increase exponentially towards the end, starting from zero uncertainty at the starting year, by construction. This is a reflection of `diverging pathways', which stems from the fact that FTT is a non-linear dynamical system (a complex system), akin to, for example, climate simulations \cite[see ][ for a discussion]{Mercure2016}. Figure \ref{fig:Sens_policy} shows differences observed between policy scenarios.

We conclude by observing the relative magnitude of combined errors, as given in table~\ref{table:sensitivity}. If the variations introduced corresponded to normally distributed measured standard errors, then they should be combined using the root of the sum of their squares, given in the bottom rows. Comparing scenarios requires a further combinations (bottom row). Here, errors are not measured, and instead, we use estimated variations that we consider reasonable. Thus, the combined error is not to be interpreted as uncertainty; however it does provide a guide to assessing model behaviour. We observe combined variation of up to 30-40\% in individual scenarios when all variations are taken into account simultaneously (when excluding changes to the turnover rate, for the reasons given above). This shows that even for extensive parameter changes, outcome scenarios are still fairly closely related to each other. This shows that the model is not dependent on any particular parameter. When comparing pairs of scenarios, up to 60\% of variations are observed. This is an indication of the uncertainty as faced by, for example, policy-makers, knowing that the current trajectory as well as the policy objectives are subject to fundamental uncertainty.

\subsection{Comparison to other modelling approaches}

Behavioural aspects and heterogeneity significantly impact the effectiveness of market-based policies \citep{Knobloch2016}. However, with the scale and time frames of IAMs, many IAMs have a simplified representation of consumer choice \citep{Pettifor2017a}. Global Integrated Assessment Models (IAMs) have been criticised for lacking behavioural realism \citep{mccollum2016improving}. Consumer choices and behaviour are not sufficiently taken into account in most global IAMs, despite the fact that there is extensive evidence that consumer preference (e.g. non financial incentives) and heterogeneity (e.g. income and driving pattern) are the key drivers in the diffusion of vehicles \citep{mccollum2016improving}. Since modelling consumer heterogeneity is crucial  in determining the rate of diffusion for technologies, the general lack of focus of consumer heterogeneity limited the general usefulness of IAMs advising policies \citep{Pettifor2017a}. 

In response to the general lack of consumer heterogeneity in the existing IAMs, several IAMs and studies have included different level of consumer diversity in their modelling approach \citep{Pettifor2017a,mccollum2016improving,karkatsoulis2016simulating,mittal2016key}. \citep{Pettifor2017a} used IMAGE and MESSAGE model to analyse the interactions between consumer preference and social influence in the transport sector. \cite{mccollum2016improving} examines the impact of introducing consumer heterogeneity and choices into the MESSAGE-transport model. \cite{karkatsoulis2016simulating} includes values that represent preferences in vehicle types within a GCE framework (GEM-E3T). \cite{mittal2016key} uses AIM/transport model (with a special focus on Asia Pacific) to assess the impact of various factors such as travel time, energy efficiency improvement, load factor mode preference along with environmental awareness factors on transport demand, energy and emissions. 

The existing studies improve the representation of consumer behaviour in global IAMs by considering some degree of heterogeneity and non-pecuniary factors that influence the consumer decisions.  However, there is a main differences in modelling approach between FTT model and optimisation models used by the above studies. Firstly, the \citep{mccollum2016improving} introduced consumer heterogeneity into IAMs by separating adopters into groups, settlement pattern and vehicle usage intensity. This introduces consumer heterogeneity into the IAM MESSAGE with 27 different consumer groups. In the FTT model, within each technology, instead of having a mean representative agent (such as in MESSAGE), consumer diversity is represented by a distribution of vehicle prices (see for example figure~\ref{fig:2-1}), which we assumed to reflect the distributed willingness to pay for different car models by consumers. Hence, the heterogeneity is represented by the the distribution of car prices, not a representative agent. While both the \citep{mccollum2016improving} and our approach improve heterogeneity in representing consumers in the IAM, there are two main differences to results regarding available of technologies and the rate of technological diffusion. To illustrate, in the FTT model, with its historical market share database for each technology, reflects existing trends in different group of consumers and the availability and popularity of particular technologies in a particular country. With a different approach to heterogeneity, the rate of technological diffusion is expected to the different, and generate different results as to the rate of  technology diffusion in each country. 

Indeed, what is observed is that technological change takes place faster in FTT than in the new version of MESSAGE \citep[as seen in ][]{Pettifor2017a}. In particular, EVs emerge in FTT with a sizeable market share in 2050, which happens in MESSAGE only nearer to 2100. We also observe that the diffusion of higher efficiency vehicles result in oil demand peaking before 2050, something not observed in MESSAGE. In FTT, the interpretation of this discrepancy is that the FTT diffusion trends are observed in historical data in recent years, and are projected, using the $\gamma_i$ parameter described in section \ref{sect:gammapractice}. This empirical observation helps ground the model into reality. We therefore argue that, while our method features inherent uncertainty that grows with simulation time span, improving optimisation models by adding ever more detailed cost data remains insufficient to explain current technology diffusion trends. Using a model grounded in empirical observation is likely to show results closer to what may unfold in the near future. 

\subsection{Validity range of the model}

In the tradition of time series econometrics, the usual rule of thumb is that a reasonable validity range in time for an econometric model is to project for as many years forwards as one has years of historical data. This may be correct for linear econometrics models. But it is not for non-linear dynamical models such as FTT. We re-state and discuss the fundamental assumptions of the model:

\begin{enumerate}
\item The diffusion trend of technologies in the model should not be broken at the start of the simulation solely because it is the start of the simulation -- real-world fleet numbers do not change so suddenly since vehicles remain used for 11-12 years on average, which generates substantial inertia. Therefore, we assume, the diffusion process (the number of vehicles and its first derivative) should be continuous across the transition between history and simulation, and this continuity can readily be observed in the data. This requires determining gamma values that ensure this is the case. This is the sole role of the gamma values.

\item The meaning of the gamma values can be interpreted in terms of the non-pecuniary value that agents ascribe to particular types of vehicles. In a pure cost-optimisation without non-pecuniary values, only the very lowest cost vehicles would diffuse, but this is of course not observed in reality (e.g. in our diffusion data). Gamma values are mostly needed to address price differences between our vehicle classes (e.g. Economic $\sim$\$15k, Mid-range $\sim$\$35k, Luxury $\sim$\$50k), expressing the value that agents ascribe to them. For the model not to converge to lowest-cost vehicles only, one expects gamma values to be of a similar order of value as prices if all vehicle classes are seen to diffuse in the data consistently with recent history. This simply reflects the fact that the non-pecuniary value ascribed by agents to e.g. luxury cars is higher than that ascribed to e.g. economic cars. Similarly, if low-carbon vehicles are observed to diffuse in the historical period, we assume that agents ascribe to them a certain non-pecuniary value consistent with that diffusion. Being constants, gamma values also do not dominate or determine the calculation; it is changes in costs that determine outcomes. It is therefore not sensible to radically change gamma values (e.g. set them to zero) in the hope of carrying out a meaningful sensitivity test, as it is clear that the model would then generate nonsense, not being true to its own definition (i.e. pure pecuniary cost-optimisation, which is not observed in reality). 

The gamma values implicitly include the effect of current policies (e.g. taxes and rebates) applied to each vehicle type. It is possible that the policy may have changed during the historical period, in which case the gamma value determination will generate an average policy value over the historical period. However, we do not observe substantial sudden changes in vehicle numbers in our historical data (except for cases in which vehicle type definitions may have changed, or if radical policies were adopted shortly before the model start date). Therefore we do not consider correct to think that the determination of gamma values is made substantially fragile by changes in the policy regime that have taken place over our historical dataset. Furthermore, policies defined in the model are additional to existing policies implicitly included in the gamma value, not absolute. 

Policies may also have changed after the start of the simulation, in which case they would need explicit representation. We acknowledge that this could influence the outcomes of our model. However, as is the case for any model, addressing this requires to survey all policy changes in every world region of the model in detail, a substantially large task which is not the subject of this work. Including all existing policies in model baselines is a large task we believe very few modellers have carried out (with the notable exception of the IEA). We are making progress towards this for some important regions and hope to publish this separately.

\item The model by construction features substantial inertia, consistent with a substantial empirical literature, which means that radical changes cannot happen even for relatively large sudden changes in the data (e.g. costs or policies). Assuming that sensible gamma values are used that ensure that the diffusion trend is not broken due to the start of the simulation (as we cannot allow that the start year of the simulation should itself generate a discontinuity in sales in that year), inertia in the model ensures that the observed diffusion trajectory is consistently followed in early simulation years (i.e. the first decade), and departs from that in subsequent years due to the policy context.

The sensitivity analysis provided in the first review round shows a relatively low influence of changes in gamma values on diffusion trajectories. This applies in particular for the first simulated decade, in which the model inertia rules out radical changes in the diffusion trajectory. This so-called `strong autocorrelation in time' (an autocorrelation time span of order 10 years) in the model implies that the model can be used to project longer time spans than what is suggested by the standard rule of thumb for linear econometric models ($\sim$10 years), perhaps three times longer than the autocorrelation time, with uncertainty increasing over time (as in climate models). We do not, however, model beyond 2050, since clearly by then new innovations not foreseen by us nor specified in the model may start diffusing, or the gamma values will have changed, and thus the uncertainty is too large. Due to our model structure, changes in gamma values that would lead to substantially different projections have to be very large (changes $>$ 20\%, e.g. setting them to zero), and would also lead to highly broken diffusion trends at the start of the simulation, something that violates our assumption (1) and substantial amounts of empirical work. In simpler words, the autocorrelation in time of around 10 years is due the fact that vehicles survive for around 10 years, which allows us to model longer time spans. This can be observed in Fig. 2 of the main text.

We finally note that the model equations provide a structure that needs to be calibrated (as any model normally does), but that is driven by internal dynamics (logistic diffusion) as much as by data and exogenous policy assumptions. The gamma value plays the role of a calibration parameter, similar to standard calibration procedures in other models, in order to match real world quantities. Meanwhile, policy assumptions are what steers model direction. We conclude by stating that the model validity range is of around 30 beyond the start date of the simulation, and therefore modelling to 2050 roughly matches the validity range, even if historical data used to calibrate the model is much shorter. The uncertainty range increases with the length of the projection; and the sensitivity analysis given above indicates what that range is. Meanwhile, uncertainty decreases when more recent historical data is added.

\end{enumerate}


\newpage
\appendix

\renewcommand\thesection{\Alph{section}}

\section{The addition of probabilistically distributed values \label{sect:AppendixA}}

\subsection{The addition is a convolution}

In this project comes periodically the problem of adding values which have either probability distributions or that are really distributed in reality. The sum of such distributed quantities must also be distributed, and we know intuitively that the resulting distribution is a function of the original distributions, and that its \emph{width} should be roughly the sum of the \emph{widths} of the initial distributions. This is not far from the truth, but it can be derived formally. 

\begin{figure}[h]
		\begin{center}
			\includegraphics[width=.4\columnwidth]{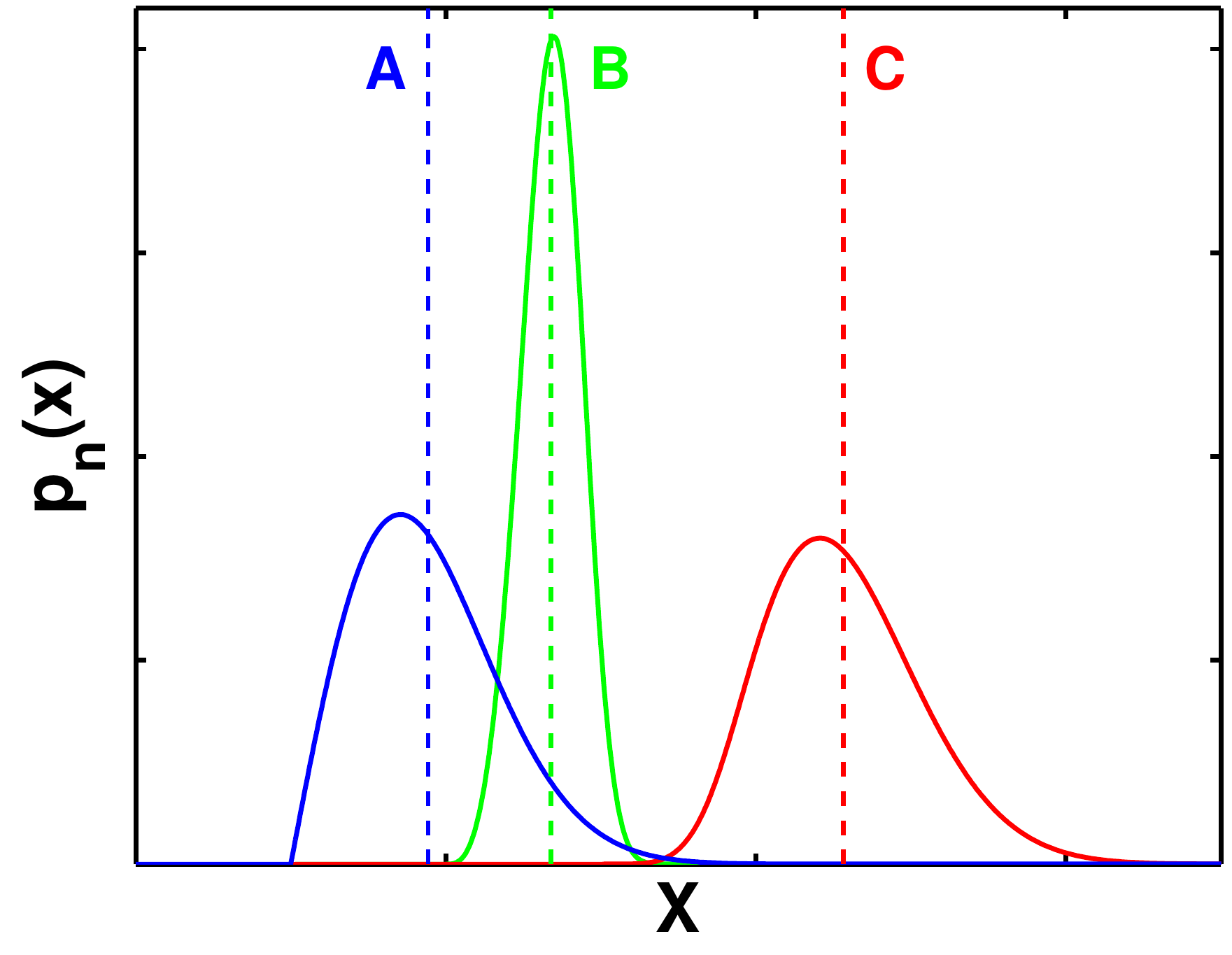}	
		\end{center}
	\caption{The addition of probabilistically distributed quantities. The red distribution is the convolution of the blue and the green, and the mean value of the sum, the value $C$, is the sum of the mean values of the other two, $A$ and $B$.}
	\label{fig:Convolutions}
\end{figure}

Imagine that we have two quantities with the same units, $A$ and $B$, distributed on the $x$ axis following the probability distributions $p_A(x)$ and $p_B(x)$ (e.g. on a cost axis, $A$ and $B$ could be prices of particular items with a stochastic probability or a real distribution corresponding to real occurrences of these prices). If the values $\overline{A}$ and $\overline{B}$ are the \emph{mean} values, and if the probability distributions are well defined, we expect the median of the sum to be $\overline{C} = \overline{A} + \overline{B}$, corresponding the the sum without distributions or uncertainty. But what about the distribution of the sum itself? And what about the width of the distribution of the sum? We can derive it here. Fig.~\ref{fig:Convolutions} shows this schematically.

To derive this, we first imagine that quantity $B$ is not distributed but has a single value $x=B$. Given that quantity $A(x)$ is distributed, the sum of $A(x)$ and $B$ must be
\beq
p_C(x) = p_A(x-B),
\eeq
which is the distribution of the quantity $A$ centred at a new value equal to $\overline{A} + B$. We can now add the uncertainty or distribution of the value $B$: each value of $p_C(x)$ as stated above has a probability $p_B(x)$ of occurring, which must be summed:
\beq
p_C(x) = \int p_A(x-B)p_B(B) dB.
\eeq
This is a convolution of $p_A(x)$ with $p_B(x)$, denoted $p_C(x) = p_A(x) \otimes p_B(x)$. Thus we see that the addition of distributed quantities is the convolution of the distributions. One can show that the mean of the result ($C$ in the graph) is equal to the sum of the means($A$ and $B$ in the graph). The width is the width of the resulting convolution, which cannot be expressed simply unless the distributions are specified. 

\subsection{Case for normal distributions: standard error analysis}

We define $p_A(x)$ and $p_B(x)$ as normal distributions with standard deviations $\sigma_A$ and $\sigma_B$ and means $\overline{A}$ and $\overline{B}$:
\beq
p_A(x) = {1\over \sqrt{2\pi} \sigma_A} e^{-\dfrac{(x-\overline{A})^2}{2 \sigma_A^2 }} \quad p_B(x) = {1\over \sqrt{2\pi} \sigma_B} e^{-\dfrac{(x-\overline{B})2}{2 \sigma_B^2 }}
\eeq
Their convolution can be shown to be a normal distribution (it's a bit tedious)\footnote{The gaussian is the only function that has the property that maintains its functional form through convolutions, except possibly the Dirac Delta function.}:
\beq
p_C(x) = {1\over \sqrt{2\pi (\sigma_A^2 + \sigma_B^2)}} e^{-\dfrac{(x-\overline{A}-\overline{B})^2}{2 (\sigma_A^2+\sigma_B^2) }}.
\eeq
We see that the new distribution $p_C(x)$ has a mean and standard deviation as follows:
\beq
\overline{C} = \overline{A} + \overline{B},
\eeq
\beq
\sigma_C = \sqrt{\sigma_A^2 + \sigma_B^2}.
\eeq
This demonstrates the standard root of the sum of the squares of the errors in uncertainty analysis, which requires the assumption of normal distributions. 

\subsection{Error propagation}

Standard error analysis in the experimental sciences dictates how one should combine uncertainty values from different sources across a function, as we do in this work for the calculation of the LCOT and its deviation $\Delta$LCOT. It also corresponds to the result of convolving a series of distributions to one another. If we have a function of several variables $f = f(x,y,z,...)$ (as in our LCOT in this work, which depends on car prices, fuel costs, O\&M costs, all of which are distributed), then the propagation of uncertainties on all variables, $\Delta x$, $\Delta y$, $\Delta z$, ... combine to an uncertainty on $\Delta f$ following:
\beq
\Delta f = \sqrt{\left({\partial f \over \partial x}\right)^2 \Delta x^2 + \left({\partial f \over \partial y}\right)^2 \Delta y^2 + \left({\partial f \over \partial z}\right)^2 \Delta z^2 + ...}
\label{eq:propagation}
\eeq

\section{List of variables \label{sect:AppendixListVar}}

\begin{tabular}{l l}
 $\Delta t$ & Time interval \\
 $t$ & Current year \\
 $D$ & Transport services demand, v$\cdot$km/y (vehicle-kilometres per year) or t$\cdot$km/y (tons-kilometres per year)\\
 $N$ & Total transport capacity for one particular service type, in v or t (vehicles or tonnes)\\
 $G_i$ & Transport services generation by technology, p$\cdot$km/y or t$\cdot$km/y\\
 $U_i$ & Transport capacity by technology, p (persons or seats) or t (tons)\\
 $FF_i$ & Filling factor, fraction of seats occupied or weight capacity used, in persons per seats (i.e. no units)\\
 $d_i$ & average distance travelled per year by one vehicle of type $i$, km/y\\
 $S_i$ & Technology share of total vehicle capacity\\
 $F_{ij}$ & Investor choice probability matrix\\
 $A_{ij}$ & Technology changeover timescales matrix\\
 $G_{ij}$ & Constraints matrix\\
 $CF_i$ & Capacity factor, km/y\\
 $\overline{CF}$ & Share weighted capacity factor, km/y\\
 $E_i$ & Emissions of pollutants in a year, t(pollutant)/y (tons of pollutants per year)\\
 $J_i$ & Fuel use by technology, in GJ/y\\
 $\alpha_i^k$ & Emission factor of technology $i$ for pollutant $k$, t(pollutant)/GJ (tons of pollutant per energy content)\\
 $\beta_i$ & Vehicle fuel consumption, GJ/km\\
 $I_i$ & Vehicle cost, \$/seat\\
 $W_i$ & Cumulative vehicle sales, in seats\\
 $B_{ij}$ & Spillover learning matrix, linking technologies with similar components\\
 $C_i$ & Generalised cost of transportation, in \$/pkm\\
 $f_i(C-C_i,\sigma_i)$dC & Distribution of vehicle purchases with average $C_i$\\
 $F_(C-C_i)$ & Cumulative distribution of vehicle purchases below price $C$\\
 $\mu_i$ & Average in normal cost space\\
 $\sigma_i$ & Standard deviation in normal cost space\\
 $m_i$ & Average in lognormal space\\
 $\nu_i$ & Standard deviation in lognormal space\\
 $LCOT_i$ & Levelised cost of transport, in \$/pkm\\
 $\Delta LCOT_i$ & STD Levelised cost of transport, in \$/pkm\\
 $\gamma_i$ & Empirical average value of the `intangibles', in \$/pkm\\
 $FT_i$ & Fuel tax in \$/pkm\\
 $OM_i$ & Operation and Maintenance costs, in \$/pkm\\
 $RT_i$ & Road tax costs, in \$/pkm\\
 $r$ & discount rate\\
 $n_i(a,t')\Delta a$ & Age distribution of the vehicle fleet of technology type $i$\\
 $a$ & Vehicle age in years\\
 $t'$ & Vehicle year of production\\
 $\ell_i$ & Survival function\\
 $p_i$ & Vehicle instantaneous force of death\\
 $\tau_i$ & Vehicle life expectancy\\
 $\xi_i$ & Sales/registrations of new vehicles, in seats\\
 $N_i$ & Total number of vehicles in the fleet of technology type $i$\\

\end{tabular}

\clearpage
\section*{References}
\bibliographystyle{elsarticle-harv}
\bibliography{FTT-TransportRefs.bib}

\end{document}